\documentclass[prb,showpacs,floatfix,superscriptaddress,twocolumn,aps,10,groupedaddress]{revtex4-2}
\usepackage[english]{babel} 
\usepackage{bm}
\usepackage{graphicx}
\usepackage{amsmath}
\usepackage{amssymb}
\usepackage{wasysym}
\usepackage{comment}
\usepackage[dvipsnames]{xcolor}
\usepackage{tikz}
\usepackage{adjustbox}
\usepackage{booktabs}
\usepackage{array}
\usepackage{multirow}
\usepackage{lipsum}   
\usepackage{makecell}
\usepackage{enumitem}   

\newcommand{\octahedron}{%
  \mathrel{\raisebox{-0.2ex}{%
    \begin{tikzpicture}[scale=0.07, line width=0.4pt, baseline]
      \coordinate (C) at (0,0);
      \coordinate (L) at (-2,1);
      \coordinate (T) at (0.7,3.4);
      \coordinate (R) at (3.2,0.6);
      \coordinate (B) at (0.7,-1.8);
      \coordinate (Ca) at (1.4,1.6); 

      \draw (L)--(T)--(R)--(B)--cycle;
      \draw (L)--(C)--(R);
      \draw (B)--(C)--(T);
      \draw[dash pattern=on 1pt off 1pt] (L)--(Ca)--(R);
      \draw[dash pattern=on 1pt off 1pt] (B)--(Ca)--(T);
    \end{tikzpicture}%
  }}%
}

\usepackage[hidelinks]{hyperref}

\begin{document}

\title{Interacting-cluster spin liquids with robust flat bands evolving\\ into higher-rank half-moon phases and topological Lifshitz transitions}

    \author{Naïmo Davier}
	\email[]{naimo.davier@u-bordeaux.fr}
    \affiliation{CNRS, Universit\'e de Bordeaux, LOMA, UMR 5798, 33400 Talence, France}

    \author{Ludovic D.C. Jaubert}
	\email[]{ludovic.jaubert@u-bordeaux.fr}
    \affiliation{CNRS, Universit\'e de Bordeaux, LOMA, UMR 5798, 33400 Talence, France}
		
	\date{\today}

\begin{abstract} 
    Classical spin liquids are disordered magnetic phases, governed by local constraints that often give rise to flat-band ground states. When constraints take the form of a zero-divergence field within a cluster of spins, the spin liquid is often described by an emergent Coulomb gauge theory. Here we introduce an interaction $\eta$ between these clusters of spins which compete with the zero-divergence field. Using a framework embracing both the connectivity matrices of graph theory and the topology of band structures, we develop a generic theory of interacting-cluster Hamiltonians. We show how flat bands remain at zero energy up to finite interaction $\eta$, until a dispersive band becomes negative, stabilizing a spiral spin liquid with a hypersurface of ground-state manifold in reciprocal space. This hypersurface can be interpreted as an effective Fermi surface in the spectrum of the parent system, acting as a tunable energy selector despite the absence of particle filling. This effective Fermi surface serves as a mold for the apparition of the half-moon patterns in the equal-time structure factor. Our generic approach enables to extend the notion of half moons to the perturbation of higher-rank Coulomb fields and pinch-line spin liquids. In particular, multi-fold half moons appear when unconventional gauge charges, such as potential fractons, are stabilized in the ground state. Finally, half-moon phases can be tuned across the equivalent of a Lifshitz transition, when the hypersurface manifold changes topology.
\end{abstract}
\maketitle

\section{Introduction}

Classical spin liquids emerge in geometrically frustrated magnets where conventional magnetic ordering is suppressed, even at zero temperature. Instead of long-range order, these systems exhibit highly degenerate ground-state manifolds governed by local constraints, which often take the form of flat bands in reciprocal space. When these constraints are written as a divergence-free condition -- analogous to a Gauss law --they give rise to an emergent gauge structure reminiscent of classical electromagnetism \cite{Henley2005, henley2010coulomb}. This leads to distinctive features such as algebraically decaying spin-spin correlations and characteristic pinch points in the structure factor, which are directly observable in neutron scattering experiments \cite{castelnovo08a, Gauss_pyro}. Paradigmatic examples include the nearest-neighbor Heisenberg antiferromagnet on the pyrochlore and kagome lattices \cite{Anderson_pyro, kagome_largeN}.

Pinch points in the equal-time structure factor are often accompanied by half-moon patterns at finite energy coming from the lowest dispersive band \cite{robert08a,guitteny13a,Yan_2018,Mizoguchi_2018,zhang19a,Yan_moon_2024}. Flat-band engineering \cite{Mizoguchi_Masafumi_2019_flat_bands,Rhim_2019,Graf_Piechon_2021_flat_bands,Maimaiti_2021,Roychowdhury_2024,Udagawa_2024,Nakai_2025} can modify the band spectrum in order to destabilize the spin liquid and make the half moons a signature of the ground state in the equal-time structure factor on kagome and pyrochlore systems \cite{Mizoguchi_2018,Mizoguchi_Masafumi_2019_flat_bands}.

Using tools recently developed for the classification of classical spin liquids \cite{Benton_Moessner_2021, Yan_2024_short, Yan_2024_long, Davier_2023,Fang_2024} together with graph-theory methods based on the connectivity matrix \cite{katsura10a,Essafi_2017,Mizoguchi_2018,Mizoguchi_Masafumi_2019_flat_bands}, we build a generic theory of interacting-cluster systems for a seemingly infinite variety of lattices and clusters in two and three dimensions (or higher). This interaction $\eta$ creates an energetic competition with the zero-divergence field required by the Coulomb gauge theory, and will be our flat-band engineering parameter. We quantify the stability of the flat-band spin liquid until $\eta$ reaches a critical value, where the energetic competition ultimately brings the lowest dispersive band below the flat bands. 
This crossing selects a hypersurface of parent-system states that are promoted to the ground state of the interacting system, defining an interaction-tunable effective Fermi level despite the absence of particle filling in spin systems. At this point, pinch points transform into half moons, the signature of a spiral spin liquid where the magnetic texture of gauge charges is stabilized in the ground state. Then we extend the notion of half moons to higher-rank U(1) gauge fields, including models with pinch lines. Since multifold half moons come from the dispersive band of a parent rank$-n$ Coulomb field, they represent the signature of rank$-n$ gauge charges, such as fractons, in the ground state. Finally, we show how half-moon phases can be adiabatically tuned across the equivalent of a Lifshitz transition separating distinct spiral spin liquids.

This paper is organized as follows.
Section \ref{Sec: Cluster Hamiltonians} defines cluster systems and characterizes them using the constraint-vector (\ref{Subsec: constraint vector fiormalism}) and connectivity-matrix (\ref{sec:connec}) formalisms. We then outline the Coulomb phase that emerges in these systems (\ref{sec:pinchpoint}) and discuss a simple criterion for when this description is expected to break down (\ref{sec:flatbandliquid}).
Section \ref{Sec: Interacting clusters Hamiltonian} introduces the interacting-cluster Hamiltonian, presenting its general properties (\ref{Subsec: Generic properties}) and the band structure of a subclass of interacting cluster systems (\ref{Sec:unifcluster}).
In Section \ref{Sec: Half moons}, we show how this band structure produces half-moon patterns in the structure factor when the cluster–cluster coupling $\eta$ is sufficiently strong (\ref{Subsec: Half moon, general theory}). This leads to the concepts of high-rank half-moons and half-moon surfaces associated with high-rank classical spin liquids (\ref{subsec: interacting cluster model for high rank h-m}), which we illustrate through two examples (\ref{Subsec: Example of the checkerboard}, \ref{Subsec: Example of the octochlore}).
Finally, we conclude with a description of the topological Lifshitz transitions that occur beyond the half-moon phase (\ref{Subsec: Topological Lifshitz transition}).



\section{Cluster Hamiltonians}
\label{Sec: Cluster Hamiltonians}

\begin{table*}[t]
    \centering
    \renewcommand{\arraystretch}{1.5} 
    \resizebox{\textwidth}{!}{
    \begin{tabular}{c c c c c c c c c c}
        \hline
        \textbf{2D Lattice} & Kagome & Checkerboard & Hexagonal & \makecell{Kagome\\ hexagonal} & \makecell{Square\\ octagon} & Ruby & Square kagome & \makecell{Decorated\\ square kagome}  & \makecell{Octagonal\\ kagome} \\
        \hline
        \noalign{\vskip 1mm}
        \makecell{ \vspace{-2.5cm} \\ \textbf{Lattice}  \\ \textbf{Scheme} } & 
        \includegraphics[height=2.5cm]{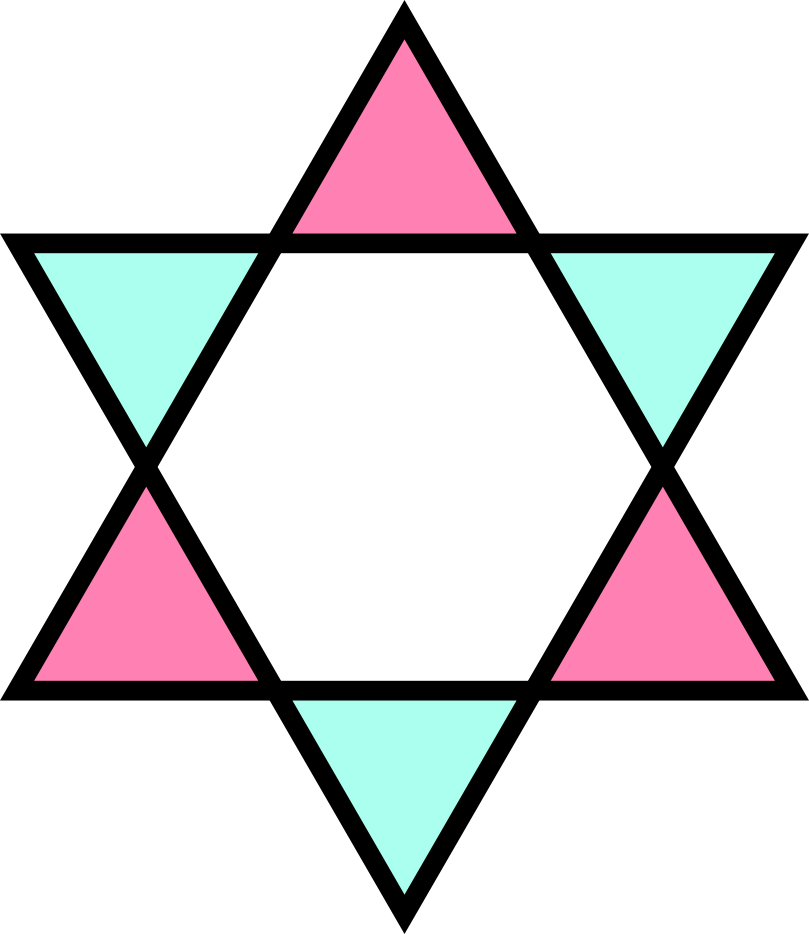} & 
        \includegraphics[height=2.5cm]{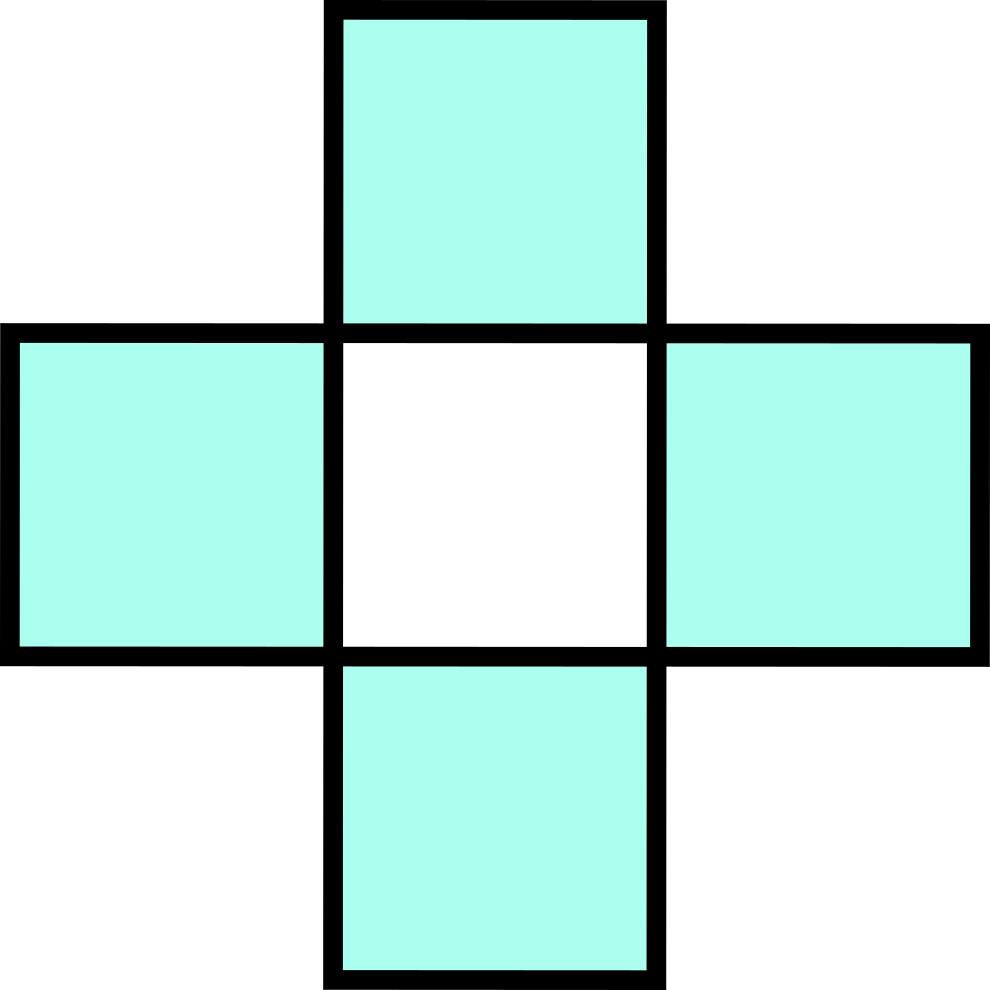} & 
        \includegraphics[height=2.5cm]{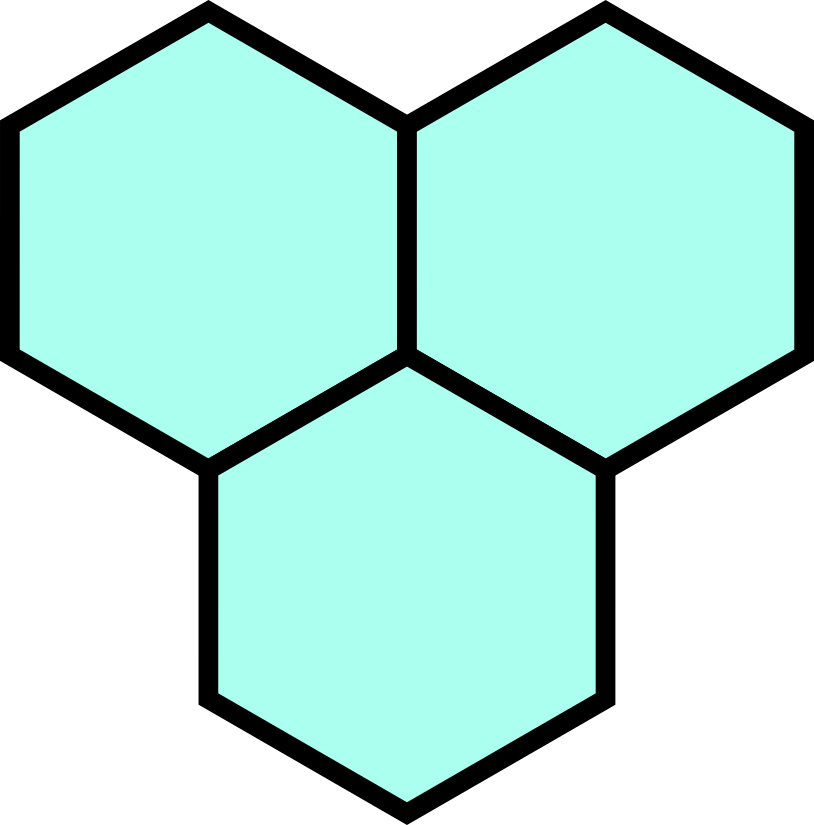} & 
        \includegraphics[height=2.5cm]{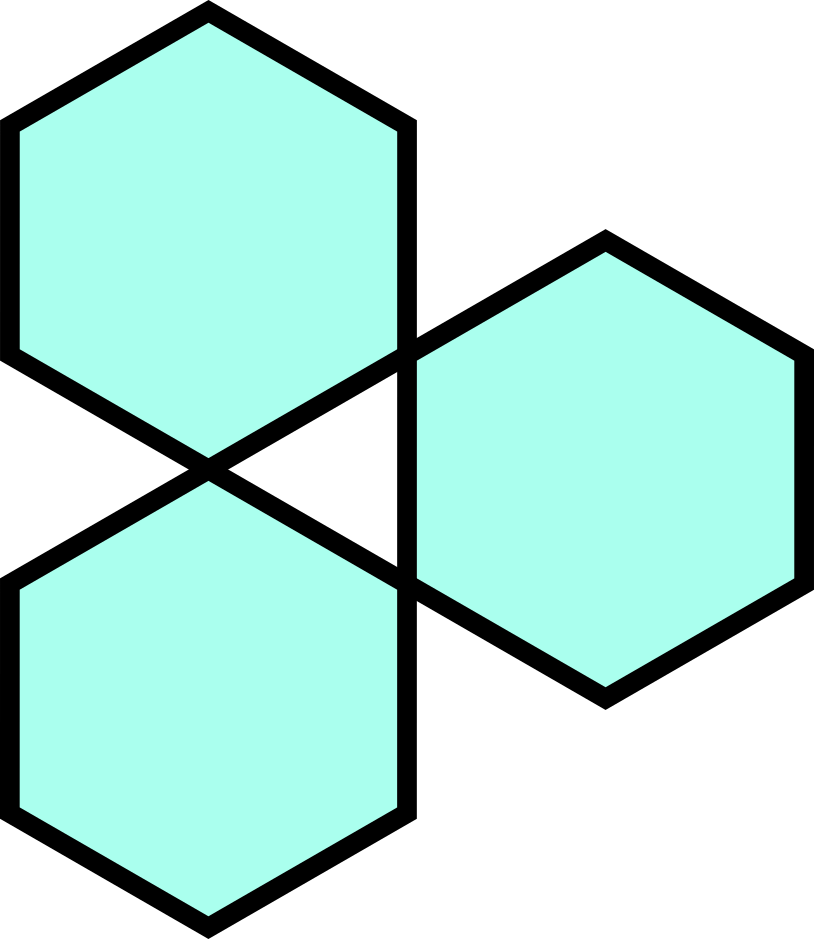} & 
        \includegraphics[height=2.5cm]{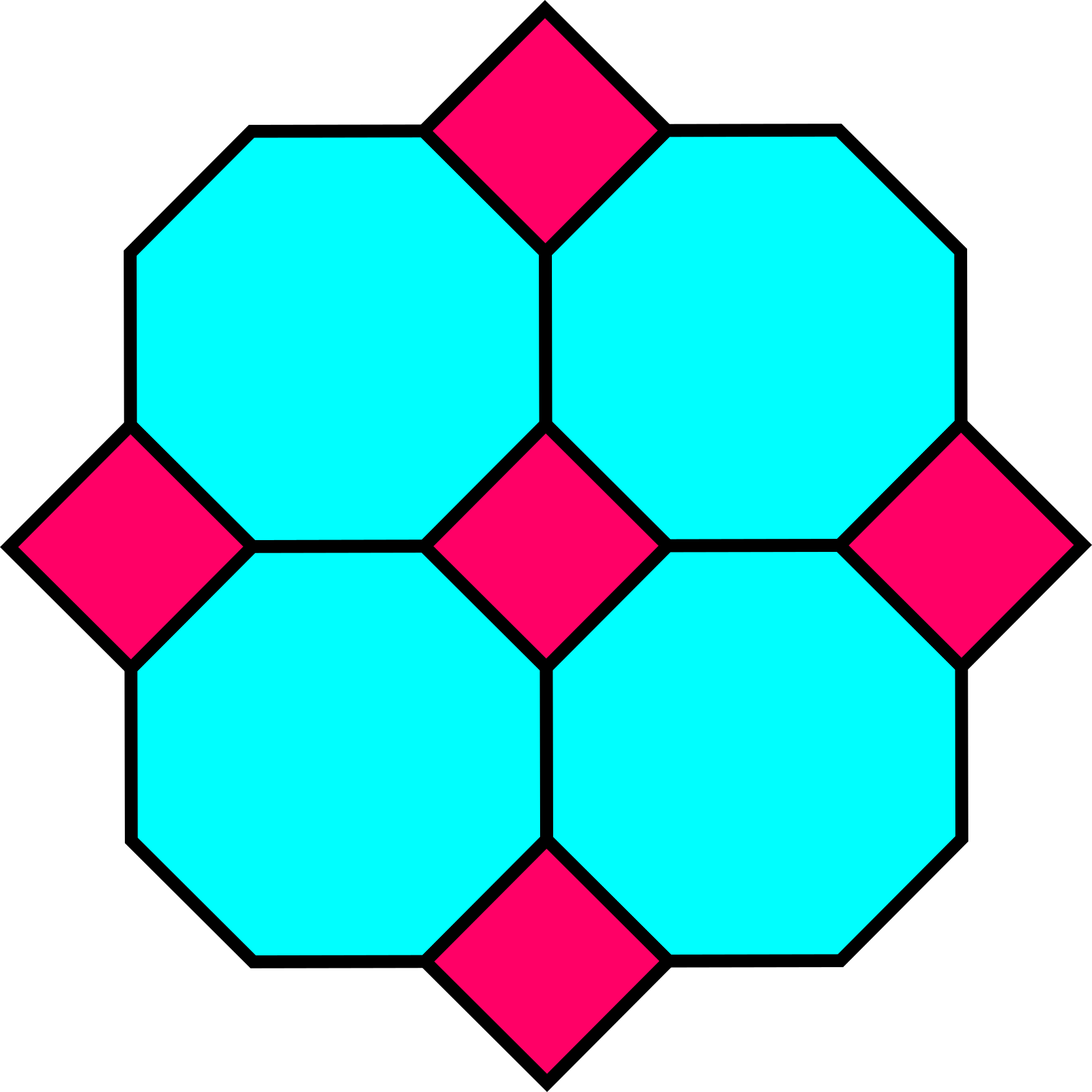} & 
        \includegraphics[height=2.5cm]{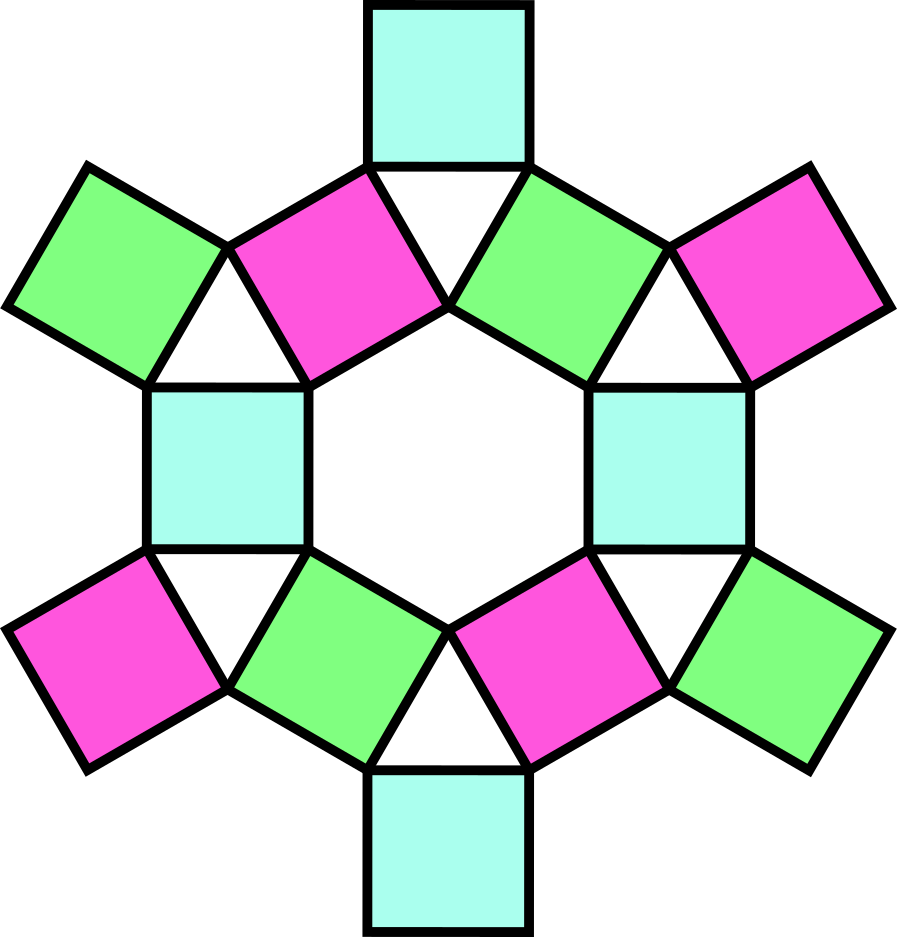} & 
        \includegraphics[height=2.5cm]{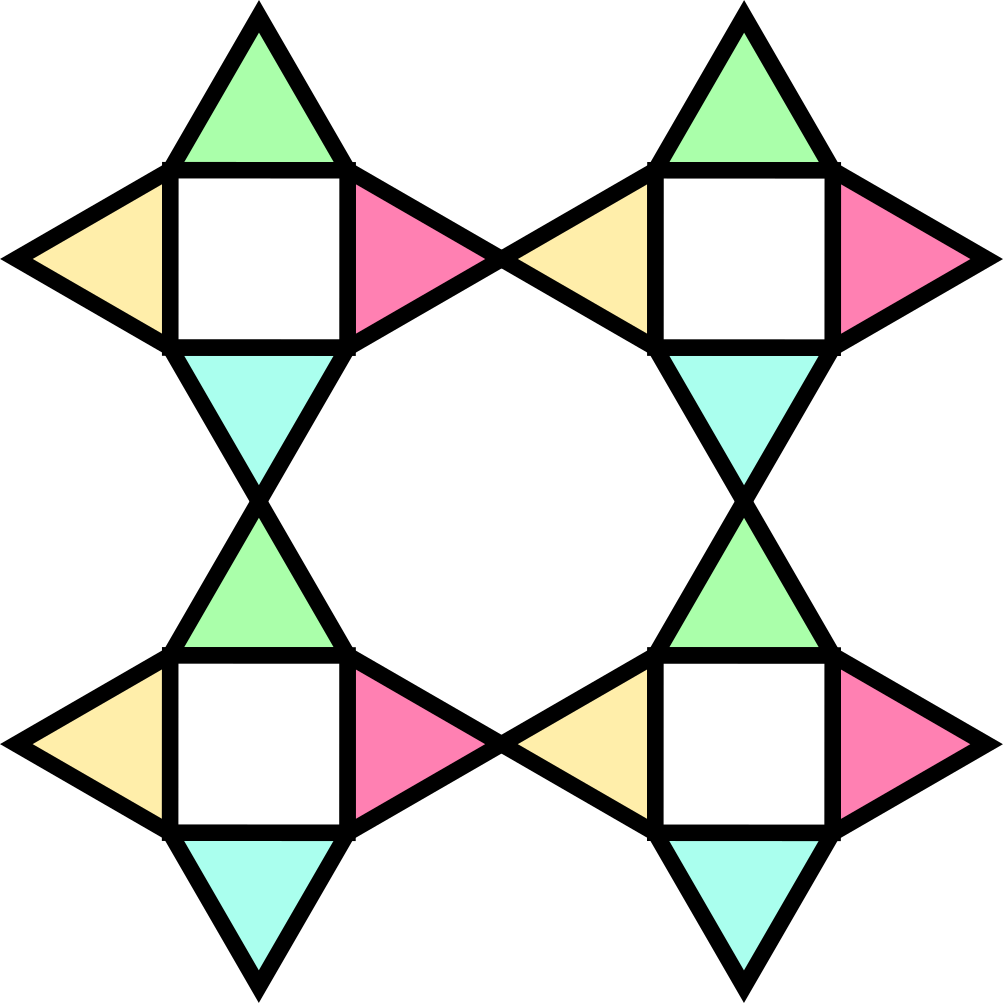} & 
        \includegraphics[height=2.5cm]{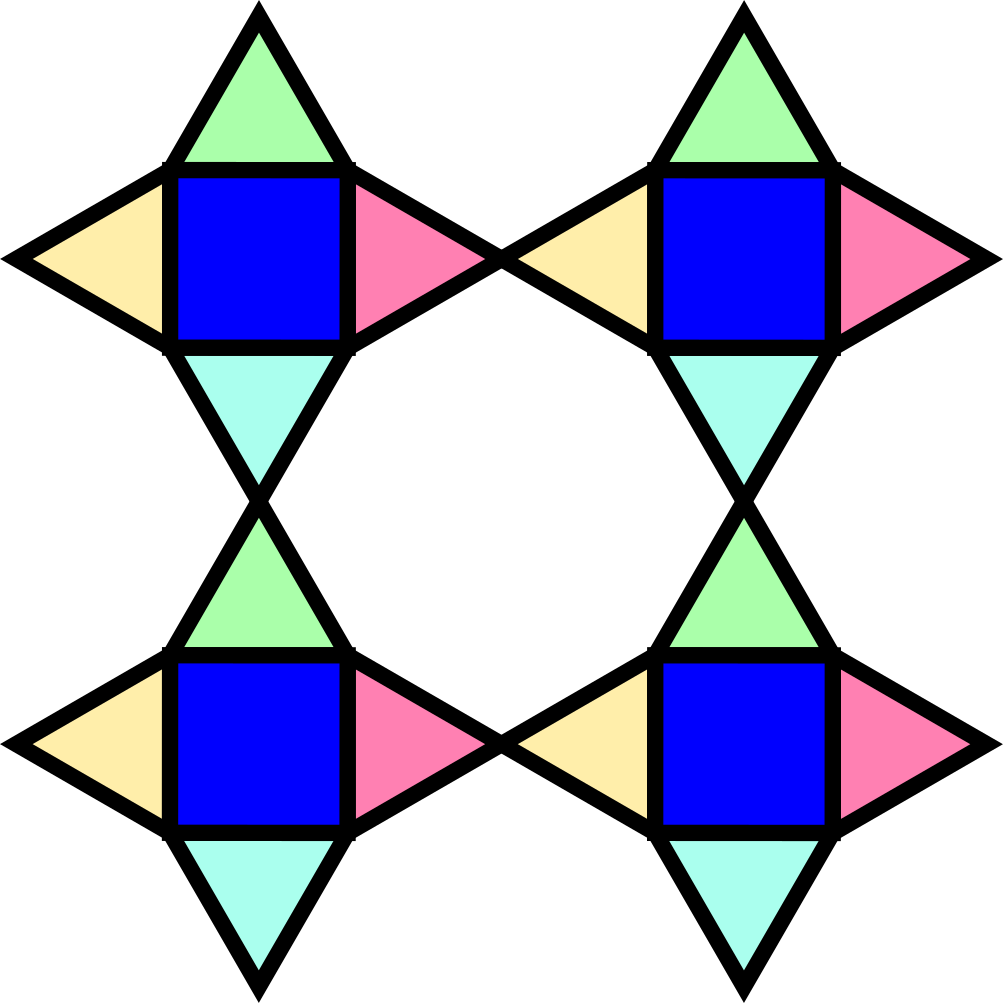}  &  
        \includegraphics[height=2.5cm]{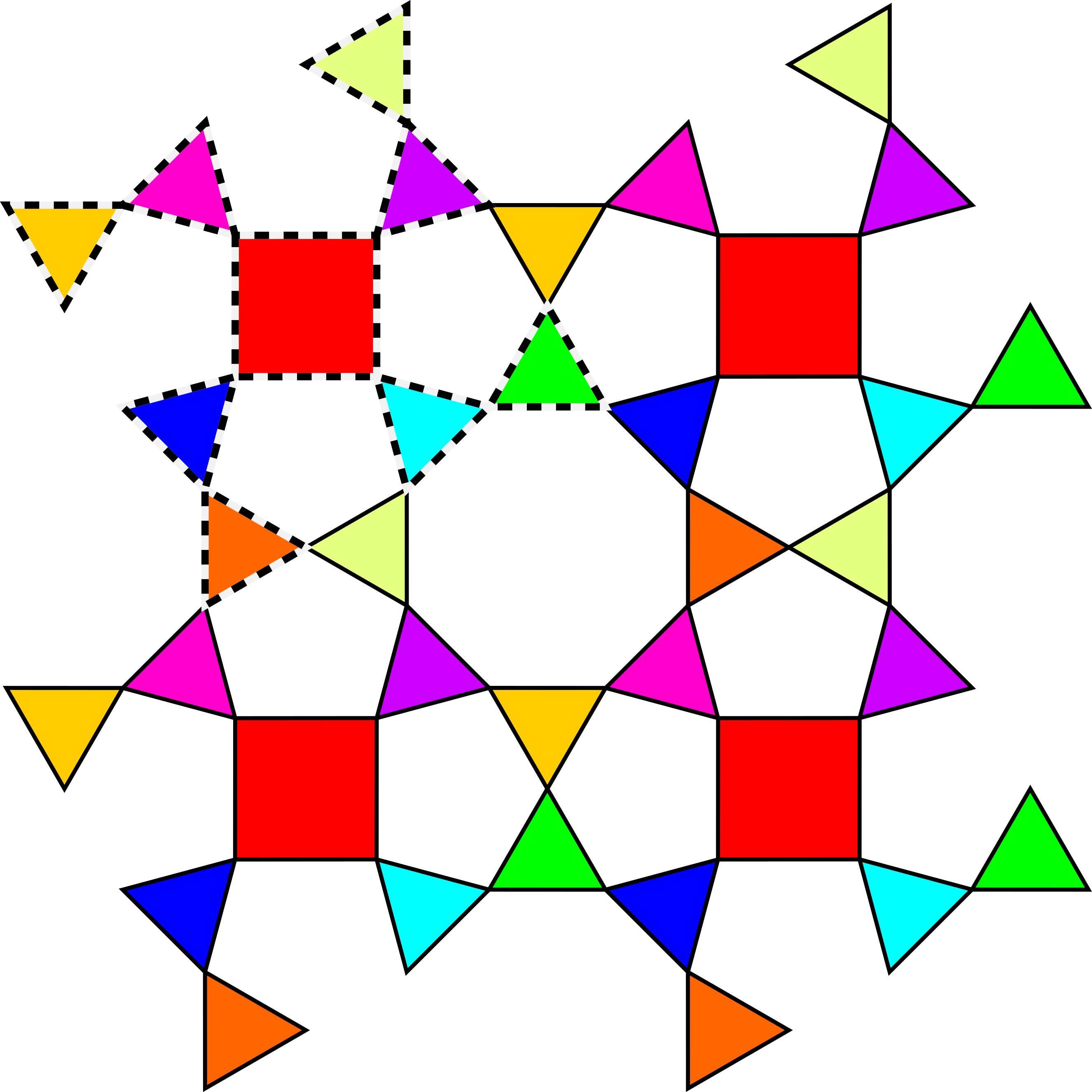} \\
        \textbf{$n_s$}     & 3 & 2 & 2 & 3 & 4 & 6 & 6 & 6 & 14 \\
        \textbf{$n_c$}     & 2 & 1 & 1 & 1 & 2 & 3 & 4 & 5 & 9 \\
        \textbf{$n_{f.b}$} & 1 & 1 & 1 & 2 & 2 & 3 & 2 & 1 & 5 \\
        \textbf{$F$}       & 0 & 1 & 1 & 3 & 2 & 3 & 0 & $-3$ & 1 \\
        \makecell{ \vspace{-2.5cm} \\ \textbf{Premedial} \\ \textbf{Lattice}  \\ \textbf{Scheme} } & 
        \includegraphics[height=2.5cm]{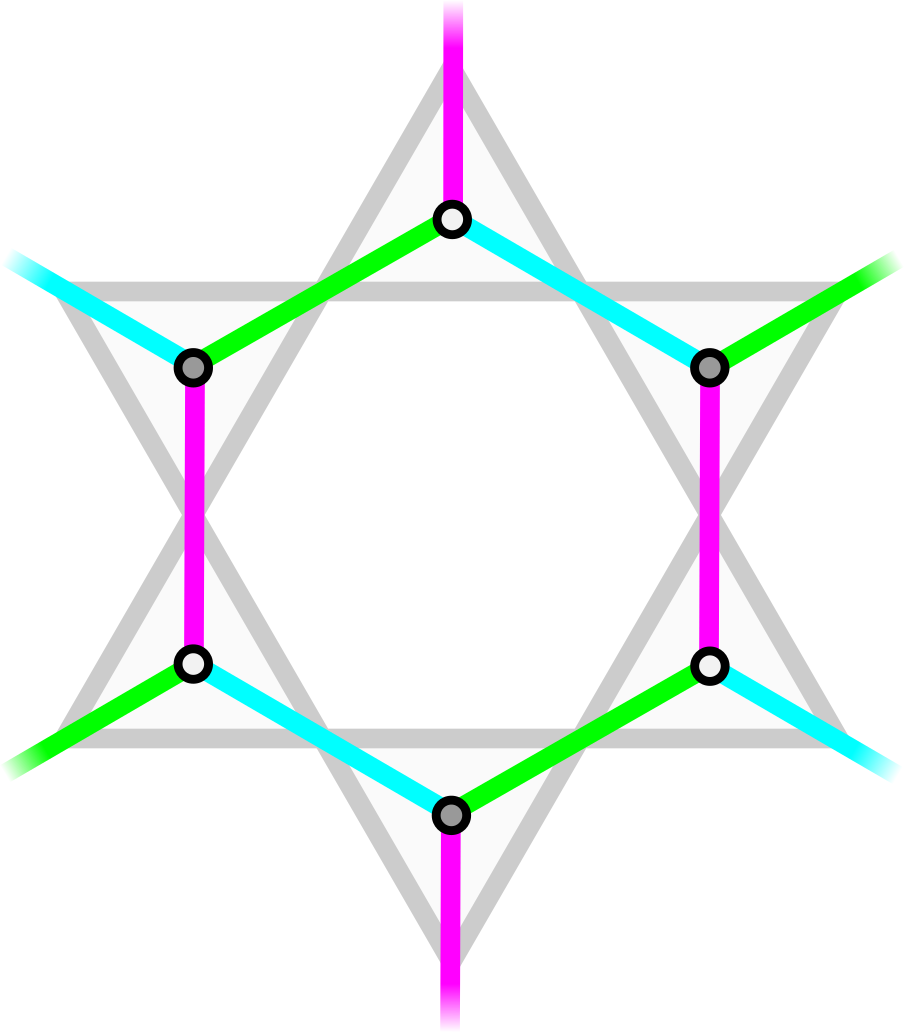} & 
        \includegraphics[height=2.5cm]{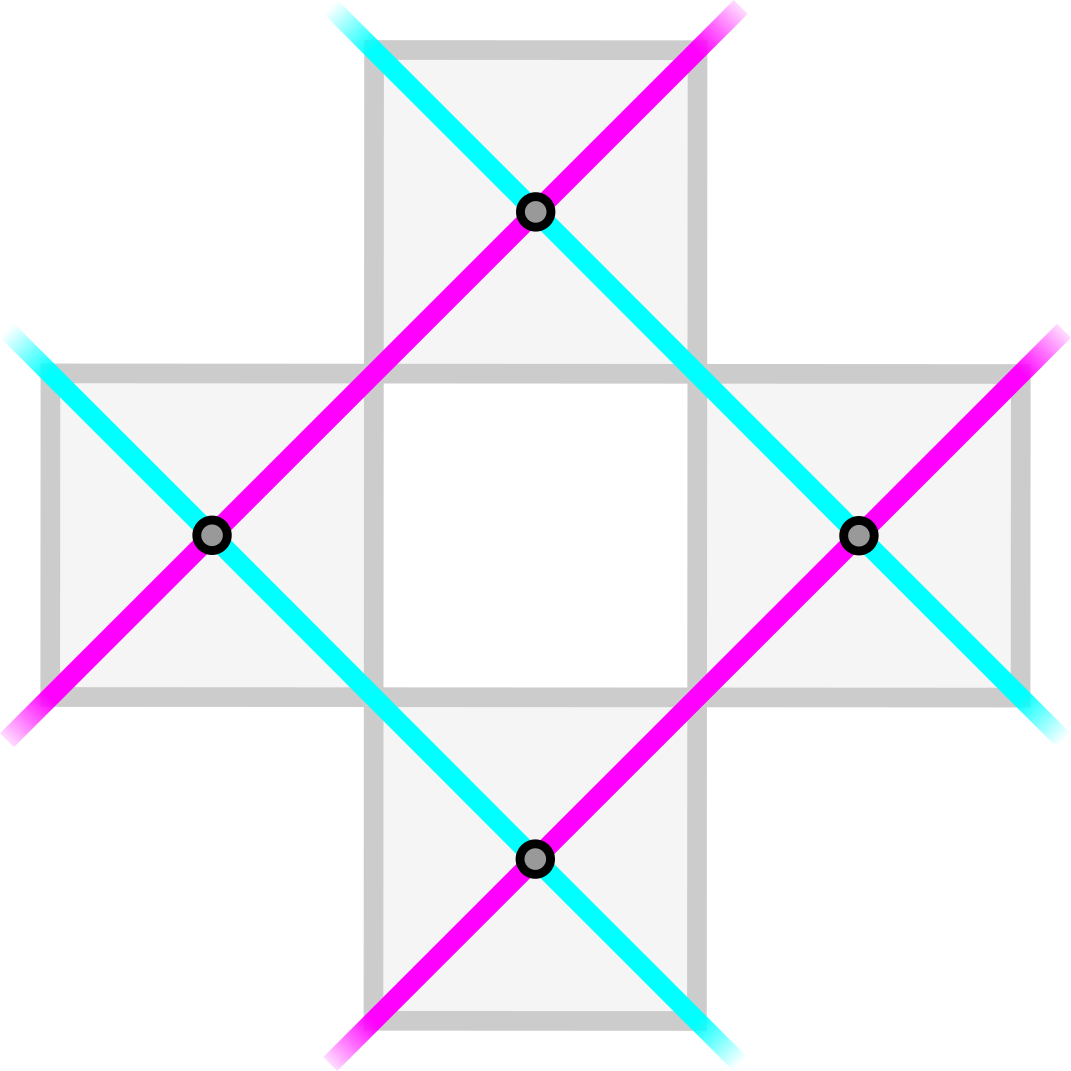} & 
        \includegraphics[height=2.5cm]{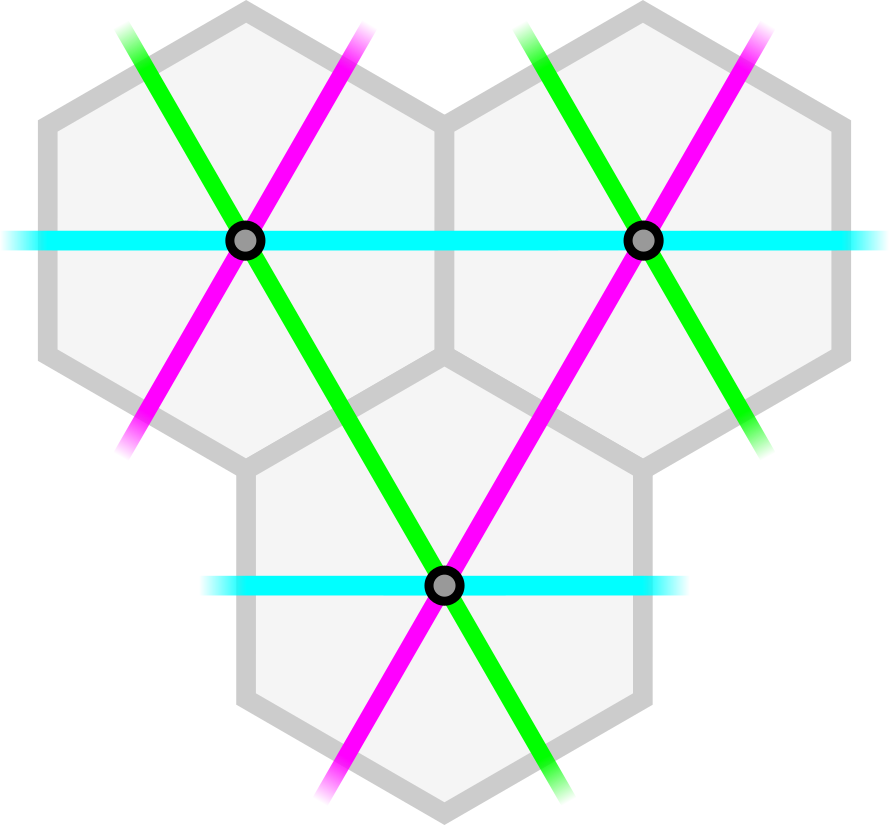} & 
        \includegraphics[height=2.5cm]{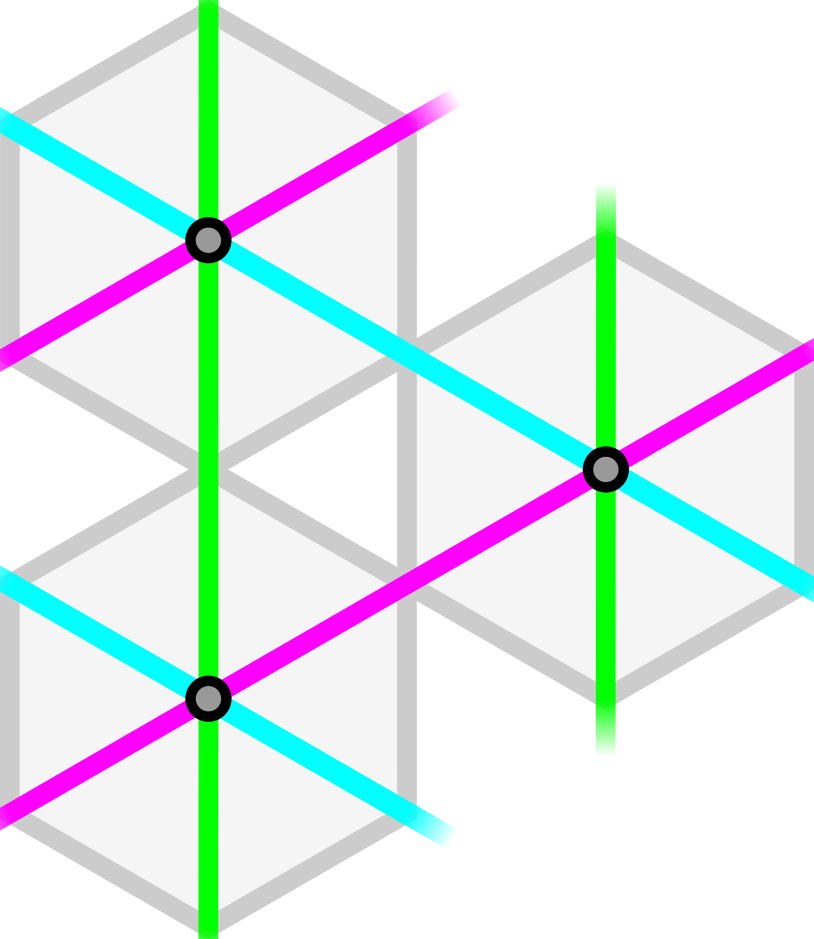} & 
        \includegraphics[height=2.5cm]{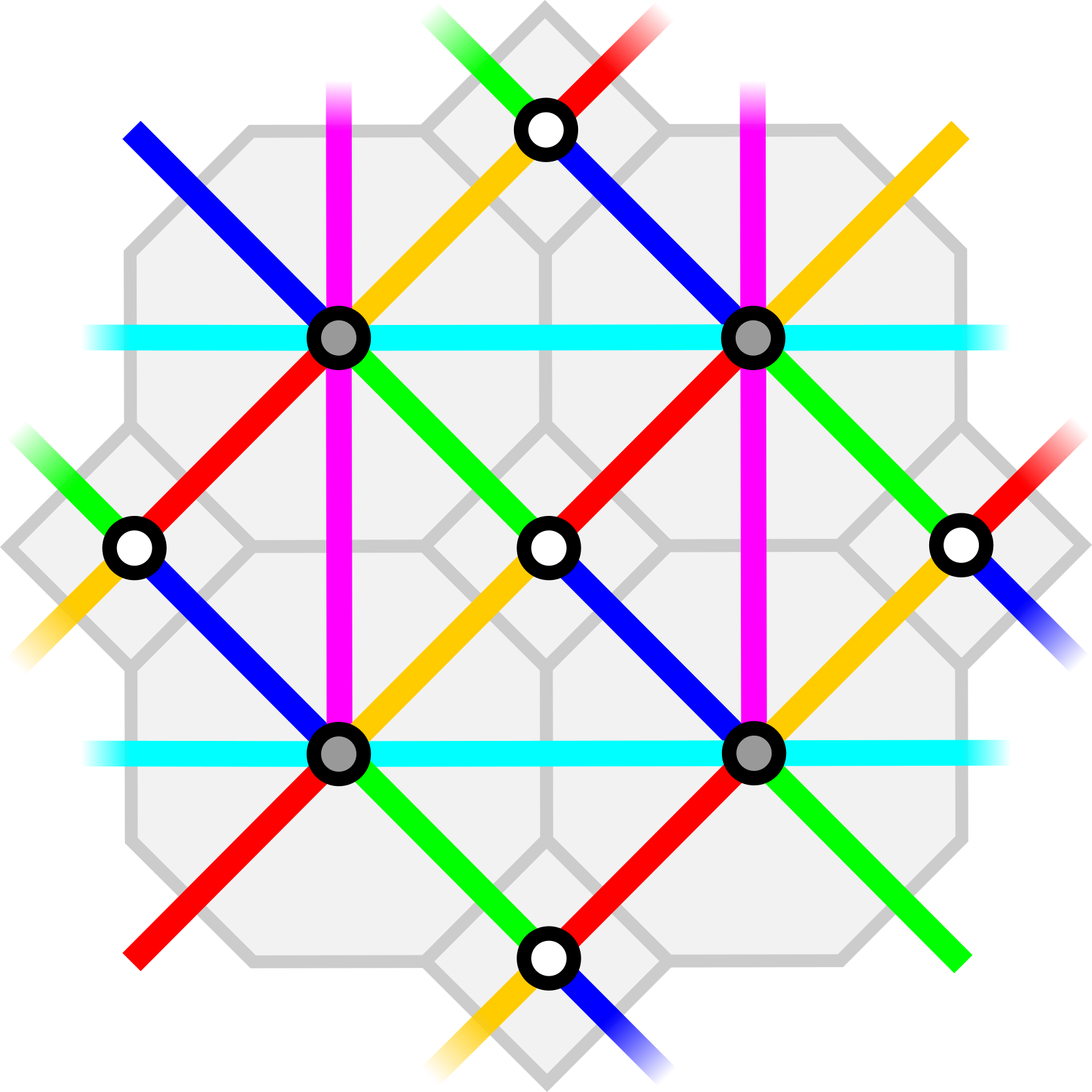} & 
        \includegraphics[height=2.5cm]{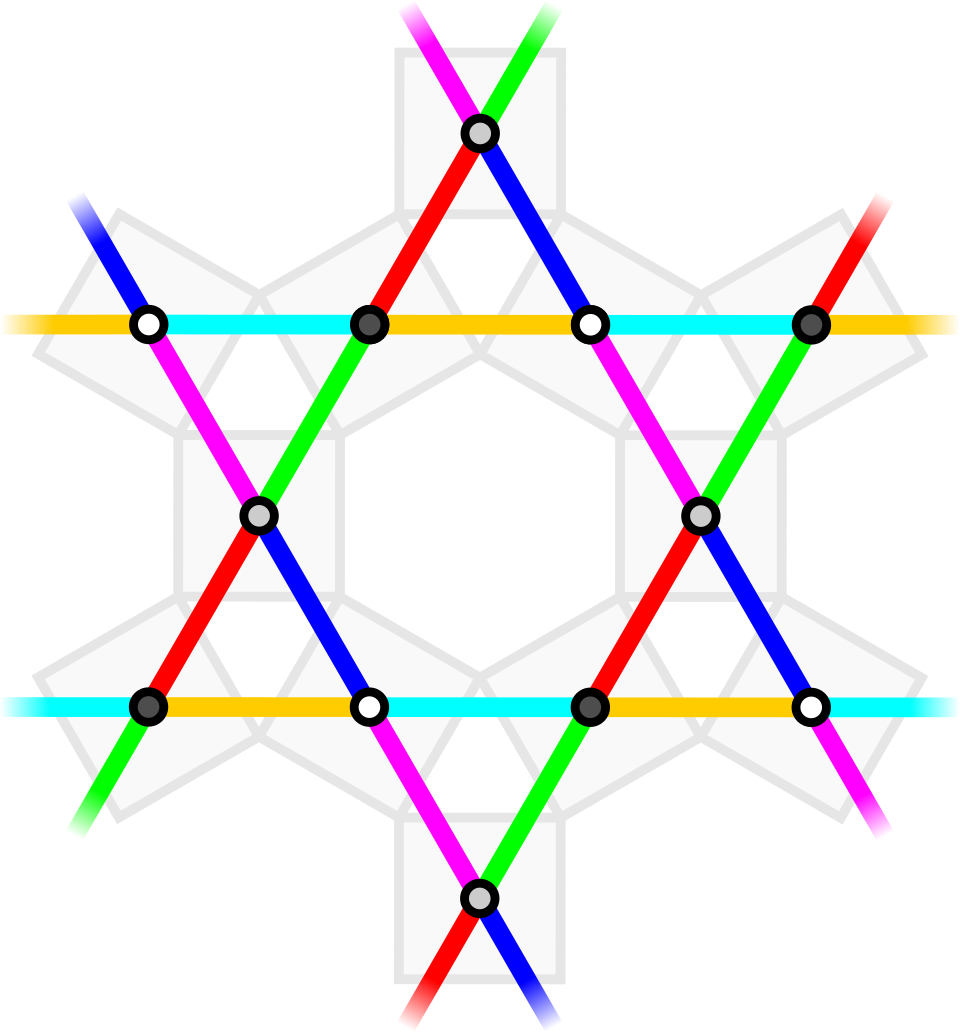} & 
        \includegraphics[height=2.5cm]{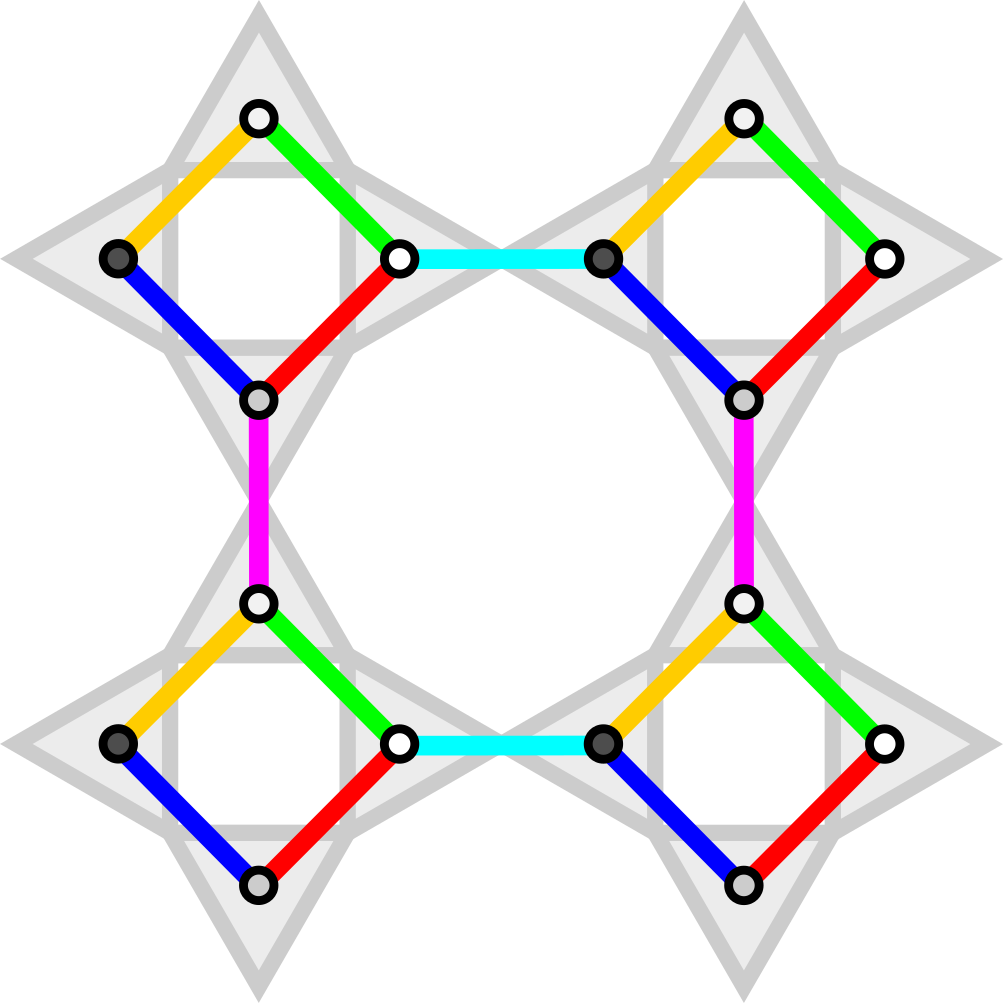} & 
        \includegraphics[height=2.5cm]{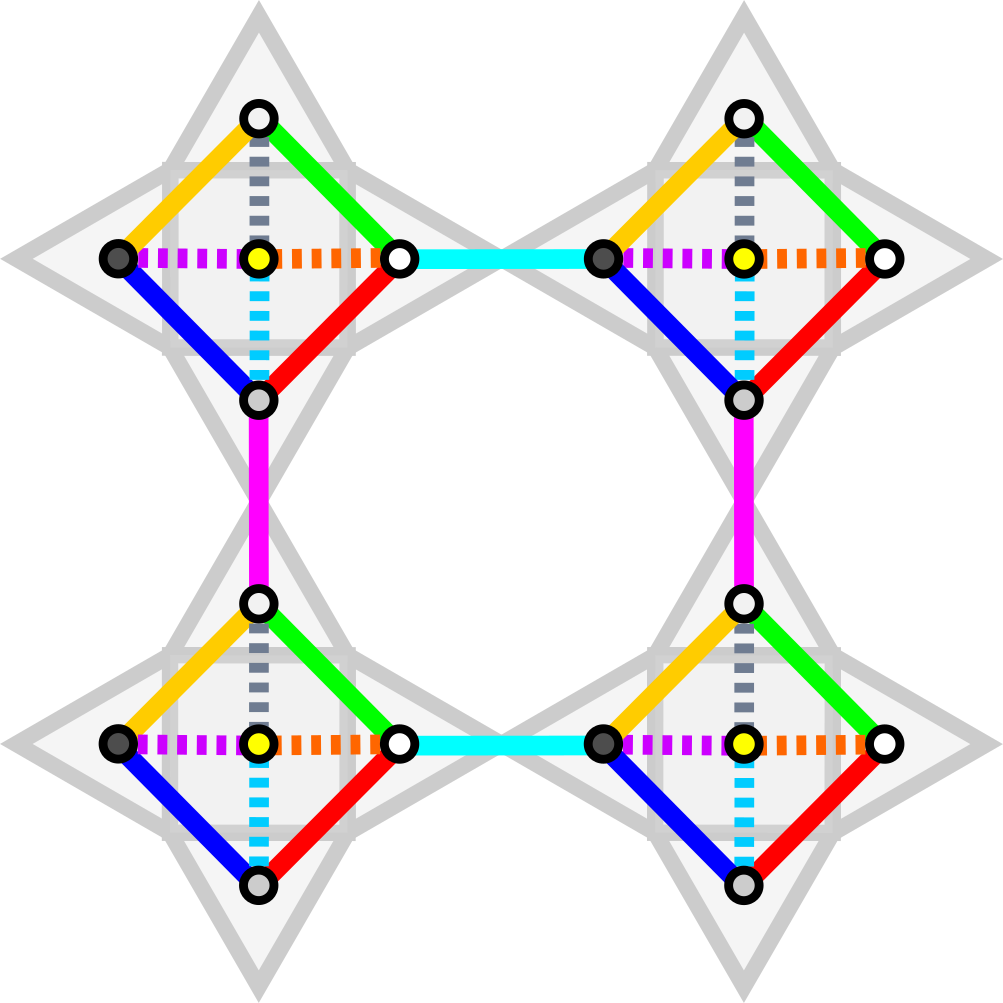} & 
        \includegraphics[height=2.5cm]{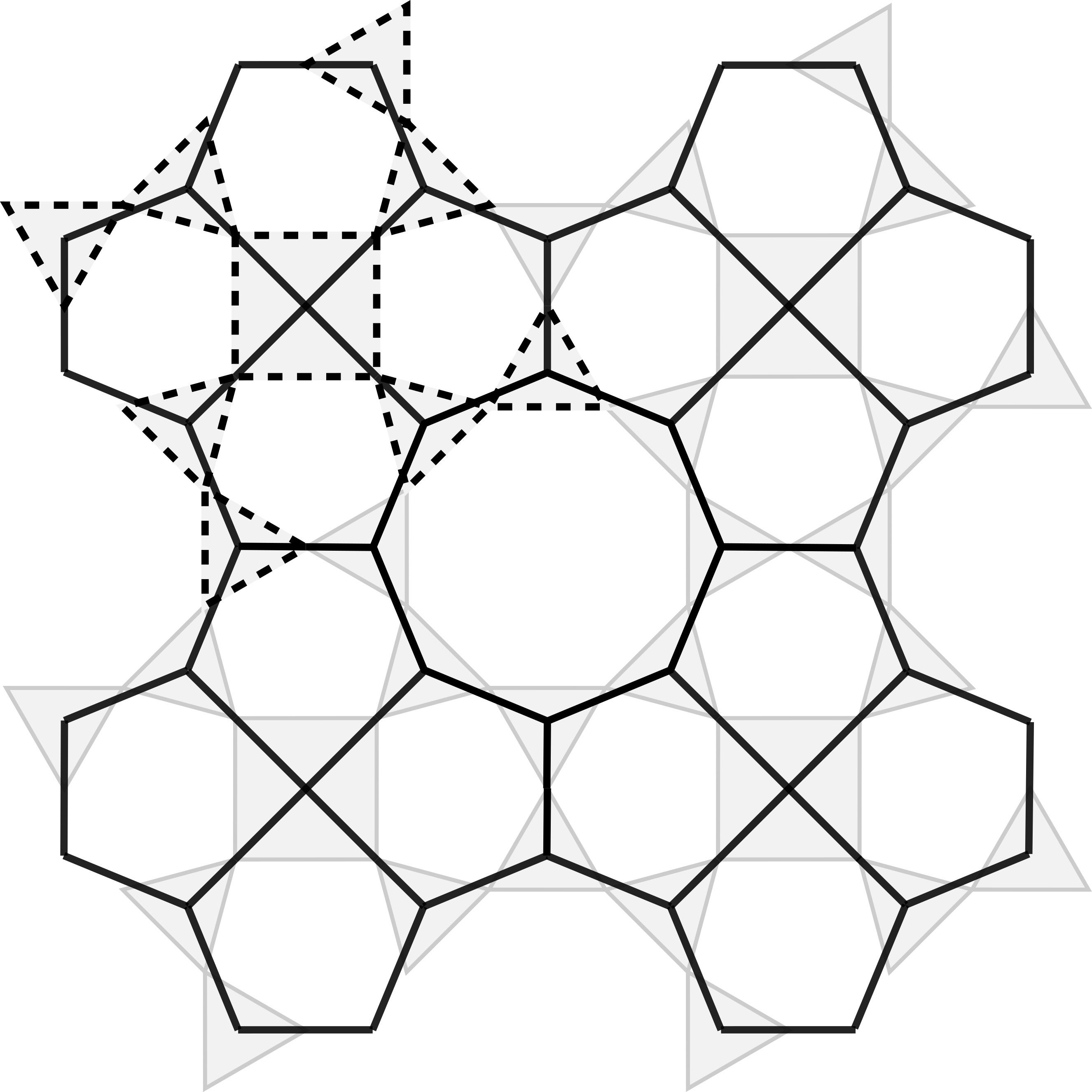} \\
        \textbf{$N_{c.l}$} & 3 & 2 & 3 & 3 & 6 & 6 & 6 & 10 & 18 \\
        \textbf{References} & \cite{Chalker1992, Davier_2023, Yan_2024_long} & \cite{Davier_2023, Yan_2024_long} & \cite{Benton_Moessner_2021, Yan_2024_long} & \cite{Davier_2023, Yan_2024_long} & \cite{Codello_2010_square_octagon} & \cite{Rehn_2017_Ruby_lattice, Verresen_2021_ruby_lattice} & \cite{Siddharthan_2001_square_kagome} & \cite{Gonzalez_2025_decorated_square_kagome} & \\
        \hline\\
    \end{tabular}}
    \resizebox{\textwidth}{!}{
    \begin{tabular}{c c c c c c}
        \hline
        \textbf{3D Lattice} & Pyrochlore & Octahedral & \makecell{Kagome\\ bipyramidal} & Quadrupahedral & Hyperkagome\\
        \hline
        \noalign{\vskip 0.5 mm}
        \makecell{ \vspace{-3cm} \\ \textbf{Lattice}  \\ \textbf{Scheme} } & 
        \includegraphics[height=3cm]{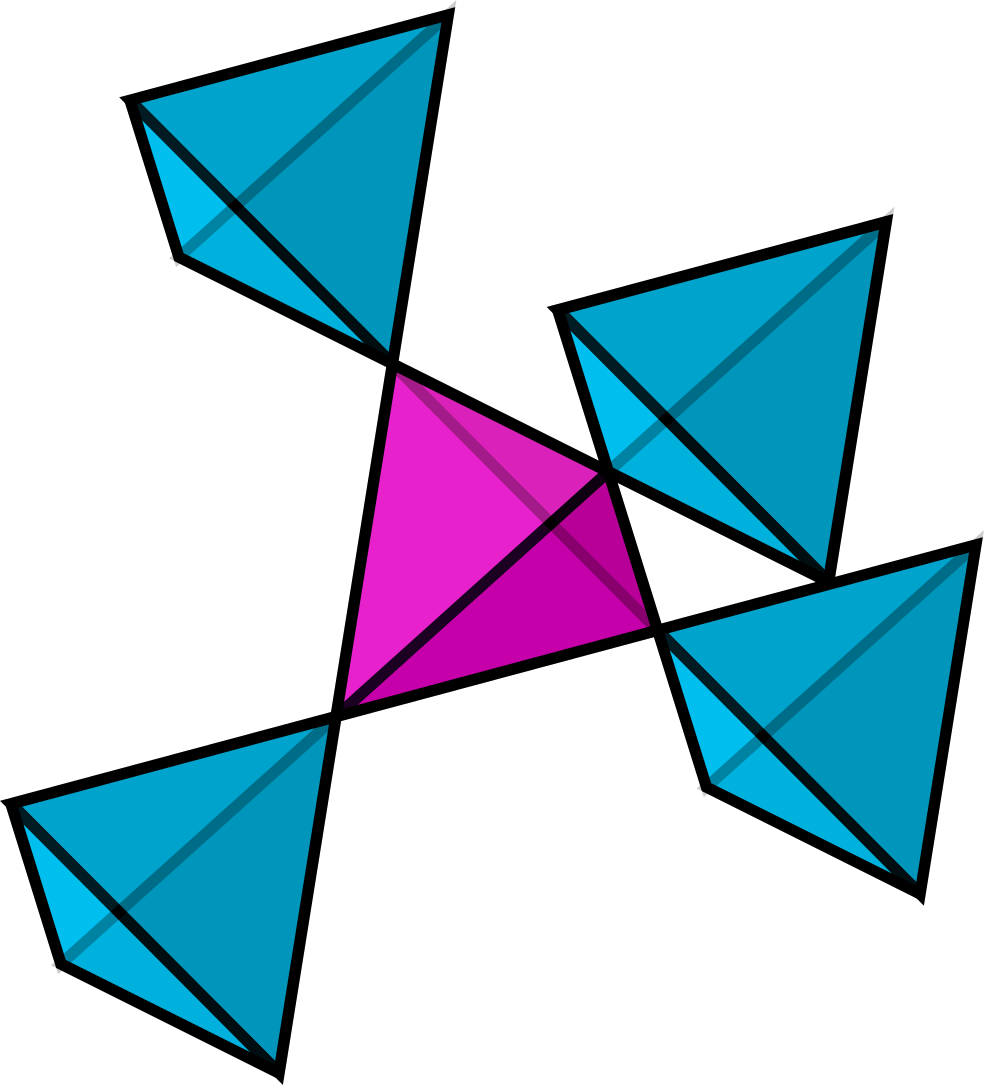} &
        \includegraphics[height=3cm]{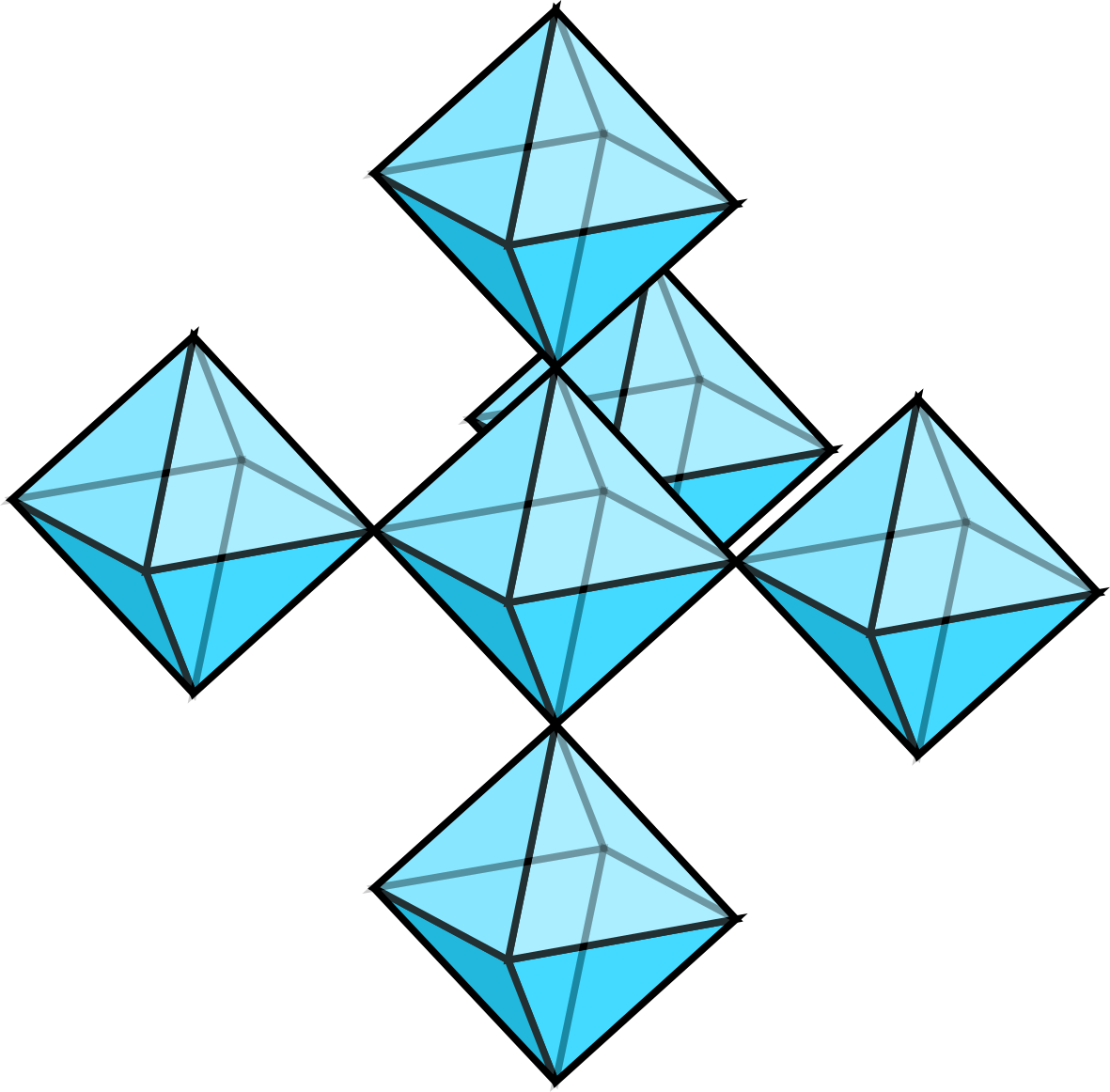} & 
        \includegraphics[height=3cm]{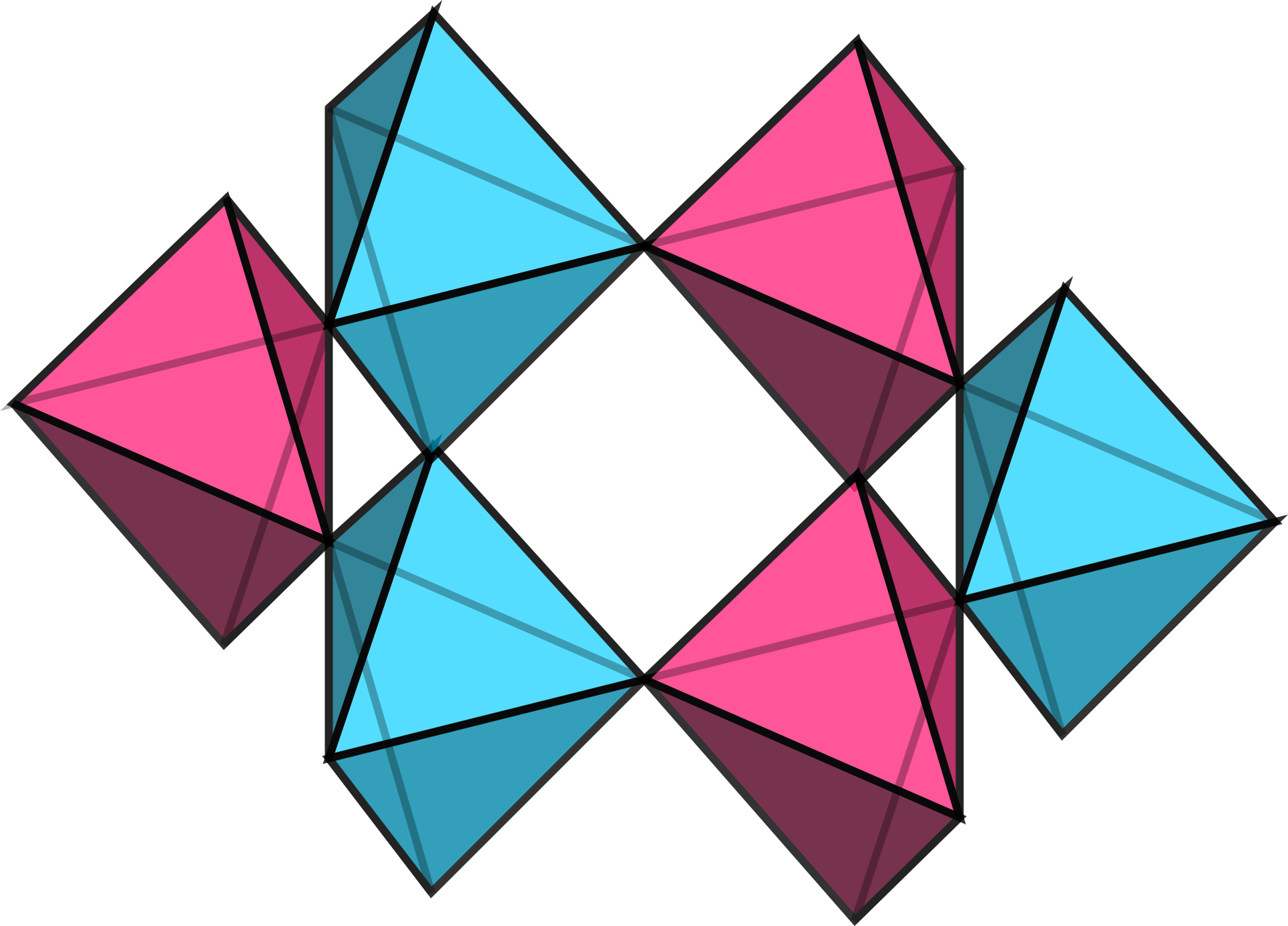} & 
        \includegraphics[height=3cm]{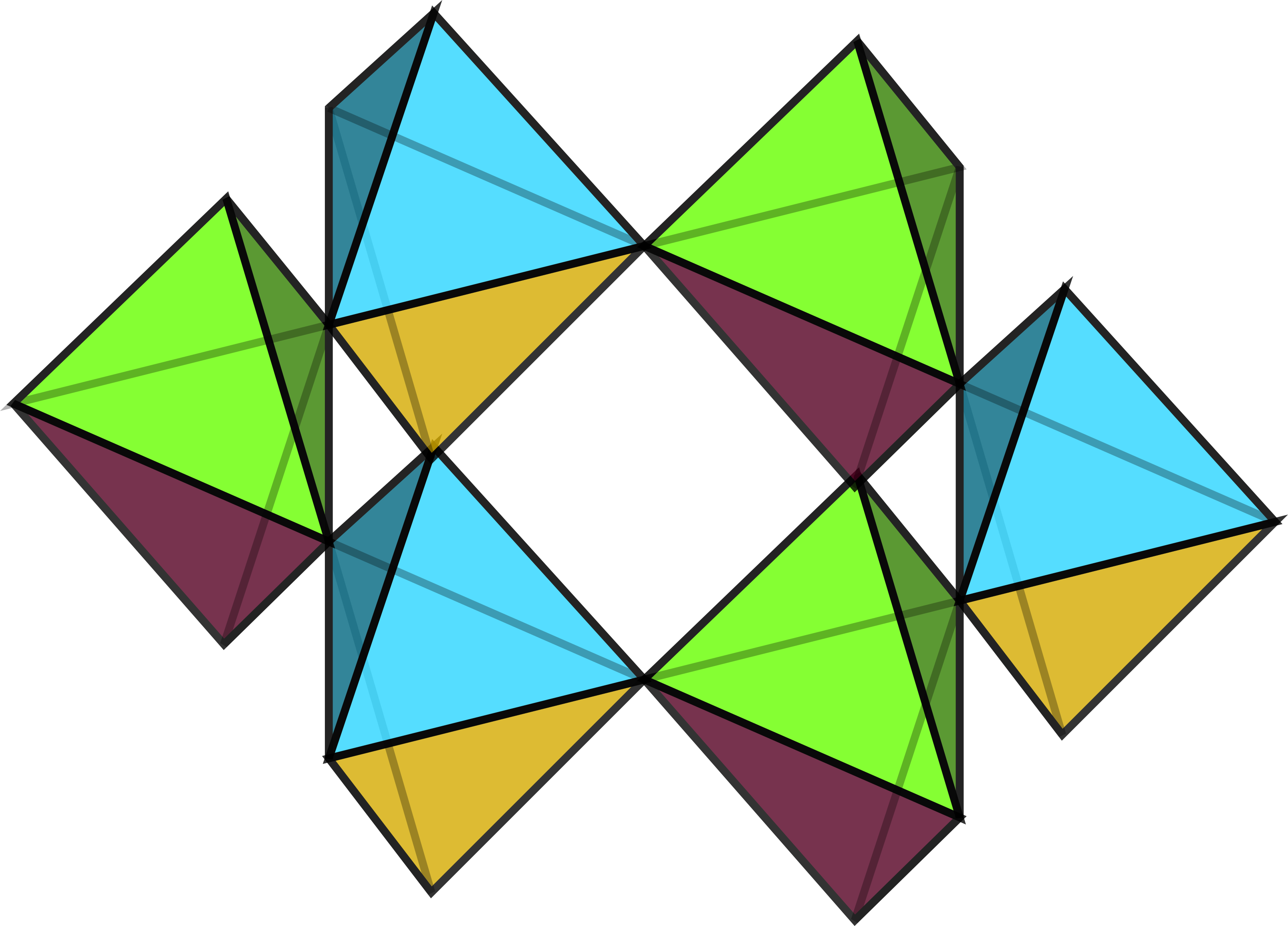} & 
        \includegraphics[height=3cm]{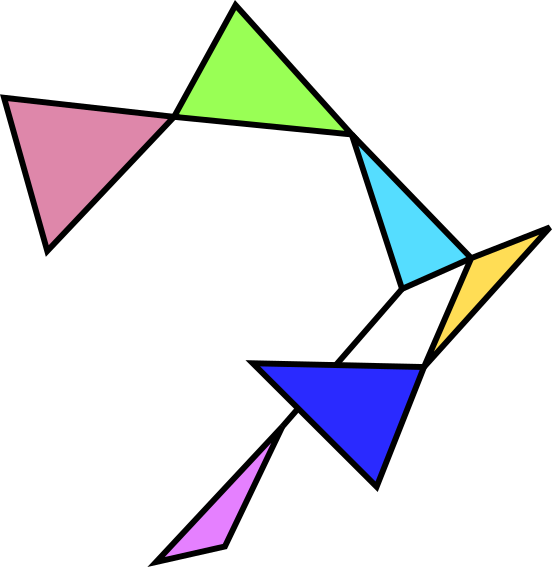} \\
        \textbf{$n_s$} & 4 & 3 & 5 & 5 & 12 \\
        \textbf{$n_c$} & 2 & 1 & 2 & 4 & 6 \\
        \textbf{$n_{f.b}$} & 2 & 2 & 3 & 1 & 6 \\
        \textbf{$F$} & 2 & 3 & 4 & $-2$ & 6 \\
        \textbf{$N_{c.l}$} & 4 & 3 & 5 & 16 & 9 \\
        \textbf{References} & \cite{Henley2005, Yan_2024_long} & \cite{Benton_Moessner_2021, Niggemann_2023} & & & \cite{Hopkinson_2007_Hyperkagome, Okamoto_2007_hyperkagome} \\
        \hline
    \end{tabular}}
    \caption{Examples of possible cluster systems in two and three dimensions. The $n_c$ different types of clusters among a single lattice are depicted with different colors. The number of flat bands $n_{f.b}$ can be computed explicitly from the number of sublattices $n_s$ and the number of cluster type $n_c$ as $n_{f.b} = n_s-n_c$. The number of zero modes per unit cell $F$ can also be derived from these two numbers for Heisenberg spins as $F = 2n_s - 3n_c$. Each cluster type corresponds to a different sublattice of the premedial lattice \cite{henley2010coulomb}, as depicted for the two-dimensional examples. There are $N_{c.l}$ inequivalent links in the premedial lattice, each one depicted with a different color (when their number allows for it). For the octagonal kagome lattice, the unit cell is composed of nine clusters and is highlighted with a dashed contour. In 3D, the kagome bipyramidal lattice is composed of bypiramidal clusters with six faces, while the quadrupahedral lattice is composed of triangular-based pyramids touching through a face. }
    \label{tab: cluster systems}
\end{table*}

\subsection{Definitions}

Cluster Hamiltonians can be decomposed as a sum of energy terms over clusters $n$ \cite{Benton_Moessner_2021, Yan_2024_short, Yan_2024_long, Davier_2023},
\begin{equation}
    \mathcal{H} = \sum_n|\bm{\mathcal{C}}_n|^2
    \label{Eq: general cluster H}
\end{equation}
where $\bm{\mathcal{C}}_n$ is called a \textit{constrainer}. This object can be seen as a weighted local magnetization for cluster $n$,
\begin{equation}
    \bm{\mathcal{C}}_n = \sum_{i \in n} \gamma_i^n \mathbf{S}_i,
    \label{Eq : general constrainer}
\end{equation}
where $\mathbf{S}_i$ are classical Heisenberg spins of unit length, $|\mathbf{S}_i| = 1$. All spins interact with each other in a given cluster via the coupling $2\gamma_i^n \gamma_j^n$, including a self-interacting term $(\gamma_i^n)^2$ that amounts to an irrelevant constant energy shift. There is a considerable variety of models, as there is a large freedom in choosing the geometry and size of each cluster, which can be connected to each other via a vertex, an edge, or a face, as illustrated in Table~\ref{tab: cluster systems}.

The ground state of Hamiltonian (\ref{Eq: general cluster H}) is obtained by minimizing the value of $|\bm{\mathcal{C}}_n|$ for all clusters $n$ in the system. In this paper, we will consider models where it is possible to impose $\bm{\mathcal{C}}_n = 0, \forall n$. Depending on the weights $\gamma_i^n$, this constraint is not always permitted, but such counter-examples usually lead to ordered ground states (see e.g. \cite{nutakki23b}).\\

Before introducing the interacting-cluster Hamiltonian and presenting our results, we shall now provide a concise review of the literature on the classification of classical spin liquids \cite{Benton_Moessner_2021, Yan_2024_short, Yan_2024_long, Davier_2023,Fang_2024} and the connectivity matrix framework in spin liquids \cite{katsura10a,Essafi_2017,Mizoguchi_2018,Mizoguchi_Masafumi_2019_flat_bands}.

\subsection{Flat-band ground state}
\label{Subsec: constraint vector fiormalism}
Let us define the Fourier transform of the spin component $\alpha\in\{x,y,z\}$ on sublattice $\mu\in\{1,..,n_s\}$ in unit cell $i\in\{1,..,N_{u.c}\}$,  
\begin{equation}
    \begin{split}
        &S^\alpha_\mu(\mathbf{q}) = \sum_{i} S_{\mu,i}^\alpha e^{-i (\mathbf{R}_i + \mathbf{r}_\mu) \cdot \mathbf{q}}, \\ 
        &S^\alpha_{\mu,i} = \frac{1}{N_{u.c}}\sum_{\mathbf{q}} S_\mu^\alpha(\mathbf{q}) e^{i (\mathbf{R}_i + \mathbf{r}_\mu) \cdot \mathbf{q}},
    \end{split}
    \label{Eq: Spin components Fourier transform}
\end{equation}
where $\mathbf{R}_i$ is the position of the unit cell and $\mathbf{r}_\mu$ is the relative position of the sublattice within the unit cell. Hamiltonian (\ref{Eq: general cluster H}) is then rewritten in Fourier space \cite{Benton_Moessner_2021}
\begin{eqnarray}
        \mathcal{H} = \frac{1}{N_{u.c}} \sum_\mathbf{q} \sum_{X, \alpha} \left| \mathbf{L}_X^\alpha(\mathbf{q}) \cdot \mathbf{S}^\alpha(\mathbf{q}) \right|^2
    \label{Eq: general cluster H : Fourier version}
\end{eqnarray}
where $X\in\{1,..,n_c\}$ refers to the sum over the $n_c$ different types of clusters in the system; e.g. for the kagome lattice [Table.\ \ref{tab: cluster systems}], there are two types of clusters corresponding to the two types of triangles $\vartriangle, \triangledown$. The constraint vector $\mathbf{L}_X^\alpha$, as introduced by Benton \& Moessner \cite{Benton_Moessner_2021}, encodes the geometric structure of the clusters,
\begin{equation}
    \left( L_X^\alpha\right)_\mu(\mathbf{q}) = \sum_{j\in \mu \cap X} \gamma_j^X e^{ i \mathbf{r}_j \cdot \mathbf{q}}
    \label{Eq: definition of the constraint vectors LX}
\end{equation}
where $\mathbf{r}_j$ denotes the vector position of site $j$ relative to the center of the cluster $X$, and the sum runs over all sites within cluster $X$ that belong to the sublattice $\mu$. Note that the dimension of vectors $\mathbf{S}^\alpha(\mathbf{q})$ and $\mathbf{L}_X^\alpha(\mathbf{q})$ is $n_s$, where each component corresponds to a different sublattice $\mu\in\{1,..,n_s\}$.\\

From now on, we shall work within the Luttinger-Tisza approximation (LTA) \cite{LT1, LT2, kaplan_2007_LTA}, where the spin length constraint is only enforced on average over the entire system, $\sum_i |\mathbf{S}_i|=N$ with $N$ the total number of spins, which are known to be suitable for cluster-system classical spin liquids, see Appendix \ref{Appendix:LTA}. A known limitation of the LTA is that it cannot always properly account for thermal order by disorder \cite{villain1980,Chalker1992}, i.e. when thermal fluctuations entropically lift the classical ground-state degeneracy and favor magnetic order at very low temperatures. It is, however, possible to estimate when order by disorder is expected or not, i.e. when flat bands are expected to lead to a magnetically disordered spin-liquid ground state (see section \ref{sec:flatbandliquid}). And even if the system ultimately orders at very low temperature $T_c$, the results of this paper remain valid at temperatures above $T_c$. Indeed, even if the LTA is an approximation, this method is particularly well suited to the study of classical spin liquids whose disordered magnetic texture accommodates well the averaged spin length constraint, as confirmed by many comparison between theory and simulations \cite{reimers91a,conlon10a,Mizoguchi_2018,Benton_Moessner_2021,Yan_2024_long, Davier_2023}. And since we shall only consider \textit{isotropic} exchange coupling here, the spin component $\alpha\in\{x,y,z\}$ is actually irrelevant and will be omitted from now on.\\

Once the Hamiltonian is written under the form (\ref{Eq: general cluster H : Fourier version}), it becomes clear that all spin modes $\mathbf{S}_{\perp}(\mathbf{q})$ orthogonal to all constraint vectors $\mathbf{L}_X(\mathbf{q})$,
\begin{equation}
    \mathbf{L}_X(\mathbf{q}) \cdot \mathbf{S}_{\perp}(\mathbf{q}) = 0, \qquad X = 1, \ldots, n_c
    \label{Eq: cluster ground state constraints}
\end{equation}
are zero energy modes. When Eq.~(\ref{Eq: cluster ground state constraints}) is valid for all wavevectors $\mathbf{q}$, then the ground state is a flat band. The number of flat bands $n_{f.b}$ is thus equal to the total number of bands -- i.e. the number of sublattices $n_s$-- minus the number of independent constraints (\ref{Eq: cluster ground state constraints}) per unit cell, which is given by the number of cluster types $n_c$, leading to $n_{f.b} = n_s - n_c$  \cite{Moessner_1998_counting,Yan_2024_long, Davier_2023}. The expected number of flat bands for various cluster systems is listed in Table \ref{tab: cluster systems}.

These flat bands solely depend on the geometry of the clusters -- namely, their type, arrangement, and how they tile the lattice -- but not on the specific values of the spin-spin interaction coefficients $\gamma_i^n$. These coefficients $\gamma_i^n$ influence the internal structure of the constraint vectors and thereby determine the detailed form of the dispersive bands, but do not generally affect these geometric flat bands. For finely tuned values, these coefficients $\gamma_i^n$ can accidentally flatten a dispersive band; a trivial example is when $\gamma_i^n=0$ for all clusters $n$ belonging to a certain type of clusters $X'$ \cite{Essafi_2017}.

Since the number of constraints -- and thus the flat-band structure -- is governed by the way spins are grouped into interacting clusters, it is natural to expect that this information can also be encoded in the connectivity of the model itself.

\subsection{The connectivity matrix} 
\label{sec:connec}

\begin{figure}[t]
    \centering
    \includegraphics[width=0.9\linewidth]{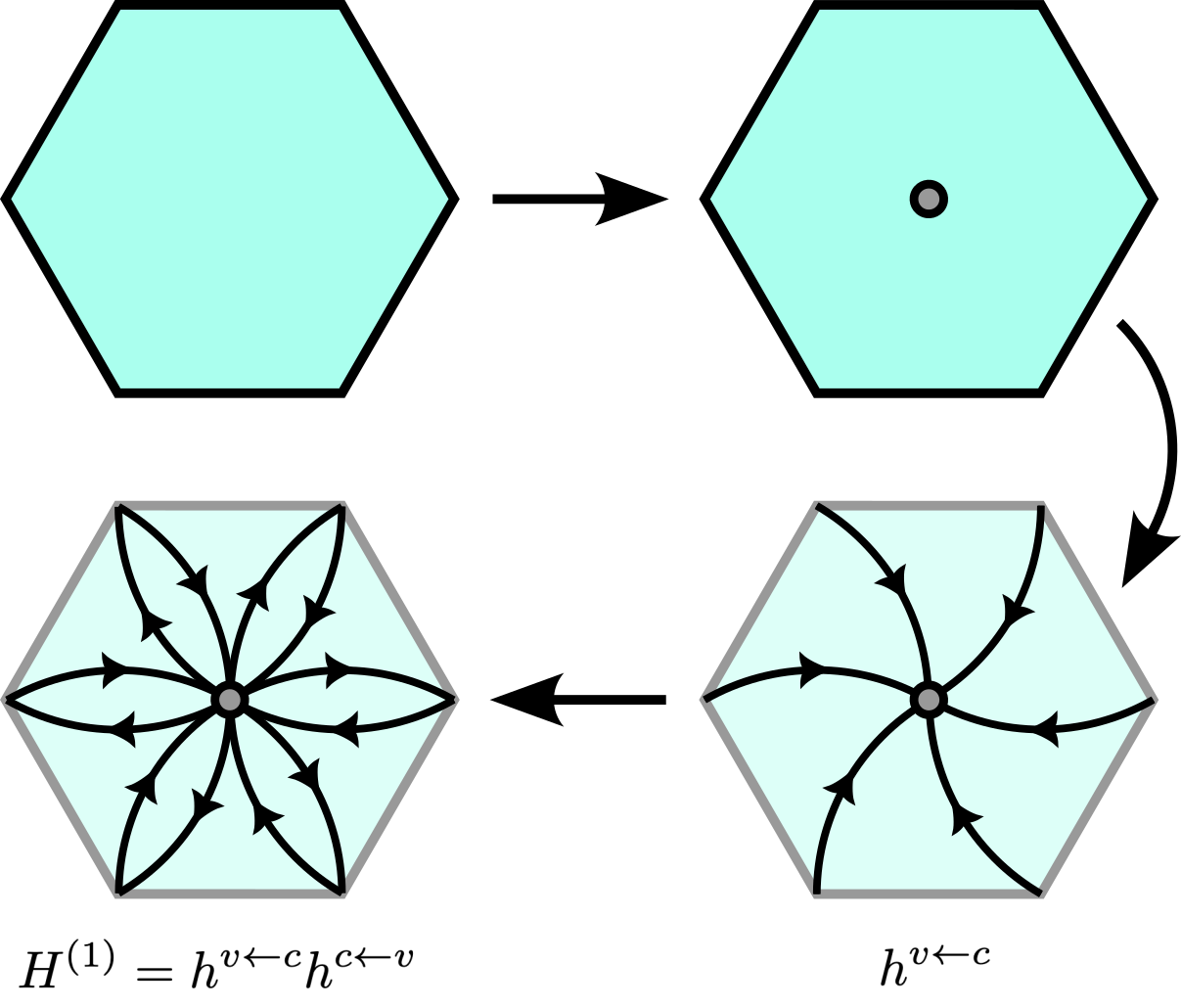}
    \caption{
    Illustration of the mapping between the cluster Hamiltonian  (\ref{Eq: general cluster H}) and the connectivity matrices of Eq. (\ref{Eq: connectivity matrix expression for H(1)}). First, we consider a cluster at random, that is arbitrarily depicted as an hexagon here. Let us place a virtual site at its center; these central sites form the premedial lattice. Second, let us link \textbf{all} cluster sites $i$ to the central site $n$ (represented as curved oriented lines here). This is precisely the definition of the connectivity matrix $h^{v \leftarrow c}$ with coefficients $h^{v \leftarrow c}_{in}=\gamma_i^n$. Finally, let us link the center site $n$ with \textit{all} cluster sites $i$ via the transpose of the connectivity matrix $h^{c \leftarrow v} = (h^{v \leftarrow c})^T$. The resulting matrix $H^{(1)} = h^{v \leftarrow c} h^{c \leftarrow v}$ thus pairs, within each cluster, all cluster sites two-by-two, and with the appropriate coupling constant as defined in Hamiltonian (\ref{Eq: general cluster H}).}
    \label{fig: Cluster connectivity matrixl}
\end{figure}

The cluster Hamiltonian (\ref{Eq: general cluster H}) can be expressed in terms of connectivity matrices\cite{katsura10a,Essafi_2017,Mizoguchi_2018}, expressing it no more as a sum over clusters, but rather as a sum over all lattice sites 
\begin{equation}
    \mathcal{H} = \sum_{i,j} H_{i,j}^{(1)} \mathbf{S}_i \cdot \mathbf{S}_j.
    \label{Eq: H(1) connectivity matrix}
\end{equation}
Since each spin interact with all other spins within the cluster, the interaction matrix $H^{(1)}$ can be simply expressed as
\begin{equation}
    H^{(1)} = h^{v \leftarrow c} h^{c \leftarrow v}
    \label{Eq: connectivity matrix expression for H(1)}
\end{equation}
where $h^{v \leftarrow c} = (h^{c \leftarrow v})^t$ denotes the $N \times N_c$ connectivity matrix that connects the $N$ lattice sites to the $N_c$ virtual cluster central sites that form the premedial lattice \cite{henley2010coulomb}. Its coefficients $h^{v \leftarrow c}_{in}$ are thus equal to $\gamma_i^n$ (see Eq.~(\ref{Eq : general constrainer})) if the spin $i$ belongs to the cluster $n$, and 0 otherwise. The mapping of Eq.~(\ref{Eq: connectivity matrix expression for H(1)}) is illustrated in Fig.\ref{fig: Cluster connectivity matrixl}.

While there is a priori no theoretical limit on the number of clusters that can be considered for a given lattice, it might be somewhat artificial to have more clusters than spins in the system. One shall thus focus on models where $N>N_c$. Under this hypothesis, and since the interaction matrix $H^{(1)}$ is the product of two $N \times N_c$ rectangular matrices, the dimension of the kernel of $H^{(1)}$ cannot be smaller than $N-N_c= N_{u.c}(n_s-n_c)$ \cite{Essafi_2017}. Hence, the connectivity matrix directly provides the same information about the number $(n_s-n_c)$ of flat bands in the system than in the previous section (since each flat band contains $N_{u.c}$ points in reciprocal space). In other words, there are less constraints than total degrees of freedom, leaving some freedom for an extensive ground state manifold. As a side note, we see that the ground-state manifold of models with more clusters than spins, $N_c > N$, should a priori be over-constrained and are thus expected to be ordered.

Using the translation invariance of the Hamiltonian matrix (\ref{Eq: connectivity matrix expression for H(1)}), $H_{\mu,\nu}(\mathbf{R}_i, \mathbf{R}_j) = H_{\mu,\nu}(\mathbf{R}_j - \mathbf{R}_i)$, one gets the Fourier transform of Hamiltonian (\ref{Eq: H(1) connectivity matrix}) \cite{Mizoguchi_2018}
\begin{equation}
    \mathcal{H} = \frac{1}{N_{u.c}}\sum_\mathbf{q} \sum_{\mu,\nu} H^{(1)}_{\mu \nu}(\mathbf{q}) \mathbf{S}_\mu(\mathbf{q}) \cdot \mathbf{S}_\nu(-\mathbf{q}),
\end{equation}
where 
\begin{equation}
    H^{(1)}_{\mu\nu}(\mathbf{q}) = \sum_{j} H^{(1)}_{(0,\mu),(j,\nu)}e^{i(- \mathbf{R}_j + \mathbf{r}_\mu-\mathbf{r}_\nu)\cdot\mathbf{q}}
\end{equation}
is the Fourier transform of the $n_s\times n_s$ Hamiltonian incident matrix for unit cell $j$ and between $\mu$ and $\nu$ sublattices. Using (\ref{Eq: connectivity matrix expression for H(1)}) to decompose
\begin{equation}
    H^{(1)}(\mathbf{q}) = h^{v \leftarrow c}(\mathbf{q}) h^{c \leftarrow v}(\mathbf{q})
\end{equation}
where
\begin{equation}
\begin{split}
    h^{v \leftarrow c}_{\mu, X}(\mathbf{q}) &= \sum_{j} h^{v \leftarrow c}_{(0,\mu),(j,X)}e^{i(- \mathbf{R}_j + \mathbf{r}_\mu -\mathbf{r}_X^c)\cdot\mathbf{q}} \\
    &= \sum_{k} h^{v \leftarrow c}_{(k,\mu),(0,X)}e^{i(\mathbf{R}_k + \mathbf{r}_\mu -\mathbf{r}_X^c)\cdot\mathbf{q}}, \\
    &= \left[h^{c \leftarrow v}(\mathbf{q})\right]^\dagger_{\mu, X}
\end{split}
    \label{Eq: h fourier transform}
\end{equation}
is the $n_s \times n_c$ connectivity matrix in reciprocal space, with $\mathbf{r}_X^c$ denoting the position of the cluster center of type $X$ among unit cell $j$. Here we recover the previous results. For $n_s > n_c$, the rectangular nature of the connectivity matrix implies at least $n_s - n_c$ flat bands associated with a zero energy. Furthermore, as the non-zero eigenvalues $\lambda_\mu(\mathbf{q})$ can be expressed as the square modulus of the singular values of $h^{v \leftarrow c}_{\mu, X}(\mathbf{q})$, i.e. $\lambda_\mu(\mathbf{q}) = |a_\mu(\mathbf{q})|^2$, all the dispersive bands are positive and thus located above the flat bands for cluster Hamiltonians. While the number of flat bands relies on the number of constraint -- i.e. the number of cluster $n_c$ -- per unit cell, the structure of the dispersive bands depends on the interaction strength associated with each bond, ruled by the coefficients $\gamma_i^n$ present in the connectivity matrix $h^{v \leftarrow c}$ and the constraint vectors, which are thus the proper tools to characterize the band structure of the system.

Finally, since $h^{v \leftarrow c}_{(k,\mu),(0,X)} \equiv \gamma_{k,\mu}^X$ in Eq.~(\ref{Eq: h fourier transform}), one automatically obtains that the columns of the connectivity matrix in Fourier space are the constraint vectors $L_X(\mathbf{q})$
\begin{equation}
    h_{\mu, X}^{v \leftarrow c}(\mathbf{q}) = \left( L_X\right)_\mu(\mathbf{q}).
    \label{eq:hL}
\end{equation}
A similar result has been obtained in \cite{Fang_2024} where the column matrix $[\mathbf{L}_1(\mathbf{q}), \mathbf{L}_2(\mathbf{q}),..., \mathbf{L}_X(\mathbf{q}),...]$ represents the hopping term between the two sublattices of a virtual bipartite construction, similar to the connectivity matrix.

\subsection{Band structure and Coulomb physics}
\label{sec:pinchpoint}

It has been established \cite{Benton_Moessner_2021,Yan_2024_short, Yan_2024_long, Davier_2023, Fang_2024} that in cluster systems exhibiting flat bands at the bottom of their spectrum, the constrainers and associated constraint vectors play a crucial role in determining the nature of the emergent spin liquid. Specifically, if the dimension of the vector space spanned by the constraint vectors --that is the one associated with dispersive bands-- decreases at some momentum point corresponding to a dispersive band touching the set of flat bands, then it implies the emergence of a Coulomb phase. This phase manifests as a pinch point in the structure factor, reflecting long-range spin-spin correlations.
The Coulomb phase is governed by a Gauss law acting on an effective electric field that emerges from the spin components. The nature of this emergent field is determined by the structure of the band-touching point. If the local dispersion near the contact point is of order $2n$, the system obeys a Gauss law of order $n$ acting on a rank-$n$ tensor field. Such higher-order laws are typically associated with $n$-fold symmetric pinch points in the structure factor\cite{Prem_2018,Yan_2020,Benton_Moessner_2021,Yan_2024_short, Yan_2024_long, Davier_2023}.

While a full discussion of the various Coulomb phases is not the purpose of this work, we introduce here the main features associated with the band structure in order to be able to discuss the impact of cluster-cluster interactions on the emerging Coulomb phase. We refer the interested reader to \cite{Yan_2024_long, Davier_2023, Fang_2024} for a detailed presentation. The key point is that the nature of the Coulomb phase is entirely determined by the structure of the vector space associated with the dispersive bands, which itself is generated by the constraint vectors introduced in Eq.~(\ref{Eq: general cluster H : Fourier version}). These constraints are encoded in the connectivity matrix $h^{v \leftarrow c}(\mathbf{q})$, whose columns are the constraint vectors.
Since ground state configurations must be orthogonal to all constraint vectors, they lie in the vector space orthogonal to the one spanned by the constraints—i.e., the space associated with the dispersive bands. Consequently, the spin-spin correlation functions in reciprocal space, encoded in the static structure factor $\mathcal{S}(\mathbf{q})$, must be proportional\cite{Henley2005} to the projection operator $\Pi$ onto this orthogonal space
\begin{equation}
    \mathcal{S}(\mathbf{q}) = \sum_{\mu, \nu} \langle \mathbf{S}_\mu (\mathbf{q}) \cdot \mathbf{S}_\nu (\mathbf{q}) \rangle  \propto \sum_{\mu, \nu} \Pi_{\mu \nu}.
\end{equation}
Because the Fourier-transformed connectivity matrix $h^{v \leftarrow c}(\mathbf{q})$ contains the constraint vectors as columns, the projector $\Pi$ can be constructed as\cite{Henley2005, Davier_2023}
\begin{equation}
    \Pi = I - h^{v \leftarrow c}\left(\left[h^{v \leftarrow c}\right]^\dagger h^{v \leftarrow c} \right)^{-1}\left[h^{v \leftarrow c}\right]^\dagger.
    \label{Eq : Projector Pi at zero temperature}
\end{equation}
This projector becomes singular when the matrix $M \equiv \left[h^{v \leftarrow c}\right]^\dagger h^{v \leftarrow c}$ is non-invertible—i.e., when its determinant vanishes. Since this determinant is the product of the dispersive band dispersions, a zero determinant indicates that a dispersive band is touching a flat band. Thus, any such band touching point necessarily results in a pinch point in the structure factor \cite{Yan_2024_short}. This occurs when one of the constraint vectors either vanishes or becomes linearly dependent on the others; see Appendix~\ref{Appendix : Band structure and Gauss laws} for further details. 
At such touching point $\mathbf{q}^\star$, there exists a critical vector $\mathbf{L}_c$ that can be expressed as a linear combination of the constraint vectors
\begin{equation}
    \mathbf{L}_c = \sum_{X = 1}^{n_c} \alpha_X \mathbf{L}_X, \quad \|\bm{\alpha}\|^2 = 1
\end{equation}
satisfying $\mathbf{L}_c(\mathbf{q}^\star) = 0$. Since all ground state configurations must satisfy the constraint equations
\begin{equation}
    \mathbf{L}_X(\mathbf{q}) \cdot \mathbf{S}(\mathbf{q}) = \sum_{\mu = 1}^{n_s} L_X^\mu(\mathbf{q}) \mathbf{S}_\mu(\mathbf{q}) = 0
\end{equation}
for every constraint vector $\mathbf{L}_X(\mathbf{q})$, the same must hold for the critical vector 
\begin{equation}
    \mathbf{L}_c(\mathbf{q}) \cdot \mathbf{S}(\mathbf{q}) = 0.
\end{equation}
However, at the contact point $\mathbf{q}^\star$, this last constraint becomes trivial since $\mathbf{L}_c(\mathbf{q}^\star) = 0$. In the vicinity of this point, the critical vector can be expanded in powers of $\delta \mathbf{q} = \mathbf{q} - \mathbf{q}^\star$, yielding the relation
\begin{equation}
    \sum_\mu \sum_{j = 0}^n a_{\mu,j} (\delta q_1)^j (\delta q_2)^{n-j} \mathbf{S}_\mu(\mathbf{q_\parallel}) = 0
    \label{Eq: Gauss law in q space}
\end{equation}
in planes $\delta\mathbf{q_\parallel} = (\delta q_1, \delta q_2)$ that contain the pinch point, where $a_{\mu,j}$ are the Taylor coefficients of $\mathbf{L}_c$ around $\mathbf{q}^\star$.
These equations correspond in real space to Gauss laws of order $n$
\begin{equation}
    \sum_\mu \sum_{j = 0}^n a_{\mu, j} (\partial_1)^j (\partial_2)^{n-j} \mathbf{S}_\mu(\mathbf{R}_\parallel) = 0,
    \label{Eq: Gauss law in real space}
\end{equation}
applying on the the coarse-grained spin fields, denoted by $\mathbf{S}_\mu( \mathbf{R}i) = \mathbf{S}{\mu,i}$. The order $n$ of the lowest nontrivial term in the expansion of the critical vector $\mathbf{L}_c$ determines the local dispersion of the touching band, which must scale as $|\delta \mathbf{q}|^{2n}$ (see Appendix~\ref{Appendix C.3}). This implies that the order of the Gauss law is dictated by the local structure of the band-touching, which in turn controls the $2n$-fold symmetry of the corresponding pinch point in the structure factor\cite{Yan_2024_long}. The constraint vectors thus provide a natural and effective framework for analyzing the emergent physics of cluster systems. 

\subsection{When flat bands lead to spin liquids}
\label{sec:flatbandliquid}

Our discussion so far relies on the Luttinger–Tisza approximation (LTA). In order for the results based on the flat band states obtained within this framework to remain valid, these modes must be appropriately combined to construct real-space ground state configurations that satisfy the spin length constraint $|\mathbf{S}_i| = 1$ for each spin. This construction requires a sufficient number of flat bands to ensure enough freedom when combining flat band modes into normalized real-space configurations. A natural question in this context is: how can one estimate -- at least qualitatively -- the minimal number of flat bands needed to preserve the extensive ground-state degeneracy and the corresponding Coulomb phase description ? This question can be addressed by evaluating, in real space, the effective number of zero modes per cluster or unit cell, this time incorporating the strict constraint of fixed spin length \cite{Moessner_1998_counting,Davier_2023}.

The effective number of zero modes per unit cell $F$ is the number of effective degrees of freedom per unit cell that remain unconstrained in the ground state. Each unit cell contains an effective number of sites that is equal to the number of sublattices $n_s$, together with a number of clusters $n_c$. On each of these sites there is a spin of dimension $n_d$ ($n_d=3$ for Heisenberg spins), which possesses $n_d$ degrees of freedom but has a constrained length $|\mathbf{S}|=1$, and is thus associated with $n_d-1$ effective degrees of freedom. As each cluster undergoes a constraint $\bm{\mathcal{C}} = 0$ of dimension $n_d$, this implies that there are $n_c \, n_d$ constraints per unit cells. The effective number of free degrees of freedom (or zero modes) per unit cell can thus simply be expressed as
\begin{equation}
    F = n_s(n_d-1) - n_c\, n_d.
    \label{Eq: Number of zero modes per u.c}
\end{equation}
If the number of zero modes is strictly positive this means that there exists some free degrees of freedom within each clusters, allowing for an extensive degeneracy of the ground state manifold, in good agreement with the existence of flat bands in the spectrum. In fact this guaranties the number of flat bands is big enough to be able to build real space ground states from flat bands states that are satisfying the spin length constraint $|\mathbf{S}|=1$.

If the number of zero modes per unit cell is zero, as it is the case for the nearest-neighbor kagome antiferromagnet \cite{Moessner_1998_counting}, this does not necessarily mean that there exists no zero modes, but that such modes do not rely on a single unit cell. For the kagome lattice for example, there exists zero modes called weather-vane modes that are relying on a group of six triangular clusters \cite{Chalker1992,Zhitomirsky2008}. In this marginal case, there exist some possibilities to combine flat band states to produce an extensive number of real space ground states satisfying the spin fixed length constraint.

If the number of zero modes obtained using Eq.~(\ref{Eq: Number of zero modes per u.c}) is negative, it means that recombining flat bands states to produce real space spins configurations while enforcing the spin length constraint amounts to losing the extensive degeneracy. In other words, restraining the LTA ground state manifold by imposing the spin length constraint reduces this manifold to a sub-manifold whose dimension does not scale with the system size; it likely leads to magnetic order at very low temperature. Note however that these systems can remain good candidates for quantum spin liquids\cite{Morita_2016_Heisenberg_Diamond_Spin_Lattices, Caci_2023_decorated_square}, as relaxing the spin length constraint can be seen as a way to incorporate quantum fluctuations of the spins\cite{Kimchi_2014}.

The good condition to look for isotropic cluster systems hosting classical spin liquid is thus to ask for $F \geq 0$, that is equivalent to ask the number of flat bands to satisfy
\begin{equation}
    n_{f.b} = n_s - n_c \geq \frac{n_s}{3}
\end{equation}
for Heisenberg spins, and where $n_s$ is to be seen here as the total number of bands. For Heisenberg spins, this corresponds to systems where at least one third of the band are flat.\\

These discussions have highlighted why cluster-based systems are excellent candidates for hosting classical spin liquid phases. This naturally raises the question: to what extent can these systems be modified while preserving the intrinsic properties responsible for the emergence of flat bands and Coulomb phases? In the following section, we propose an extension of the cluster Hamiltonian that modifies both the interaction matrix while preserving the band structure that underlies the emergent classical spin liquid phase of the parent system.

\section{Interacting clusters Hamiltonian}
\label{Sec: Interacting clusters Hamiltonian}
Let us consider a generalization of the cluster Hamiltonian (\ref{Eq: general cluster H}) with an additional term mediating an interaction between different clusters,
\begin{equation}
    \mathcal{H} = \alpha \sum_n|\bm{\mathcal{C}}_n|^2 + 2\eta \sum_{\langle m,n \rangle} \bm{\mathcal{C}}_m\cdot \bm{\mathcal{C}}_n.
    \label{Eq: general H(2)}
\end{equation}
where the constrainers $\bm{\mathcal{C}}_n$ are defined as before by Eq.~(\ref{Eq : general constrainer}). For $\eta=0$, one recovers the parent cluster Hamiltonian with $\alpha$ the intra-cluster interaction strength. The second sum runs over pairs of neighboring clusters with $\eta$ a real coefficient fixing the interaction strength of neighboring clusters. Specific examples of Hamiltonian (\ref{Eq: general H(2)}) have been studied on the checkerboard, pyrochlore and kagome lattices \cite{rau16b,udagawa16a,mizoguchi17a,Mizoguchi_2018,Tokushuku2019,Lugan22b} where the first term was the standard nearest-neighbor antiferromagnetic term on tetrahedral and triangular units respectively, while the second term was due to second- and third-neighbor couplings between spins (with $J_2=J_3$). It means that despite its apparent complexity due to its generic nature, relatively simple and realistic models can be described by Hamiltonian (\ref{Eq: general H(2)}).

\subsection{Generic properties} 
\label{Subsec: Generic properties}

\subsubsection{Constraint vector framework} 

Let us define $z_c$ the coordination number of a spin cluster, i.e. the number of clusters linked to a given cluster. The interacting cluster Hamiltonian (\ref{Eq: general H(2)}) can then be rewritten as
\begin{equation}
    \mathcal{H} = \left(\alpha - z_c \eta \right) \sum_n|\bm{\mathcal{C}}_n|^2 + \eta \sum_{\langle m,n \rangle} \left( \bm{\mathcal{C}}_m + \bm{\mathcal{C}}_n \right)^2.
    \label{Eq: general H(2) Cm + Cn}
\end{equation}
Note that if this coordination number is unique for most cluster lattices, it is not necessarily always true. One could take for example the square octagon lattice of Table \ref{tab: cluster systems} where octagonal cluster have coordination number $z_{c,o} = 8$ while square ones have $z_{c,s} = 4$. In this case the correct Hamiltonian would be
\begin{equation}
    \mathcal{H} = \sum_n \left(\alpha - z_{c,n} \eta \right) |\bm{\mathcal{C}}_n|^2 + \eta \sum_{\langle m,n \rangle} \left( \bm{\mathcal{C}}_m + \bm{\mathcal{C}}_n \right)^2.
\end{equation}
This does not change the outcome of our discussion though, and we shall consider a unique $z_c$ for pedagogical reasons. In particular, Hamiltonian (\ref{Eq: general H(2) Cm + Cn}) makes it clear that configurations satisfying $\bm{\mathcal{C}}_n = 0$ for all clusters $n$ in the system have zero energy, and that for $\alpha-z_c \eta > 0$ these are the ground state configurations. In reciprocal space we get
\begin{equation}
    \begin{split}
        \mathcal{H} &= \frac{\alpha - z_c \eta}{N_{u.c}}\sum_\mathbf{q} \sum_{X} \left| \mathbf{L}_X(\mathbf{q}) \cdot \mathbf{S}(\mathbf{q}) \right|^2 \\
    &+ \frac{\eta}{N_{u.c}} \sum_\mathbf{q} \sum_{(X Y)} \left| \bm{\mathcal{L}}_{XY}(\mathbf{q}) \cdot \mathbf{S}(\mathbf{q}) \right|^2 
    \end{split}
    \label{Eq: general H(2) Fourier}
\end{equation}
where the sum $\sum_{(X Y)}$ runs over the $N_{c.l}$ types of cluster to cluster links; see Table \ref{tab: cluster systems} for the value of $N_{c.l}$ for different lattices together with illustrations of the premedial-lattice sites depicting these different types of links. Cluster-pair constraint vectors $\bm{\mathcal{L}}_{XY}(\mathbf{q})$ can be defined from cluster constraint vectors $\mathbf{L}_X$ as
\begin{equation}
    \bm{\mathcal{L}}_{XY}(\mathbf{q}) = e^{-i \mathbf{r}_{XY}\cdot\mathbf{q}} \mathbf{L}_X(\mathbf{q}) + e^{i \mathbf{r}_{XY}\cdot\mathbf{q}} \mathbf{L}_Y(\mathbf{q})
    \label{Eq: Cluster-pair constraint vectors definition}
\end{equation}
where the vector $\mathbf{r}_{XY}$ is half the vector that links the center of a cluster of type $X$ to a neighboring cluster of type $Y$. From the expression (\ref{Eq: general H(2) Fourier}) of the Hamiltonian it appears that zero energy modes forming flat bands are spin modes that satisfy both 
\begin{equation}
    \mathbf{L}_X(\mathbf{q}) \cdot \mathbf{S}(\mathbf{q}) = 0 \qquad \& \qquad  \bm{\mathcal{L}}_{XY}(\mathbf{q}) \cdot \mathbf{S} (\mathbf{q}) = 0.
        \label{Eq: Cluster-pair constraint vectors zero}
\end{equation}
Since the cluster-pair constraint vectors $\bm{\mathcal{L}}_{XY}$ are simply linear combinations of cluster constraint vectors $\mathbf{L}_X$, this flat band manifold turns out to be identical to the one of the parent cluster Hamiltonian when $\eta=0$, i.e. the vector space orthogonal to all cluster constraint vectors $\mathbf{L}_X$. This implies that the number of flat bands associated with zero energy, as well as their attached eigenstates, will remain the same when switching on the inter cluster interaction $\eta$. Note that this result remains valid even if we attach a specific coefficient $\eta_{XY} = \eta_{YX}$ for each type of cluster pairs $X-Y$ among the system.

As introduced in section \ref{sec:pinchpoint}, ground-state flat bands can support pinch points in the structure factor at wavevectors $\mathbf{q}^\star$ where a dispersive band becomes gapless \cite{Yan_2024_short}. Since the dispersive bands are entirely defined by the cluster constraint vectors, Eqs.~(\ref{Eq: Cluster-pair constraint vectors definition}) and (\ref{Eq: Cluster-pair constraint vectors zero}) imply not only the persistence of flat bands up to $ \eta <\alpha/z_c$, but also that the structure and number of pinch points, if any, are expected to be conserved. We shall now make this result more transparent using connectivity matrices. We shall focus on $\eta\geq 0$ even if many of our results also apply to the negative case.

\begin{figure}[t]
    \centering
    \includegraphics[width=0.9\linewidth]{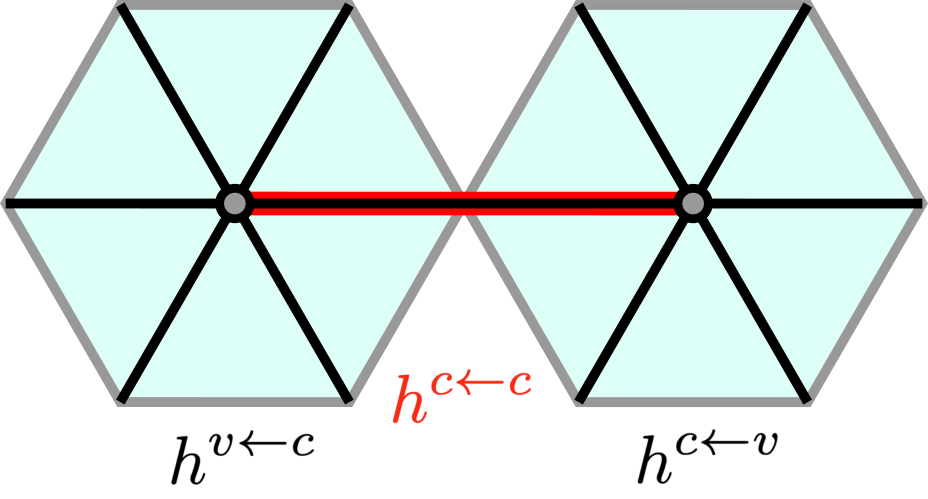}
    \caption{Schematic representation of Eq. (\ref{Eq: connectivity matrix expression for H(2)}). The scalar product of two constrainers associated with two neighboring clusters can be translated in term of connectivity matrices by introducing two different types of virtual bonds. The first types are the one already used for cluster Hamiltonians. These virtual bonds that link all vertices sites from a cluster to its virtual central site, are here depicted as black lines, and are associated with the connectivity matrix $h^{v \leftarrow c}$ as already discussed. The second type of virtual links are bonding the central sites from neighboring clusters together, they are depicted in red and can be encoded in the incident matrix $h^{c \leftarrow c}$ which coefficients $h^{c \leftarrow c}_{mn}$ are unity if the central sites $m$ and $n$ belong to two interacting clusters and zero otherwise. }
    \label{fig: cluster to cluster interaction matrix H(2)}
\end{figure}

\subsubsection{Connectivity-matrix formalism} 

The connectivity-matrix framework can also be directly applied to analyze the cluster interacting Hamiltonian (\ref{Eq: general H(2)}) as the second part of this Hamiltonian, proportional to $\eta$, naturally translates in this formalism as 
\begin{equation}
    2\sum_{\langle m,n \rangle} \bm{\mathcal{C}}_m\cdot \bm{\mathcal{C}}_n \quad \to \quad H^{(2)} = h^{v \leftarrow c}  h^{c \leftarrow c}  h^{c \leftarrow v}
    \label{Eq: connectivity matrix expression for H(2)}
\end{equation}
where the matrix $h^{c \leftarrow c}$ is a $N_c \times N_c$ connectivity matrix linking cluster centers to centers of neighboring clusters. Its coefficients $h^{c \leftarrow c}_{mn}$ are thus equal to 1 if the clusters $m$ and $n$ are neighboring clusters, and zero otherwise. The factor $2$ in front of the left term comes from double counting when the connectivity matrix $H^{(2)}$ is inserted into Hamiltonian (\ref{Eq: H(1) connectivity matrix}). The interacting-cluster Hamiltonian thus writes as
\begin{equation}
    H = h^{v \leftarrow c} \left(\alpha  I_{N_c} + \eta   h^{c \leftarrow c} \right) h^{c \leftarrow v} \equiv h^{v \leftarrow c} g^{c \leftarrow c} h^{c \leftarrow v}
\end{equation}
with $I_{N_c}$ the identity matrix of dimension $N_c$, and where $g^{c \leftarrow c}$ is a square matrix of dimension $N_c \times N_c$. Once expressed in Fourier space this Hamiltonian becomes 
\begin{equation}
    H(\mathbf{q}) = h^{v \leftarrow c}(\mathbf{q}) g^{c \leftarrow c}(\mathbf{q}) h^{c \leftarrow v}(\mathbf{q})
    \label{Eq : H general decomposition hgh}
\end{equation}
where 
\begin{equation*}
    g^{c \leftarrow c}_{X, Y}(\mathbf{q}) = \sum_{j} g^{c \leftarrow c}_{(0,X),(j,Y)}e^{i(- \mathbf{R}_j + \mathbf{r}_X^c -\mathbf{r}_Y^c)\cdot\mathbf{q}}
\end{equation*}
is a $n_c \times n_c$ matrix. In this situation the general rank relation 
\begin{equation}
    \text{Rank}(A B) \leq \text{Min} \left(\text{Rank}(A), \text{Rank}(B)\right)
\end{equation}
can be used to justify that, because $g^{c \leftarrow c}_{X, Y}(\mathbf{q})$ has a rank bounded by its dimension $n_c$, and as already stated $h^{v \leftarrow c}(\mathbf{q})$ has a rank inferior or equal to its smallest dimension $n_c$, the product (\ref{Eq : H general decomposition hgh}) must also have a rank smaller or equal to $n_c$.  This imposes that the Hamiltonian of an interacting cluster system must in general be associated with a minimal number $n_{f.b} = n_s - n_c$ of flat bands, as for the parent cluster system; a useful property for flat-band engineering of tight-binding models \cite{Mizoguchi_Masafumi_2019_flat_bands,Rhim_2019,Graf_Piechon_2021_flat_bands,Maimaiti_2021,Udagawa_2024,Nakai_2025}.

\subsection{Band dispersion of uniform interacting-cluster models} 
\label{Sec:unifcluster}

For a family of interacting-cluster models respecting the following conditions, the bands dispersions can be derived directly from the ones of the parent cluster system
\begin{enumerate}[label=(\roman*)]
    \item There is a unique type of cluster with $\xi$ spins per cluster and weighting coefficients $\{\gamma_i\}_{i=1,..,\xi}$.
    \item Only nearest-neighbor clusters share spins, whose number $\omega$ is fixed: all clusters are either corner-sharing ($\omega = 1$), or bond-sharing ($\omega = 2$ a priori), or face-sharing ($\omega \geq 3$).
    \item For any neighboring clusters $n$ and $n'$, the quantity
    \begin{equation}
        \Omega = \sum_{\substack{i=1\\ i\in n \cap n'}}^\omega  \gamma_i^n \gamma_i^{n'},
        \label{eq:Omega}
    \end{equation}
    summed over all spins $i$ belonging to both clusters, is a constant.
\end{enumerate}
Even if condition (iii) looks somewhat complex, it is actually a natural property for a cluster of spins, as long as equivalent spins within a cluster have the same weight $\gamma_i$. Our point here is that we do not need this full equivalence within a cluster, but simply a weaker version in the form of condition (iii). For convenience, we shall refer to a Hamiltonian respecting these three conditions as a uniform interacting-cluster model.\\

For such models, squaring the cluster Hamiltonian connectivity matrix leads to 
\begin{equation}
    \begin{split}
        \left(H^{(1)}\right)^2 &= h^{v \leftarrow c} \left(h^{c \leftarrow v} h^{v \leftarrow c} \right) h^{c \leftarrow v} \\
        &= h^{v \leftarrow c} \left( \Omega h^{c \leftarrow c} + \Xi I_{N_c} \right)h^{c \leftarrow v} \\
        &= \Omega H^{(2)} + \Xi H^{(1)}
        \label{Eq. H(1) square}
    \end{split}
\end{equation}
as illustrated in Fig.~\ref{fig: cluster to cluster interaction matrix H(2)}. The identity contribution $\Xi I_{N_c}$ counts all two–step paths that start at a cluster center, visit a single vertex, and return to the same center. Since the path $n\!\to\! i\!\to\! n$ carries weight $\gamma_i^n$ on both hops, the coefficient is
\begin{equation}
    \Xi = \sum_{i \in n} \left( \gamma_i^n\right)^2. 
\end{equation}
If condition (i) was not satisfied, this sum would depend on the cluster type, yielding a coefficient $\Xi_X$ for each type of cluster $X$, and the corresponding term in Eq.~(\ref{Eq. H(1) square}) would no longer be proportional to the identity.

The term $\Omega\, h^{c \leftarrow c}$ counts two–step paths connecting neighboring cluster centers via spins shared by both clusters. If conditions (ii) and (iii) were not satisfied, this coefficient would depend on the pair, $\Omega_{nn'}$, and the corresponding contribution in Eq.~(\ref{Eq. H(1) square}) would no longer be proportional to $h^{c \leftarrow c}$.

Using Eq.~(\ref{Eq. H(1) square}) to express $H^{(2)}$ in terms of $H^{(1)}$ and substituting into the Hamiltonian (\ref{Eq: general H(2)}) yields
\begin{equation}
    H =  \left(\alpha - \eta \frac{ \Xi }{\Omega }\right) H^{(1)} + \frac{\eta}{\Omega} \left(H^{(1)}\right)^2.
    \label{Eq: H(2) connectivity matrix expression}
\end{equation}
This connectivity matrix is a polynomial of the matrix $H^{(1)}$ and thus possesses the same eigenbasis as each eigenvector of $H^{(1)}$ is automatically an eigenvector of $H$. Denoting the eigenvalues of the parent cluster Hamiltonian $H^{(1)}$ as $\lambda_\mu(\mathbf{q})$ in reciprocal space, the energy spectrum of the interacting cluster Hamiltonian is
\begin{equation}
    \Lambda_\mu(\mathbf{q}) = \left(\alpha - \eta \frac{ \Xi }{\Omega }\right) \lambda_\mu(\mathbf{q}) + \frac{\eta}{\Omega} \lambda_\mu(\mathbf{q})^2. 
    \label{Eq: Dispersion H(2)}
\end{equation}
This dispersion imposes the conservation of the flat bands manifold as $\lambda_\mu(\mathbf{q}) = 0$ naturally imposes $\Lambda_\mu(\mathbf{q}) = 0$, meaning the number of flat bands and their eigenvectors are conserved. This means that for this class of systems, not only the eigenbasis of the dispersive bands is conserved, but each of the Hamiltonian eigenspaces are individually conserved.

This derivation can be straightforwardly generalized to cluster systems composed of inequivalent cluster types and involving multiple kinds of geometrical intersections between clusters, such as the two variants of the square kagome lattice shown in Table~\ref{tab: cluster systems}. Details of this generalization are presented in Appendix~\ref{Appendix:Generalized_cluster_systems}.

The definition of Eq.~(\ref{eq:Omega}) leaves the sign of $\Omega$ undefined a priori. But there is a hidden gauge degree of freedom in Hamiltonian (\ref{Eq: H(2) connectivity matrix expression}) since $\Omega$ always appears as part of the ratio $\eta/\Omega$. We shall from now on consider that
\begin{eqnarray}
\eta/\Omega>0,
\end{eqnarray}
which corresponds to the interesting regime where the first term of Hamiltonian (\ref{Eq: H(2) connectivity matrix expression}) can change sign as a function of $\eta$. Since we chose to study the regime $\eta\geq 0$, it means that we restrict our analysis to positive $\Omega$.

The dispersion relation of Eq.~(\ref{Eq: Dispersion H(2)}) reveals that in interacting-cluster systems, the lowest dispersive band becomes ground state when
\begin{eqnarray}
    \Lambda_{ld}(\mathbf{q}) \leq 0 \Leftrightarrow \lambda_{ld}(\mathbf{q}) \leq \Xi \left(1-\frac{\alpha}{\eta}\frac{\Omega}{\Xi}\right),
    \label{Eq:dispersiveGS}    
\end{eqnarray}
where $ld$ denotes the lowest dispersive band of the interacting–cluster model ($\Lambda_{ld}$) and of the parent model ($\lambda_{ld}$). The relevant control parameter is the ratio
\begin{equation}
    \zeta \;\equiv\; -\frac{\alpha}{\eta}, \qquad \zeta_c \;=\; -\frac{\Xi}{\Omega}<0,
\end{equation}
so the lowest dispersive band becomes the ground state when $\zeta>\zeta_c$. If the parent model has a gap $\Delta$ between the flat band(s) and $\lambda_{ld}$, then the interacting system retains a flat–band ground state up to
\begin{equation}
    \zeta < \zeta_c(\Delta) \;=\; \Big(1 - \frac{\Delta}{\Xi}\Big)\zeta_c,
    \label{eq:zetacdelta}
\end{equation}
with $\zeta_c(0)=\zeta_c$. Thus a finite gap requires stronger inter–cluster coupling to destabilize the flat ground state, as detailed in Sec.~\ref{subsubsec : gapped phase}.\\

If the parent model is gapless ($\Delta=0$), at the threshold $\zeta = \zeta_c$, the local band dispersion around the band-touching point $\mathbf{q}^\star$ is squared compared to the rank order $n$ of the parent cluster Hamiltonian
\begin{eqnarray}
\Lambda_{ld}(\mathbf{q}^\star+\delta \mathbf{q}) = \frac{\eta_c}{\Omega} \lambda_\mu(\mathbf{q}^\star+\delta \mathbf{q})^2\propto\frac{\eta_c}{\Omega} (\delta \mathbf{q})^{2n},
\end{eqnarray}
at lowest order in $\delta \mathbf{q}$ and with $\eta_c = -\alpha / \zeta_c$. Naively, this observation would suggests that the critical point $\zeta = \zeta_c$ may be generically associated with the emergence of an exotic Coulomb phase, characterized by higher-rank Gauss laws and multifold pinch points (see explanations in section \ref{sec:pinchpoint}).
However, the correspondence between the dispersion order $2n$ near a contact point and the order $n$ of the associated Gauss law crucially depends on the Hamiltonian being a Gram matrix constructed from the constraint vectors; we refer the interested reader to Appendix~\ref{Appendix : Band structure and Gauss laws} for a technical explanation. This condition no longer holds in the case of the interacting cluster Hamiltonian, which instead takes the form of a second-order polynomial of such a Gram matrix. It means that the nature of the Coulomb phase remains unchanged for any $\zeta \leq \zeta_c$, including $\zeta_c$, as long as the flat bands remain the lowest-energy states and the eigenvector space of the dispersive bands is preserved.\\

From Eqs.~(\ref{Eq: general H(2) Cm + Cn}) and (\ref{Eq: general H(2) Fourier}), one might naively conjecture that the critical value of $\zeta$ takes the form $\zeta_c=-z_c$, which in turn would suggest
\begin{eqnarray}
z_c \;=\; \frac{\Xi}{\Omega}
\;=\; \frac{\sum_{i \in n} \left( \gamma_i^n\right)^2}{\sum_{i \in n \cap n'}  \gamma_i^n \gamma_i^{n'}}.
\label{eq:zcXiOmega}
\end{eqnarray}
This identity can be enforced by fine–tuning the interaction weights $\gamma_i^n$, but it does not hold generically. Nevertheless, Eq.~(\ref{eq:zcXiOmega}) is valid under certain geometric conditions. Assuming $\gamma_i^n=\mathrm{cst}$ for all sites $i$ and clusters $n$, the weights factor out of Eq.~(\ref{eq:zcXiOmega}), yielding $\Xi/\Omega=\xi/\omega$. If, in addition, each site belongs to exactly two neighboring clusters, then $\xi=\omega\,z_c$ follows automatically, and hence $\zeta_c=-z_c$. This condition is naturally satisfied by all corner–sharing lattices (e.g., kagome and pyrochlore, as derived in \cite{Mizoguchi_2018}, as well as trillium, hyperkagome, checkerboard, ruby); it also holds for bond–sharing lattices provided the two sites of each bond belong to exactly two clusters (e.g., the square–octagon lattice in Table~\ref{tab: cluster systems} when only the octagonal clusters are considered).\\

Beyond these specific models, the failure of Eq.~(\ref{eq:zcXiOmega}) in generic interacting–cluster systems indicates that the minima of the second terms in Eqs.~(\ref{Eq: general H(2) Cm + Cn}) and (\ref{Eq: general H(2) Fourier}) are not generally zero. Consequently, within the constraint–vector formalism one must explicitly evaluate
\begin{equation}
    \sum_{(X Y)} \left| \bm{\mathcal{L}}_{XY}(\mathbf{q}) \cdot \mathbf{S}(\mathbf{q}) \right|^2
    \label{Eq : sum of cluster-cluster constraint vectors}
\end{equation}
to determine the threshold $\zeta_c$ above which the flat bands of the interacting–cluster system cease to be the ground state. This is illustrated by systems with a single type of cluster—and thus a single constraint vector—for which the interaction–matrix formalism allows an explicit computation of the sum. In such cases the cluster–pair constraint vectors are proportional to the unique constraint vector, implying
\begin{equation}
    \sum_{(X Y)} \left| \bm{\mathcal{L}}_{XY}(\mathbf{q}) \cdot \mathbf{S}(\mathbf{q}) \right|^2
    \;=\; b(\mathbf{q})\,\left|\mathbf{L}(\mathbf{q})\cdot \mathbf{S}(\mathbf{q}) \right|^2,
\end{equation}
with a real, non–negative coefficient $b(\mathbf{q})$ encoding the cluster–cluster connectivity. Combining this with Eq.~(\ref{Eq: general H(2) Fourier}) gives the dispersion
\begin{equation}
    \Lambda_\mu(\mathbf{q})
    \;=\; (\alpha - z_c \eta)\, |\mathbf{L}(\mathbf{q})|^2 \;+\; \eta\, b(\mathbf{q})\, |\mathbf{L}(\mathbf{q})|^2.
    \label{Eq : second dispersion relation}
\end{equation}
Since $\lambda(\mathbf{q})=|\mathbf{L}(\mathbf{q})|^2$, a direct comparison between Eqs.~(\ref{Eq: Dispersion H(2)}) and (\ref{Eq : second dispersion relation}) yields
\begin{equation}
    \sum_{(X Y)} \left| \bm{\mathcal{L}}_{XY} \cdot \mathbf{S} \right|^2
    \;=\; \left( z_c - \frac{\Xi}{\Omega} + \frac{\lambda(\mathbf{q})}{\Omega} \right)
    \left|\mathbf{L}\cdot \mathbf{S} \right|^2.
    \label{Eq: Sum LXY}
\end{equation}
Thus, this term does not, in general, have zero as its minimum value. Hence, one cannot infer the value of $\zeta_c$ from Eq.~(\ref{Eq: general H(2) Cm + Cn}) alone. For uniform interacting-cluster systems where the dispersion relation (\ref{Eq: Dispersion H(2)}) applies, the interaction–matrix formalism is therefore the appropriate framework to determine the critical parameter $\zeta_c$ governing the transition from pinch points to, as we shall now see, half–moons.

\section{Half moons}
\label{Sec: Half moons}

From now on, we shall focus on gapless cluster models where $\Delta=0$ (with the exception of section \ref{subsubsec : gapped phase}), whose structure factor bears pinch points at $\alpha=1, \eta=0$ corresponding to the limit $\zeta \to -\infty$. What the previous discussion tells us is that these pinch points persist for $\zeta \leq \zeta_c$, and something changes as $\zeta>\zeta_c$. The first term of Hamiltonian (\ref{Eq: general H(2) Cm + Cn}) is ultimately expected to break the $\bm{\mathcal{C}}_n = 0$ constraint as $\zeta$ increases, implying that the flat bands cannot be ground state anymore. In parallel, minimizing $|\bm{\mathcal{C}}_n + \bm{\mathcal{C}}_m|$ induces strong correlations between neighboring clusters that are known to be responsible for half-moon patterns in the equal-time structure factor of kagome and pyrochlore lattices with further-neighbor spin couplings \cite{rau16b,udagawa16a,mizoguchi17a,Mizoguchi_2018,Yan_2018}. Here, it is this formation of zero-energy half moons that we shall generalize beyond the canonical kagome and pyrochlore lattices and, noticeably, beyond regular two-fold pinch points.\\


But to describe half moons, we need to know what an equal-time structure factor is. We refer the reader to Appendix \ref{Appendix : SCGA} for its definition and for a presentation of the Self-Consistent Gaussian Approximation\cite{Garanin_SCGA, kagome_largeN} (SCGA) used to compute it. For the sake of the paper, one simply needs to understand that, in the SCGA formalism, the static structure factor takes the form
\begin{equation}
    \mathcal{S}(\mathbf{q}) = 
    \sum_{\kappa = 1}^{n_s}
    \frac{W_\kappa(\mathbf{q})}{\lambda + \beta \varepsilon_\kappa(\mathbf{q})}, 
\end{equation}
where 
\begin{equation}
    W_\kappa(\mathbf{q}) \equiv \sum_{\mu, \nu} \left[\psi_\kappa ^*(\mathbf{q})\right]_\mu \left[\psi_\kappa(\mathbf{q})\right]_\nu
\end{equation}
represent the weight of the eigenstates $\psi_\kappa$, and thus obey the sum rule over all bands $\kappa$
\begin{equation}
    \sum_{\kappa = 1}^{n_s} W_\kappa(\mathbf{q}) = n_s.
\end{equation}
%

\subsection{General theory of half moon patterns on uniform interacting-cluster systems}
\label{Subsec: Half moon, general theory}

\begin{figure*}[t]
    \centering
    \includegraphics[width=\linewidth]{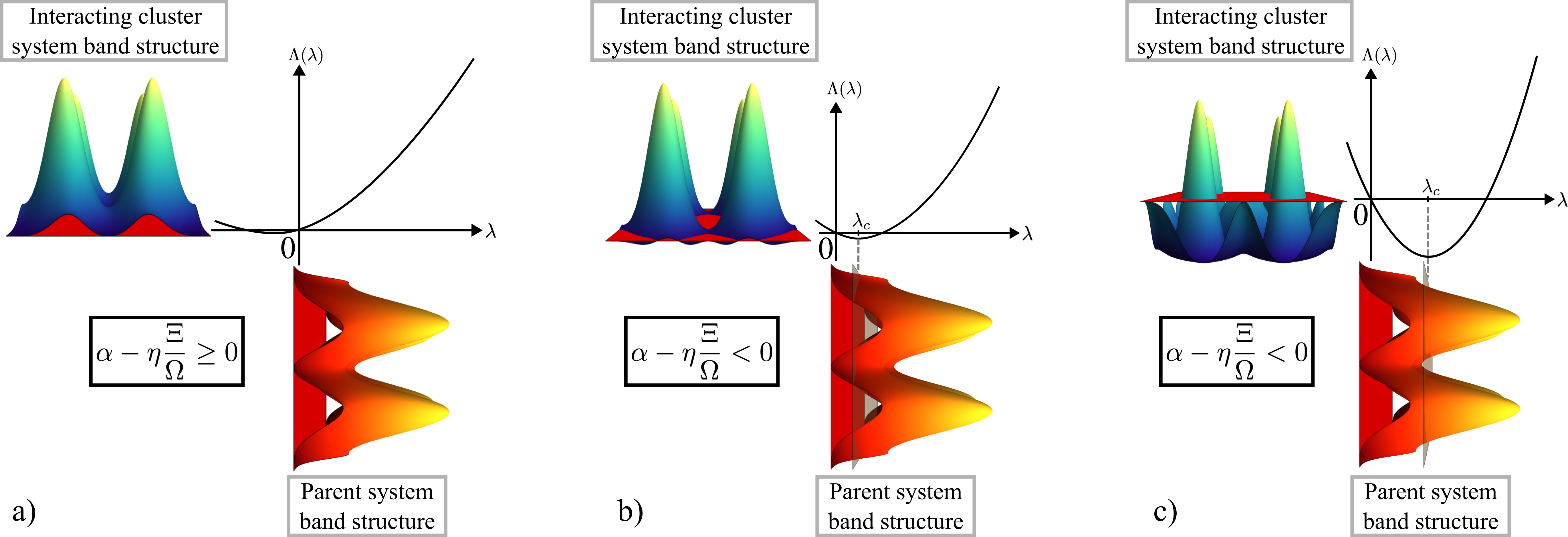}
    \caption{Effective Fermi-level illustration. The band structure of an interacting-cluster system can be written as a polynomial function $\Lambda(\lambda)$ of the band structure of the parent cluster system (see Eq.~(\ref{Eq: Dispersion H(2)})). 
    (a) Since the parent spectrum contains only positive energies, a positive value of the linear coefficient $\alpha - \eta \Xi / \Omega$ results in a simple, non-uniform rescaling of the parent band structure. 
    (b-c) When the linear coefficient $\alpha - \eta \Xi / \Omega$ becomes negative, the polynomial ceases to be strictly increasing, and the minima of the polynomial select the intersections with the parent band structure as the new ground-state manifold. 
    (b) Just beyond the transition, the new ground-state manifold $\mathcal{Q}$ consists of closed contours encircling the contact points of the parent system.
    (c) When the polynomial minimum $\lambda_c$ crosses a saddle point of the parent-system band structure, the topology of the ground-state manifold changes, corresponding to a Lifshitz transition.
    Note that the flat bands of the parent system (depicted in red), which are located at zero energy, remain pinned at zero energy in the interacting-cluster spectrum, since the polynomial $\Lambda(\lambda)$ has no constant term.
  }
    \label{fig:Fermi_level}
\end{figure*}

It is possible to derive a theory for the apparition of half-moon patterns in the structure factor of uniform interacting-cluster models thanks to the known form of the dispersive band given by Eq.~(\ref{Eq: Dispersion H(2)}). For $\zeta > \zeta_c= -\frac{\Xi}{\Omega}$, the energy minima respects the equation
\begin{equation}
    \bm{\nabla}_\mathbf{q} \Lambda_{ld} = \left[ \alpha + \frac{\eta}{\Omega}\left( -\Xi + 2 \lambda_{ld}  \right) \right]  \bm{\nabla}_\mathbf{q} \lambda_{ld} = 0.
    \label{eq:gradlambda}
\end{equation}
The condition $\bm{\nabla}_\mathbf{q} \lambda_{ld} = 0$ is the same as for the parent cluster Hamiltonian ($\alpha=1, \eta=0$) where flat bands were ground state; its physics thus does not apply here where $\zeta > \zeta_c$. On the other hand, canceling the prefactor of Eq.~(\ref{eq:gradlambda}) defines the condition on $\lambda_{ld}$ for the energy minimum of the interacting-cluster system. In reciprocal space, the position of this energy minimum is given by the manifold $\mathcal{Q}$
\begin{equation}
    \left\{\mathbf{q}\in\mathcal{Q} \;\Big|\; \lambda_{ld}(\mathbf{q}) = \frac{\Xi}{2}\left(1-\frac{\zeta}{\zeta_c}\right) 
    \equiv \lambda_c \right\}.
    \label{Eq: Energy minima}
\end{equation}
Since we are working on a gapless parent cluster model ($\Delta=0$), the eigenvalue $\lambda_{ld}$ is greater or equal than 0, and the manifold $\mathcal{Q}$ is thus properly defined for $\zeta > \zeta_c$. On this manifold, the minimum energy of the interacting-cluster model is always negative, i.e. below the flat band,
\begin{equation}
    \begin{split}
        \Lambda_{ld}(\mathbf{q}\in\mathcal{Q}) =-\frac{\alpha \Xi }{4}\frac{\zeta_c}{\zeta} \left(\frac{\zeta}{\zeta_c}-1\right)^2 <0.
        \label{Eq: Lambda(lambda_c)}
    \end{split}
\end{equation}
Since $\lambda_{ld}(\mathbf{q}^\star)=0$ by definition of the the band-touching point $\mathbf{q}^\star$, $\mathbf{q}^\star$ remains a band-touching point for all values of $\zeta$, even when the dispersive band $\Lambda_{ld}$ becomes negative. By analytical continuity of Eq.~(\ref{Eq: Dispersion H(2)}), the manifold $\mathcal{Q}$ typically forms a closed contour in 2D or a closed surface in 3D, in other words a hypersurface, around $\mathbf{q}^\star$ in reciprocal space. 
The evolution of the band structure as $\zeta$ increases can be understood visually by noting that the dispersion relation in Eq. (\ref{Eq: Dispersion H(2)}) is given by a polynomial $\Lambda(\lambda)$ acting on the band structure of the parent system, as illustrated in Fig. \ref{fig:Fermi_level}. The spectrum of the interacting system is therefore entirely determined by the behavior of $\Lambda(\lambda)$ over the spectrum of the parent system. In particular, the minimum $\lambda_c$ of the polynomial $\Lambda(\lambda)$ plays the role of an effective Fermi level. Indeed, in spin systems there is no notion of filling, and the low-temperature behavior is therefore fully determined by the ground state. However, the introduction of cluster–cluster interactions allows parent-system states at energy $\lambda_c$ to become ground states of the interacting system. This mechanism enables one to probe the parent-system energy states located around $\lambda_c$. In this sense, $\lambda_c$ acts as an effective Fermi level that can be continuously tuned by varying the strength of the cluster–cluster interactions controlled by $\zeta$.

At low temperature, the eigenstates attached to $\Lambda_{ld}(\mathbf{q}\in\mathcal{Q})$ dominate the spin-spin correlations responsible for the structure factor. If the density of these eigenstates were uniform along $\mathcal{Q}$, one would obtain an approximately circular and spherical contour in 2D and 3D respectively. However, the sum rule imposes that the density of states must be uniform once summed up over \textit{all} bands. Hence, since the highest intensity region of the flat band(s) is the bow-tie shape of the pinch point, the highest intensity region of the dispersive band(s) must be the complementary region of the bow tie in reciprocal space \cite{Mizoguchi_2018,Yan_2018}. Ultimately, it is the intersection between the complementary region of the bow tie and the hypersurface $\mathcal{Q}$ that forms the half-moon patterns of the structure factor.\\

In absence of extensively degenerate flat bands, this half-moon phase is not a traditional spin liquid. But since the manifold $\mathcal{Q}$ is sub-extensive and isomorphic to an $(d-1)-$sphere within a given Brillouin zone, the half-moon phase corresponds to a so-called spiral spin liquid \cite{rastelli79a,bergman07a,Balla19a,Pohle21a,yao_2021,Gao_2022} with unusual, non-local, spin dynamics \cite{Yan_2022}. 
At very low temperatures, beyond the Luttinger–Tisza approximation, thermal order-by-disorder (ObD) may entropically lift the degeneracy of the manifold $\mathcal{Q}$, particularly in $d=3$ dimensions. In $d=2$ dimensions, however, the Hohenberg–Mermin–Wagner theorem forbids the emergence of long-range magnetic order, although the breaking of discrete lattice rotational symmetries remains possible \cite{Mulder10a}. Such symmetry breaking, potentially associated with a Berezinskii–Kosterlitz–Thouless–type transition in the case of $XY$ spins, can only occur at very low temperatures. At intermediate temperatures, thermal fluctuations are instead expected to restore the effective degeneracy of $\mathcal{Q}$, even slightly away from fine tuning. As a consequence, the half-moon features in the equal-time structure factor are expected to remain robust above the temperature regime where ObD takes place.

In any case, for $\zeta>\zeta_c$, the energy minimum of the interacting-cluster Hamiltonian ($\Lambda_{ld}$) comes from the lowest excited dispersive band of the parent cluster Hamiltonian ($\lambda_{ld}$). Since the flat bands support an emergent Coulomb gauge theory, the lowest dispersive band supports a gapless form of the gauge-charge excitations of the Coulomb theory \cite{Mizoguchi_2018,Yan_2018}. 
\emph{From a real-space perspective}, the appearance of half-moon features therefore signals that these gauge charges are no longer merely thermal excitations but become an integral part of the ground-state manifold. For two-fold half moons, these gauge charges take the form of magnetic monopoles, corresponding to flux defects that locally violate the flux-conservation constraint imposed by \(\bm{\mathcal{C}}=0\). Note that for spins of dimension $n_s$, there are $n_s$ equivalent copies of such scalar gauge charges. In simple systems where the constrainer corresponds to the local cluster magnetization, combinations of the $n_s$ types of gauge charges can be interpreted as effective magnetic moments localized on individual clusters. 
For multifold half moons, the situation is more subtle. In this case, the emergent gauge field is no longer vectorial, and the corresponding gauge charges cannot be interpreted as simple magnetic dipoles. Nevertheless, they remain associated with local violations of a tensorial Gauss law, and thus correspond to higher-rank, spatially structured charge defects in real space. These objects represent generalized gauge charges whose correlations define the characteristic magnetic textures underlying multifold half-moon patterns.
The type of flat-band engineering we have introduced via the cluster interaction $\eta$ thus stabilizes magnetic textures where gauge charges have become part of the ground state. At low temperature, fluctuations are restricted to perturbations close to the hypersurface manifold. But at intermediate temperatures, when the flat bands start to be populated, there is a temperature window where a dense phase of gauge charges co-exist with the vacuum of the flat bands. Although this regime may appear reminiscent of a conventional Coulomb phase, the cluster--cluster interaction introduces an additional dominant interaction between gauge charges. This interaction sets a characteristic length scale for their spatial organization, directly related to the inverse radius of the ground-state hypersurface. Because spins are shared between neighboring clusters, it is generally impossible to simultaneously maximize the norms of oppositely oriented neighboring constrainers. This geometric frustration leads to nontrivial charge arrangements in real space, where a maximally charged cluster is typically surrounded by oppositely charged clusters carrying smaller individual charges, as previously observed in Ref.~\cite{Mizoguchi_2018}. 
By varying the parameter $\eta$, one can include more or less of the dispersive band below the flat bands, i.e. qualitatively more or less gauge charges in the low-energy regime.

For models where the thermal order-by-disorder selects a long-range magnetic order at very low temperatures, the ObD transition would correspond to the crystallization of these gauge charges \cite{brooks14a,Mizoguchi_2018}.

\subsection{How to extend the theory to higher-rank half moons}
\label{subsec: interacting cluster model for high rank h-m}

As presented in section \ref{sec:pinchpoint}, higher-rank U(1) spin liquids support emergent Coulomb gauge fields of rank$-n$ where the emergent electric field can be a vector ($n=1$), a matrix ($n=2$), or a tensor of higher rank \cite{Pretko_2_2017}. Their higher-rank structure confers additional conserved quantities on the spin texture, where excitations may naturally take the form of fractons \cite{Pretko_2017}. The most famous signature of higher-rank U(1) spin liquids is probably the $2n-$fold pinch points in the structure factor \cite{Prem_2018,Yan_2020}. At the band-touching point $\mathbf{q}^\star$, the bow-tie shape of regular pinch points splits into $2n$ branches. 

To the best of our knowledge, half-moon patterns have only been studied out of regular (rank$-1$) Coulomb gauge theories \cite{Mizoguchi_2018,Yan_2018}. It is thus tempting to see if our approach could account for multifold half moons, and the physics of higher-rank gauge charges stabilized at zero temperature.\\

There is, however, an issue. In absence of anisotropic exchange \cite{Yan_2020,Lozano24a,lozanogomez_2024_preprint,chung_2024_preprint}, higher-rank spin liquids are only known on extended cluster systems \cite{Benton_Moessner_2021, Yan_2024_short, Yan_2024_long, Davier_2023}, i.e. where there is a large overlap between adjacent clusters, and spins belong to more than 2 clusters. This is because the higher-rank nature requires to cancel the first $(n-1)$ order terms in the expansion of each constraint vector component $\mathbf{L}_\mu(\mathbf{q})$ (see Eq.~(\ref{Eq: Gauss law in q space})). The only degrees of freedom available to enable this cancellation are the couple of parameters $(\gamma_{i, \mu}, \mathbf{r}_{i,\mu})$ associated with each site belonging both to the cluster and to sublattice $\mu$. To be able to enforce this cancellation for $n>1$ thus requires to have multiple sites \textit{from a same sublattice} encapsulated in a same cluster, corresponding to the case of an extended cluster. 

The inconvenience is that extended–cluster systems do not satisfy condition (ii) of section \ref{Sec:unifcluster}. Consequently, their lowest dispersive band $\Lambda_{ld}$ does not, in general, take the convenient polynomial form of Eq.~(\ref{Eq: Dispersion H(2)}). We usually have
\begin{equation}
    \Lambda_{ld}(\mathbf{q}) = \alpha\lambda_{ld}(\mathbf{q}) + \eta g(\mathbf{q})
    \label{eq:lambdagq}
\end{equation}
with a given function $g(\mathbf{q})$. The minimum of $\Lambda_{ld}(\mathbf{q})$ has no reason to form a closed contour in reciprocal space, nor to be related to the band-touching point $\mathbf{q}^\star$ of the pinch point. Instead its gradient a priori cancels on a discrete set of $\mathbf{q}$ points only, since $\bm{\nabla}_\mathbf{q} \Lambda_{ld} =0$ corresponds to a set of three equations with three unknown variables $(q_x,q_y,q_z)$. The intersection between the bow tie of a pinch point and the manifold of minimum energy is not expected to produce any remarkable feature. A priori, this would suggest to preclude observing higher–rank half moons within the simple interacting–cluster models governed by Hamiltonian~(\ref{Eq: general H(2)}).\\

Fortunately, there is a way to circumvent the problem. The strategy is to engineer a model in which the lowest dispersive eigenvalue of the interacting system, $\Lambda_{ld}$, becomes a simple function of the parent eigenvalue $\lambda_{ld}$. For simplicity, let us consider parent clusters that satisfy condition (i), i.e. that admit a single constraint vector $\mathbf{L}$ and a single set of weights $\gamma_i^n=\gamma_i \;\forall n$. We then seek an interacting model with a dispersive band $\Lambda$ that is a quadratic polynomial of $\lambda \equiv |\mathbf{L}|^2$,
\begin{equation}
    \Lambda(\mathbf{q}) = \left[\,\alpha +  \eta \left( a + b \lambda(\mathbf{q})\right)\,\right] \lambda(\mathbf{q})
    \label{eq:Lambdaab}
\end{equation}
with $a$ and $b$ some real scalars coefficients. In other words, the function $g(\mathbf{q})$ of Eq.~(\ref{eq:lambdagq}) would be a linear function of $\lambda(\mathbf{q})$, and the Fourier transform of the Hamiltonian would take the form
\begin{equation}
    \mathcal{H} = \frac{1}{N_{u.c}}\sum_{\mathbf{q}}
    \Big[\,\alpha +  \eta \left( a + b|\mathbf{L}|^2 \right)\,\Big] |\mathbf{L}|^2\,
    \left|\mathbf{l}\cdot\mathbf{S}\right|^2,
    \label{eq:hammfhm}
\end{equation}
where $\mathbf{l}(\mathbf{q})=\mathbf{L}(\mathbf{q})/|\mathbf{L}(\mathbf{q})|$ is the dispersive–band eigenvector. We shall soon explain why this type of Hamiltonian supports higher-rank half moons in the equal-time structure factor. But before that, let us justify our strategy by explaining how to design a model with Hamiltonian (\ref{eq:hammfhm}). To this end, we generalize the model with cluster interactions beyond nearest neighbors, that can vary between different pairs
\begin{equation}
    \mathcal{H} = \alpha\sum_n |\bm{\mathcal{C}}_n|^2
    + \eta \sum_{n,p} K_{np}\,\bm{\mathcal{C}}_n\!\cdot\!\bm{\mathcal{C}}_p,
    \label{Eq : general interacting clusters H}
\end{equation}
with translation invariance of the scalar kernel $K_{np}=K_{\mathbf{R}_{pn}}$. $\mathbf{R}_{pn}$ is the vector connecting the centers of clusters $n$ and $p$. Since we only consider one type of cluster, there must be a cluster associated with each unit cell (by definition of a unit cell). $\mathbf{R}_{pn}$ thus also connects the centers of unit cells. In Fourier space, Hamiltonian (\ref{Eq : general interacting clusters H}) becomes
\begin{equation}
    \mathcal{H} = \frac{1}{N_{u.c}}\sum_{\mathbf{q}}
    \Big[\,\alpha +  \eta K(\mathbf{q})\,\Big] \left|\mathbf{L}(\mathbf{q})\cdot\mathbf{S}(\mathbf{q})\right|^2, 
\end{equation}
with 
\begin{equation}
    K(\mathbf{q}) = \sum_{p} K_{\mathbf{R}_{cp}} e^{ - i \mathbf{q} \cdot \mathbf{R}_{cp}}
\end{equation}
summed over all clusters $p$ with respect to a reference cluster $c$. Since the system is translation invariant, $K(\mathbf{q})$ is independent of the reference cluster. The simplest way to reach the target form of Eq.~(\ref{eq:hammfhm}) is to choose $K_{\mathbf{R}}$ such that
\begin{equation}
    K(\mathbf{q}) \equiv |\mathbf{L}(\mathbf{q})|^2,
\end{equation}
where $a=0$ and $b=1$. This is achieved by 
\begin{equation}
    K_\mathbf{R} = \sum_{\mu}\sum_{i\in c\cap\mu}\sum_{j\in c\cap\mu}
    \gamma_{i,\mu}\gamma_{j,\mu}\,\delta_{\mathbf{R},\ \mathbf{r}_{i,\mu}-\mathbf{r}_{j,\mu}} 
    \label{eq:Kansatz1}
\end{equation}
for which
\begin{equation*}
    \begin{split}
        &K(\mathbf{q}) = \sum_\mu \sum_{i\in c\cap\mu}\sum_{j\in c\cap\mu}
        \gamma_{i,\mu}\gamma_{j,\mu} e^{i (\mathbf{r}_{i,\mu}-\mathbf{r}_{j,\mu})\cdot \mathbf{q}} \\
        &= \sum_\mu \left(\sum_{i\in c\cap\mu} \gamma_{i,\mu}  e^{i \mathbf{r}_{i,\mu}\cdot \mathbf{q}}\right)
        \left(\sum_{j\in c\cap\mu} \gamma_{j,\mu} e^{-i\mathbf{r}_{j,\mu}\cdot \mathbf{q}}\right) \\
        &= |\mathbf{L}(\mathbf{q})|^2.
    \end{split}
\end{equation*}

\begin{figure*}[t]
    \centering
    \includegraphics[width=\linewidth]{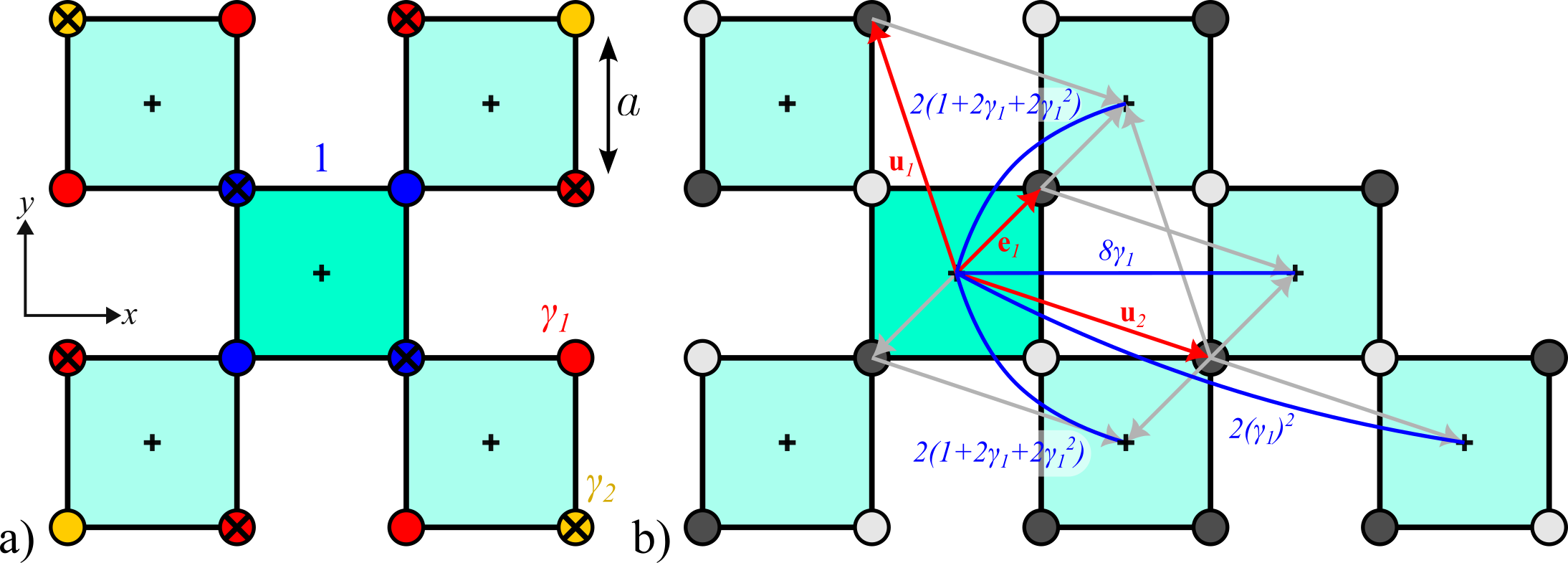}
    \caption{
    (a) Cluster structure of the generalized checkerboard model. The plain and crossed circles depict the two sublattices, while blue, red and yellow circles respectively appear with coefficient 1, $\gamma_1$ and $\gamma_2$ in the constrainer $\bm{\mathcal{C}}_\square$ of Eq.~(\ref{Cckb}). (b) Effective strength of the cluster interactions associated with the kernel $K_\mathbf{R}$ (\ref{Eq: Relevant Kernel}) for the generalized checkerboard, assuming $\gamma_2 = 0$ for clarity of the figure. The vector between two clusters is defined as $\mathbf{R}_{np}$. The strength of each cluster interaction $(n,p)$ is determined by all the combinations of pairs of sites $i$ and $j$ belonging to sublattice $\mu$, whose positions in the cluster respect $\mathbf{r}_{i,\mu} - \mathbf{r}_{j,\mu} = \mathbf{R}_{np}$. Each of these combinations is associated with the products of the weights $\gamma_{i, \mu} \gamma_{j, \mu}/\omega$, with $\omega = 1$ here. Since the two sublattices are equivalent, let us apply our reasoning on one arbitrary sublattice. One has a weight $1$ associated with vectors $\pm \mathbf{e}_1$, and $\gamma_1$ associated with vectors $\pm \mathbf{u}_1, \pm \mathbf{u}_2$. The weight of a cluster-to-cluster link is thus always a combination of terms proportional to either $1^2$, $1\times\gamma_1$ or $(\gamma_1)^2$ as depicted in blue.
    Because of the two possible orientations of the bond vector $\mathbf{R}_{np}$ each cluster-to-cluster link weight must be proportional to two. 
    Note that some links weights can get contributions from two sublattices, as it is the case for the bonds linking first or second neighbors clusters.
    }
    \label{fig:gencheck}
\end{figure*}

This construction relies on the requirement that, within a given sublattice, all relative position vectors $\mathbf{r}_{i,\mu}-\mathbf{r}_{j,\mu}$ are lattice translations $\mathbf{R}_n$, which is naturally satisfied when clusters encapsulate only vertex spins. The resulting interactions between pairs of clusters are illustrated in Fig.~\ref{fig:gencheck}(b) for the example of the generalized checkerboard model.

The resulting dispersive band of Hamiltonian (\ref{Eq : general interacting clusters H}) is given by Eq.~(\ref{eq:Lambdaab}): $\Lambda(\mathbf{q}) = \left[\,\alpha +  \eta \lambda(\mathbf{q})\,\right] \lambda(\mathbf{q})$, whose minimum is either the same as $\lambda(\mathbf{q})$ (i.e. at the band-touching point $\mathbf{q}^\star$) or for $\lambda(\mathbf{q})=\zeta/2$. Since $\lambda(\mathbf{q})$ is always positive (or zero), the second solution only exists for $\zeta>0 \Rightarrow \alpha<0$. While negative values of $\alpha$ will be considered in a following section, let us push our reasoning a little further to see if we can find half moons while preserving the positivity of $\alpha$.

The problem is that condition $K(\mathbf{q}) = |\mathbf{L}(\mathbf{q})|^2$ is too simple. The Ansatz of Eq.~(\ref{eq:Kansatz1}) includes a troublesome, and somewhat unphysical, self-interaction which cancels out the $\eta-$contribution to the prefactor of $\sum_n |\bm{\mathcal{C}}_n|^2$ in the Hamiltonian. Going from Hamiltonian (\ref{Eq: general H(2)}) to its rewritten form (\ref{Eq: general H(2) Cm + Cn}), the prefactor of $\sum_n |\bm{\mathcal{C}}_n|^2$ changes from $\alpha$ to $(\alpha - z_c \eta)$. Thanks to this negative term, $-z_c \eta$, the $\bm{\mathcal{C}}_n=0$ condition ultimately breaks for large enough $\eta$, which is what makes the dispersive band ground state. Since the Ansatz of Eq.~(\ref{eq:Kansatz1}) does not ``renormalize'' $\alpha$, the flat band always remains ground state. This problem is easily solved as it is enough to remove the on–site ($\mathbf{R}=0$) term from the Ansatz.
\begin{eqnarray}
    K_\mathbf{R} &=& \frac{1}{\omega} \left(\sum_{\mu}\sum_{i, j\in c\cap\mu}     \gamma_{i,\mu} \gamma_{j,\mu} \,(\delta_{\mathbf{R},\mathbf{r}_{i,\mu}-\mathbf{r}_{j,\mu}} - \delta_{\mathbf{R},0})\right)\nonumber\\
    &=& \sum_{\mu}\sum_{i, j\in c\cap\mu}
    \frac{\gamma_{i,\mu}\gamma_{j,\mu}}{\omega}\,\delta_{\mathbf{R},\mathbf{r}_{i,\mu}-\mathbf{r}_{j,\mu}} - \frac{\Xi}{\omega} \, \delta_{\mathbf{R},0}.
    \label{Eq: Relevant Kernel}
\end{eqnarray}
The division by $\omega$ in the above Ansatz is not necessary, but it makes for easier comparison with previous results on models with nearest–neighbor cluster interaction only, as it avoids multiple countings of the same nearest–neighbor link. Here $\omega$ is the number of spins shared between adjacent unit cells. In Fourier space we get
\begin{equation}
    K(\mathbf{q}) = \frac{|\mathbf{L}(\mathbf{q})|^2}{\omega} - \frac{\Xi}{\omega}.
\end{equation}
and
\begin{equation}
    \mathcal{H} = \frac{1}{N_{u.c}}\sum_{\mathbf{q}}
    \Big[\,\alpha +  \frac{\eta}{\omega} \left( -\Xi +|\mathbf{L}|^2 \right)\,\Big] |\mathbf{L}|^2\;
    \left|\mathbf{l}\cdot\mathbf{S}\right|^2,
    \label{Eq: relevant generalized cluster interacting Hamiltonian}
\end{equation}
which matches Eq.~(\ref{Eq: Sum LXY}) for simple systems. 
This generalized interacting–cluster model thus has a dispersive band
\begin{equation}
    \Lambda(\mathbf{q}) = \left[\,\alpha +  \frac{\eta}{\omega} \left( -\Xi + \lambda(\mathbf{q})\right)\,\right] \lambda(\mathbf{q}).
    \label{Eq: Dispersion H(2) bis}
\end{equation}
%
\begin{figure*}[t]
    \centering
    \includegraphics[width=0.95\linewidth]{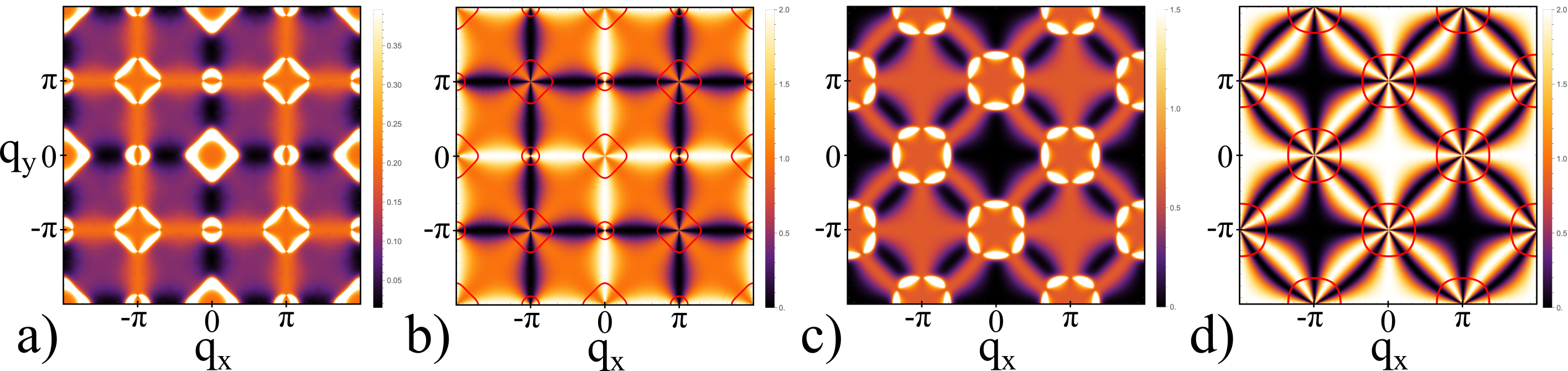} 
    \caption{
    \textbf{Generalized checkerboard model} Figures (a,b) are obtained with constrainer parameters $\gamma_1 = -1/2, \gamma_2 =0, \alpha=1$, which fixes $\Xi = 6$ and thus $\zeta_c = -6$. The parameter $\zeta$ is thus set to $\zeta = -4 > \zeta_c$. 
    The figure (a) presents the structure factor, showing fourfold half moons, in good agreement with figure (b) representing the weight $W_2(\mathbf{q})$ of the dispersive band and the minimum of this dispersive band depicted as a red line. 
    Figures (c-d) are obtained with $\gamma_1 = 1, \gamma_2 = 1/3$, imposing $\Xi = 112/9$ and $\zeta_c = -112/9 \simeq -12.4 $, suggesting the choice $\zeta = -10$. The structure factor (c) presents sixfold half moons, in good agreement with the weight $W_2(\mathbf{q})$ depicted on panel (d) that shows sixfold pinch points. 
    For both cases calculations of the structure factors are performed via SCGA (see Appendix \ref{Appendix : SCGA}) with inverse temperature $\beta = 20$ and with the $q_i$ in $a^{-1}$ units.
    }
    \label{fig: ckb 4f half moon}
\end{figure*}
%
If we redefine the critical parameter as
\begin{equation}
    \zeta_c = -\frac{\Xi}{\omega},
    \label{Eq: eta_c generalized interacting clusters model}
\end{equation}
then, using the same reasoning as for the simple interacting–cluster model, the ground–state manifold for $\zeta>\zeta_c$ again forms a hypersurface,
\begin{equation}
    \left\{\mathbf{q}\in\mathcal{Q} \;\Big|\; \lambda(\mathbf{q}) = \lambda_c = \frac{\Xi}{2}\left(1-\frac{\zeta}{\zeta_c}\right)\right\},
    \label{Eq: Energy minima bis}
\end{equation}
surrounding the band-touching point $\mathbf{q}^\star$ of the parent model. We obtain a spiral spin liquid, and since this strategy is valid for parent models made of extended clusters that can support multifold pinch points at $\mathbf{q}^\star$, the generalized interacting-cluster Hamiltonian can support multifold half moons, as illustrated in the next section.\\

Half moons can thus take exotic, multifold, shapes indicating the proximity of higher-rank U(1) gauge fields in parameter space. The lines of minimum intensity in these multifold half moons always mark the position of the pinch-point branches in the related tensorial spin liquid, which has an exciting consequence. Multifold half moons belong to the lowest excited dispersive band of the parent cluster Hamiltonian ($\lambda_{ld}$). Since the flat bands support a higher-rank Coulomb gauge theory, the lowest dispersive band supports a gapless form of the higher-rank gauge-charge excitations, which, in some models, are expected to take the form of fractons. In that sense, multifold half moons are the smoking gun for fractons in the ground state, albeit confined to the $\mathcal{Q}-$manifold vicinity. As discussed in the previous section for regular half moons, one can expect here a regime of higher-rank gauge charges of tunable density, co-existing with the vacuum of the flat bands at intermediate temperatures. And if order by disorder takes place, which is probable in three dimensions, then one could observe the crystallization of fractonic matter at very low temperature.

\begin{figure*}
    \centering
    \includegraphics[width=\linewidth]{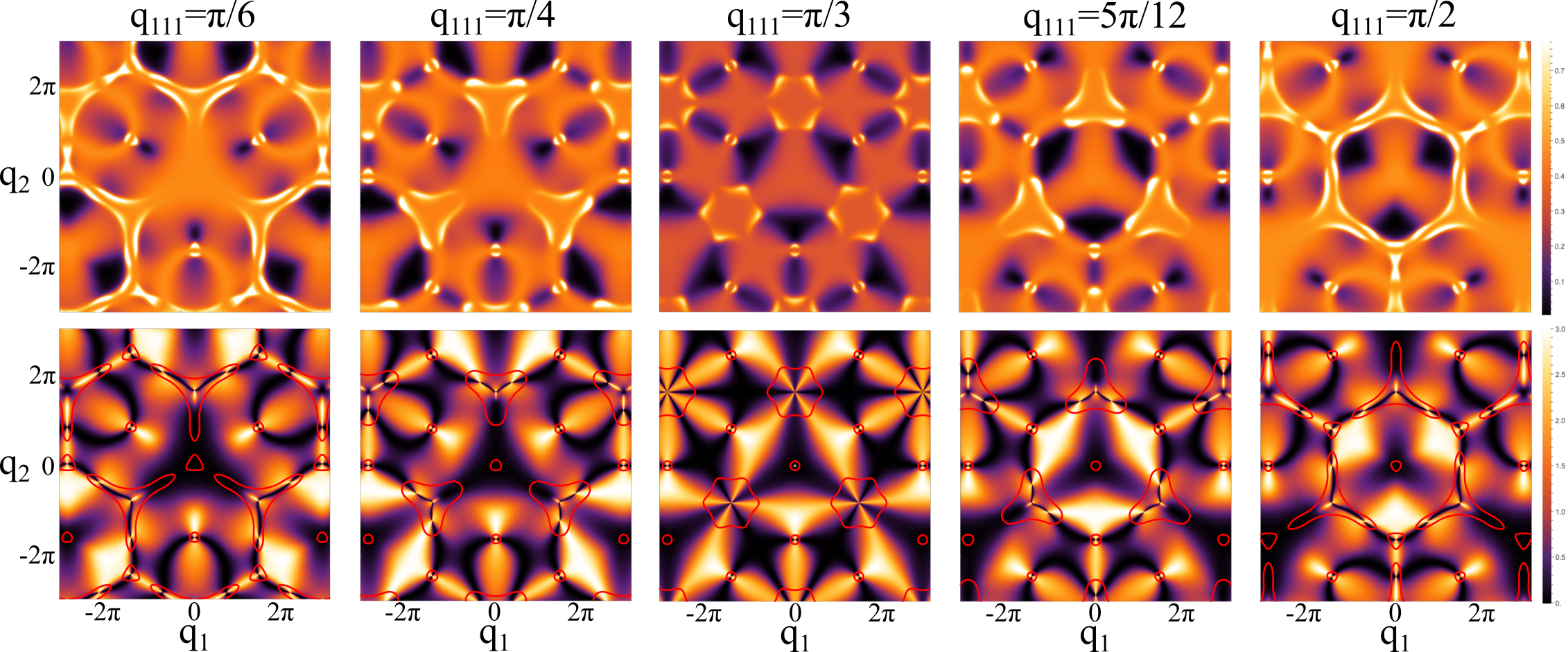} 
    \caption{
    \textbf{Generalized octochlore model:} The generalized interacting cluster Hamiltonian is set with parameters, $\gamma_1 = -1/2, \gamma_2 = 1$ to sit at the parameter point with a pinch-line singularity when $\eta=0$. In this case $\Xi = 18$ and thus $\zeta_c = -18$ suggesting to use $\zeta = -17$ to be slightly above the transition.
    The first row presents the structure factor, in good agreement with the second row representing the weight $W_2(\mathbf{q})$ of the dispersive band whose minimum is given by the red lines. 
    Each column corresponds to a different planar cut orthogonal to the [111] axis in reciprocal space: $q_1 = \mathbf{q}\cdot(1,-1,0)/\sqrt{2}$, $q_2 = \mathbf{q} \cdot (1,1,-2)/\sqrt{6}$, $q_{111} = \mathbf{q}\cdot(1,1,1)$. Calculations of the structure factors are performed using SCGA (see Appendix \ref{Appendix : SCGA}) with $\alpha =1$ and inverse temperature $\beta = 300$ and with the $q_i$ in $a^{-1}$ units.
    }
    \label{fig: octochlore half moon line}
\end{figure*}
\begin{figure}[t]
    \centering
    \includegraphics[width=0.7\linewidth]{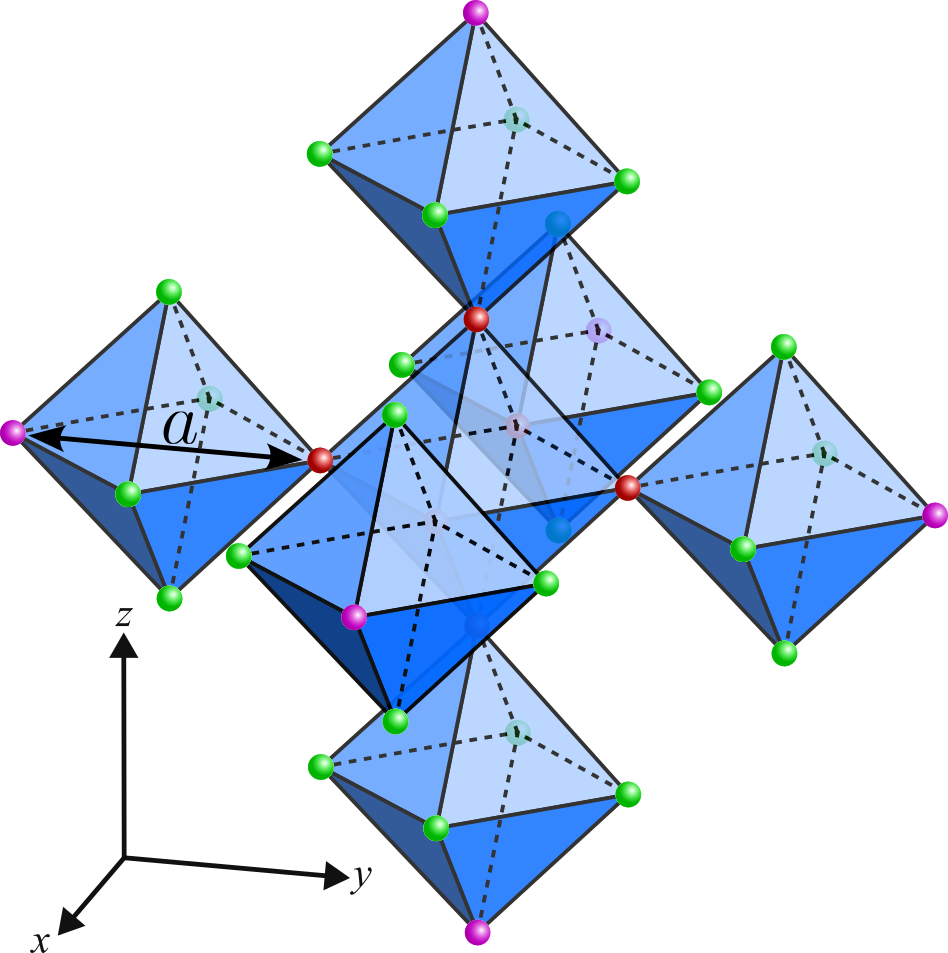}
    \caption{
    Cluster structure of the generalized octochlore model. The red dots indicate site counted with a coefficient 1 in the constrainer definition, while green and magenta dots are encapsulated with coefficients $\gamma_1$ and $\gamma_2$ (see Eq.~(\ref{Coct})).
    }
    \label{fig: generalized octochlore}
\end{figure}

\subsection{Example: higher-rank half moons on the generalized checkerboard model}
\label{Subsec: Example of the checkerboard}

The generalized checkerboard model supports several distinct spin liquids \cite{Davier_2023}, with the following constrainer
\begin{equation}
\bm{\mathcal{C}}_{\square} = \sum_{i\in\textcolor{blue}{\square}} \mathbf{S}_i + \gamma_1\sum_{i\in\textcolor{red}{\langle\square\rangle}} \mathbf{S}_i + \gamma_2 \sum_{i\in\textcolor{Goldenrod}{\langle\langle\square\rangle\rangle}} \mathbf{S}_i.
\label{Cckb}
\end{equation}
as illustrated in Fig.~\ref{fig:gencheck}. 
This model hosts fourfold pinch points in the structure factor when $\gamma_2=-1-2\gamma_1$, and sixfold pinch points for $\gamma_2=(1-2\gamma_1)/3$. It is therefore well suited to study how higher–rank pinch points evolve into half–moon patterns for $\zeta>\zeta_c$.

The extended–cluster geometry imposes $\Xi=4\,(1+2\gamma_1^2+\gamma_2^2)$ and $\omega=1$. At low temperature, for $\zeta>\zeta_c$ with $\gamma_1=-1/2$ and $\gamma_2=-1-2\gamma_1=0$, the structure factor exhibits fourfold half–moons [Fig.~\ref{fig: ckb 4f half moon}(a)]. The corresponding dispersive–band weight $W_2(\mathbf{q})$ [Fig.~\ref{fig: ckb 4f half moon}(b)] shows fourfold pinch points that are the complement of the flat–band weight $W_1(\mathbf{q})$ (see the dark regions in Fig.~\ref{fig: ckb 4f half moon}(b)). Because the dispersive band attains local minima along the $\mathcal{Q}$ manifold depicted as red contour in Fig.~\ref{fig: ckb 4f half moon}(b), those fourfold singularities lie at finite energy and are thus absent from the low–temperature structure factor. It is only the intersection between the intensity of $W_2(\mathbf{q})$ and the $\mathcal{Q}$ manifold that appears at low energy, and is thus responsible for a half-moon pattern with fourfold symmetry. Note that we also have regular twofold pinch points in this system, giving rise to regular half moons.

Tuning to $\gamma_1=1$ and $\gamma_2=(1-2\gamma_1)/3=-1/3$ produces the expected sixfold pinch points in $W_2(\mathbf{q})$ [Fig.~\ref{fig: ckb 4f half moon}(d)], which, upon intersecting with $\mathcal{Q}$, generate sixfold half–moons in the structure factor [Fig.~\ref{fig: ckb 4f half moon}(c)].

\begin{figure*}
    \centering
    \includegraphics[width=\linewidth]{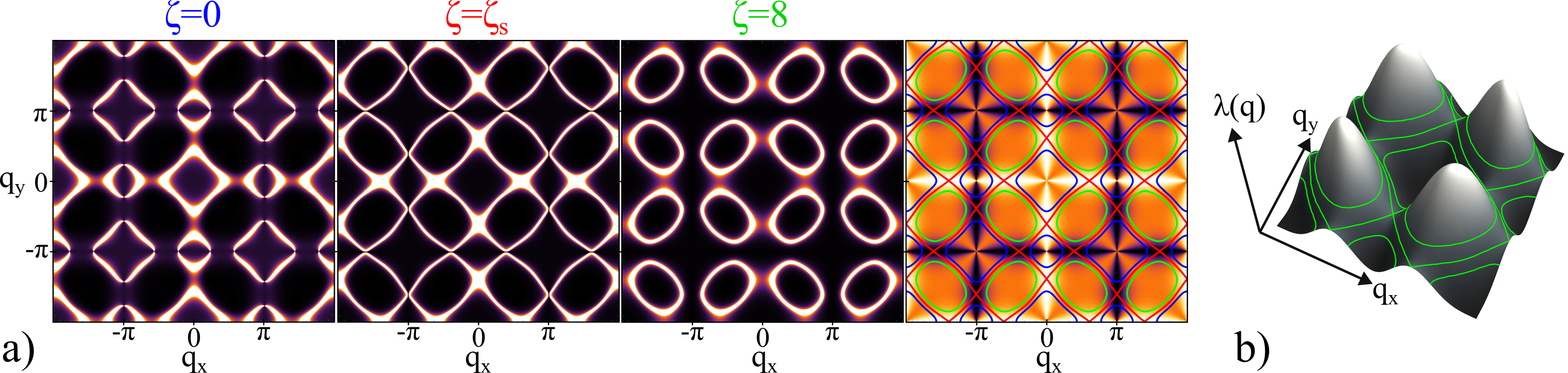} 
    \caption{
    \textbf{Generalized checkerboard model:} (a) The interacting cluster Hamiltonian is set with parameters $\gamma_1 = -1/2, \gamma_2 =0$, for varying values of $\zeta$ while fixing $\eta =1$. The right panel represents the weight $W_2(\mathbf{q})$ of the dispersive band with contours corresponding to the energy minimum of this dispersive band for $\zeta=0$ (blue), $\zeta=\zeta_s\approx 3.48$ (red) and $\zeta=8$ (green). The respective structure factors are plotted on the three left panels using SCGA (see Appendix \ref{Appendix : SCGA}) for inverse temperature $\beta=20$ and with the $q_i$ in $a^{-1}$ units.
    (b) 3D representation of $\lambda(\mathbf{q})$, with consecutive intersections with planes located at height $\lambda_c(\zeta)$. For $\zeta < \zeta_s$, the energy minima form closed line surrounding the position of pinch points located at $\Gamma$ and $M$ special points. When $\zeta$ reaches the critical value $\zeta_s$ that is such that $\lambda_c$ reaches the saddle energy, these closed lines enter in contact to form, for $\zeta > \zeta_s$, another set of lines that are centered around $\lambda$ maxima.
    }
    \label{fig:topotransition}
\end{figure*}

\subsection{Example: from pinch line to half-moon surfaces on the generalized octochlore model}
\label{Subsec: Example of the octochlore}

But not all higher-rank spin liquids take the form of multifold pinch points. In three dimensions, the singularity may extend into a one-dimensional line in reciprocal space, forming a so-called pinch line \cite{Benton2016,Niggemann_2023, Yan_2024_long,Fang_2024,Davier_2025}. How does this string singularity evolve when the interaction between clusters is strong enough to transform the flat-band gauge field fluctuations into excitations ?

Here we shall consider the generalized octochlore model \cite{Benton_Moessner_2021}, composed of corner-sharing octahedra, where the cluster is extended to encapsulate the sites of the six neighboring octahedra (see Fig.~\ref{fig: generalized octochlore}(a)). Its constrainer is defined as 
\begin{equation}
    \bm{\mathcal{C}}_{\octahedron} = \sum_{i\in\textcolor{red}{\octahedron}} \mathbf{S}_i 
    + \gamma_1 \hspace{-5pt}\sum_{i\in\textcolor{green}{\left\langle\octahedron\right\rangle}} \hspace{-5pt} \mathbf{S}_i 
    + \gamma_2 \hspace{-8pt} \sum_{i\in\textcolor{magenta}{\left\langle\left\langle\octahedron\right\rangle\right\rangle}} \hspace{-8pt} \mathbf{S}_i.
\label{Coct}
\end{equation}
as illustrated in Fig.~\ref{fig: generalized octochlore}. 
This system supports a pinch line for $\gamma_1 = -1/2, \gamma_2 = 1$ \cite{Niggemann_2023}. This parent cluster system can be turned into a generalized interacting cluster model following the procedure of Sec.~\ref{subsec: interacting cluster model for high rank h-m}. Using parameters $\gamma_1 = -1/2, \gamma_2 = 1$ imposes $\zeta_c = -\Xi/\omega = -18$. Choosing $\zeta_c = -17$ should therefore allow to observe the half moon pattern associated with the presence of a pinch line in the dispersive band weight $W_2(\mathbf{q})$.
The pinch lines on the generalized octochlore model are along the [111] axes (and equivalent ones). It means that in order to visualize the lines, one needs to look at multiple cuts orthogonal to a given [111] axis in reciprocal space. This is visible in the second row of Fig.~\ref{fig: octochlore half moon line} which presents the weight $W_2(\mathbf{q})$ of the dispersive band. Keeping in mind that these figures are approximately the ``photographic negative'' of the flat-band weight $W_1(\mathbf{q})$, the persistence of pinch points along [111] is a signature of the (negative of the) pinch line. In particular, the plane at $q_{111} = \pi/3$ is a high-symmetry plane where multiple pinch lines cross, giving rise to apparent multifold pinch points \cite{chung_2024_preprint,Davier_2025}. Since the second line of Fig.~\ref{fig: octochlore half moon line} presents the weight of the dispersive band, we can delimit the $\mathcal{Q}$ manifold of lowest energy in reciprocal space (see red contours), which corresponds to the region of high-intensity in the structure factor displayed in the first row.

When the singularity in $W_2(\mathbf{q})$ is well isolated, one sees clear half moons in the structure factor. Keeping in mind that the half moon propagates along the [111] direction, the resulting pattern is actually an elongated surface in the 3D reciprocal space, similar to a long cylinder cut in two along its length. In order to emphasize the link between these patterns and the original half moons, we shall call them ``half-moon surfaces''. However, since pinch lines can cross at specific $\mathbf{q}$ points, it means that some of the singularities are actually close to each other. At the high-symmetry point $q_{111} = \pi/3$ where they cross, we get multifold half moons, that deform continuously as one moves along the [111] axis. Looking at those patterns at $q_{111} = 5\pi/12$ and $\pi/2$, one could be forgiven not to recognize any half-moon pattern; those patterns are nonetheless the signature of the proximity, in parameter space, of extended one-dimensional singularities, and thus of the proximity of a particular form of higher-rank spin liquid.

\subsection{Topological Lifshitz transition at the edge of half-moon phases}
\label{Subsec: Topological Lifshitz transition}

For interacting–cluster systems, the post–threshold behavior for $\zeta>\zeta_c$ depends crucially on whether the parent system is gapless or gapped. We therefore discuss these two cases separately.

\subsubsection{Gapless systems}

For gapless parent models with pinch points, the system first transitions from a Coulomb spin liquid to a half–moon phase when $\zeta$ crosses $\zeta_c$ (see section \ref{Sec:unifcluster}). Then, a second, topological, transition may follow. On kagome and pyrochlore, it had been noticed that a change of the dispersive band topology could transform half–moon contours into so-called “star” patterns in the equal-time structure factor~\cite{Mizoguchi_2018,buessen16a,Lugan22b,Kiese23a}. Now we can rationalize this change in the context of the present generic theory, based on the physics of the parent Hamiltonian.

For models with a single cluster type (e.g.\ checkerboard, octochlore), one has $\lambda(\mathbf{q})\equiv|\mathbf{L}(\mathbf{q})|^2$. If the parent system dispersive band $\lambda(\mathbf{q})$ possesses a saddle point, then, as the threshold $\lambda_c(\zeta)$ rises past that saddle energy $E_s$ (see Eqs.~(\ref{Eq: Energy minima},\ref{Eq: Energy minima bis})), the topology of the ground–state manifold $\mathcal{Q}$ changes, as illustrated in Figs.~\ref{fig:Fermi_level}(b,c) and ~\ref{fig:topotransition}(b) for the generalized checkerboard lattice. The key point to understand is that the manifold $\mathcal{Q}$ varies continuously with $\zeta$. For small $\zeta$, the manifold $\mathcal{Q}$ encircles the pinch points as expected for half moons; see the blue contours on the weight $W_2(\mathbf{q})$ of the dispersive band (right panel of Fig.~\ref{fig:topotransition}(a)) and the half moons in the structure factor (left panel).

\begin{figure*}
    \centering
    \includegraphics[width=\linewidth]{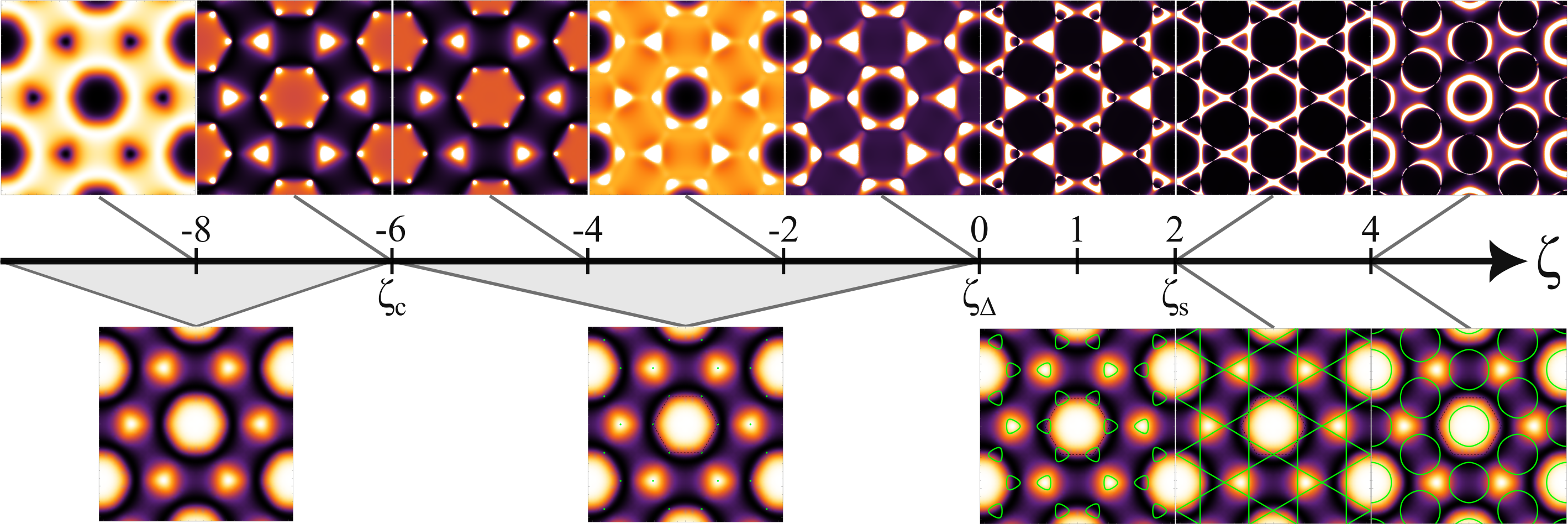} 
    \caption{
    \textbf{Hexagonal checkerboard model:} The top line contains structure factors computed for different values of $\zeta$ using the SCGA (see Appendix \ref{Appendix : SCGA}) with $\eta = 1$ and inverse temperature $\beta = 100$. The bottom line shows the dispersive band weight $W_3(\mathbf{q})$ together with the ground state manifold $\mathcal{Q}$ depicted as green dots or lines. In both cases the plots are obtained with $q_x, q_y \in [-2\pi, 2\pi]$ in inverse units of a cluster edge. The structure factors show four different phases as increasing $\zeta$. For $\zeta < \zeta_c$ the structure factor is similar to the one of the non interacting system, presenting no pinch point as the band structure is gapped. When $\zeta$ becomes larger than $\zeta_c$ the flat bands cease to be the ground state, which is now constituted of the points $\mathbf{q}_\Delta$ for which $\lambda(\mathbf{q}_\Delta) = \Delta$. The structure factor thus depicts a selection of the dispersive band weight around these minima, producing bright regions increasing when $\zeta$ approaches $\zeta_\Delta$. At this point $\lambda_c(\zeta)$ reaches the value $\Delta$, and the ground state manifold becomes an hypersurface, producing bright lines in the structure factor. When $\zeta$ finally reaches $\zeta_s$ the structure of the ground state manifold change, corresponding to a Lifshitz transition, that appears in the structure factor as the bright lines changing of pattern.
    }
    \label{fig: gapped topo transition}
\end{figure*}
As $\zeta$ increases, the contour of $\mathcal{Q}$ gets bigger and bigger, until eventually neighboring contours touch each other at a critical value $\zeta_s$ 
\begin{eqnarray}
\lambda_c(\zeta_s) = \frac{\Xi}{2}\left(1-\frac{\zeta_s}{\zeta_c}\right) = E_s 
\Leftrightarrow \zeta_s=\zeta_c\left(1-\frac{2 E_s}{\Xi}\right)
\end{eqnarray}
Since we find that $E_s>\Xi/2$ in our model, we have $\zeta_s>0$; more precisely, $\zeta_s\approx 3.48$ \footnote{As a side note, the positivity of $\zeta_s$ was found on all models we have tried (not shown here), which suggests a possible relation between the saddle-point energy $E_s$ and $\Xi$.}. Contours touching at the saddle point are depicted as red lines in the right panel of Fig.~\ref{fig:topotransition}(a) and will give rise to stronger soft-mode excitations at the saddle-point wavevectors. If not protected by the Hohenberg-Mermin-Wagner theorem, one can expect order by disorder at those wavevectors for $\zeta=\zeta_s$.

For $\zeta>\zeta_s$, the contours disconnect from each other, surrounding a different region in reciprocal space, away from the pinch-point singularity; see the green contours in the right panel of Fig.~\ref{fig:topotransition}(a) and the corresponding structure factor for $\zeta=8$. One gets a distinct spiral spin liquid without half moons, but with closed rings. Since the manifold $\mathcal{Q}$ does not encircle a pinch point anymore, it does not intersect with its branches, and the weight $W_2(\mathbf{q})$ of the dispersive band has thus no reason to cancel along $\mathcal{Q}$.\\

Making an analogy between the contour formed by manifold $\mathcal{Q}$ and the Fermi surface in the electronic structure of a material, we have here the equivalent of a topological Lifshitz transition \cite{Lifshitz60a,Balla19a}. At the level of the band structure, there is no symmetry breaking between the two phases, but simply a change of the topology of the contour. This Lifshitz transition thus delimits two distinct regions of the phase diagram that can be distinguished from a clear change in the structure factor. If the parent $\lambda(\mathbf{q})$ hosts multiple minima and saddle points, multiple Lifshitz transitions can occur as $\zeta$ is increased. 


\subsubsection{Gapped systems}
\label{subsubsec : gapped phase}

While pinch points and half moons are coming from gapless band spectrum, this is not necessary for Lifshitz transitions. What is necessary though, is that the minimum energy manifold $\mathcal{Q}$ forms a hypersurface in reciprocal space; this way, this hypersurface can change its topology (closing around distinct $\mathbf{q}$ points) across the phase diagram. As discussed in section \ref{subsec: interacting cluster model for high rank h-m}, not all interacting-cluster models respect that condition. But many do, such as the family of generalized interacting-cluster systems (see section \ref{subsec: interacting cluster model for high rank h-m}) which, when restricted to nearest-neighbor cluster interaction, includes the uniform cluster models of section \ref{Sec:unifcluster}. 

Let us consider a parent cluster system with a finite energy gap $\Delta>0$ at $\mathbf{q}_\Delta$; hence, $\lambda_{ld}(\mathbf{q}_\Delta)=\Delta$. The gap requires a stronger interaction $\eta$, i.e. a higher value of $\zeta=\zeta_c(\Delta)>\zeta_c$, for the dispersive band to make contact with the flat band (see Eq.~(\ref{eq:zetacdelta})). But the main difference with gapless parent systems is that, by definition, there is no fixed point $\mathbf{q}^\star$ where $\lambda_{ld}(\mathbf{q}^\star)=0$ would have imposed $\Lambda_{ld}(\mathbf{q}^\star)=0$ (see Eqs.~(\ref{Eq: Dispersion H(2)}) and (\ref{Eq: Dispersion H(2) bis})). As $\zeta$ increases above $\zeta_c(\Delta)$, the dispersive band $\Lambda_{ld}(\mathbf{q})$ simply goes through the flat band and preserves its minimum energy at $\mathbf{q}_\Delta$ since
\begin{eqnarray}
    \bm{\nabla}_\mathbf{q} \lambda_{ld}(\mathbf{q}_\Delta) = 0 \quad\Rightarrow\quad \bm{\nabla}_\mathbf{q} \Lambda_{ld}(\mathbf{q}_\Delta) = 0.
\end{eqnarray}
The ground–state “manifold’’ is therefore a finite set of points in reciprocal space; we expect the system to long-range order at wave-vector $\mathbf{q}_\Delta$.

But Eq.~(\ref{eq:gradlambda}) is still valid and the manifold $\mathcal{Q}$ -- as defined by Eqs.~(\ref{Eq: Energy minima}) and (\ref{Eq: Energy minima bis}) -- remains an extremum of the dispersive band $\Lambda_{ld}(\mathbf{q})$. The gap simply shifted its domain of existence to larger values of $\zeta$
\begin{eqnarray}
    \lambda_c&=&\frac{\Xi}{2}\left(1-\frac{\zeta}{\zeta_c}\right)\geq\Delta\nonumber\\
    \Leftrightarrow \zeta&\geq& \zeta_\Delta=\zeta_c\left(1-\frac{2\Delta}{\Xi}\right) =\zeta_c(\Delta)- \zeta_c\frac{\Delta}{\Xi}\\
    \Leftrightarrow \zeta&\geq& \zeta_\Delta> \zeta_c(\Delta) \nonumber
\end{eqnarray}
When $\zeta=\zeta_\Delta$, we have $\lambda_c=\Delta$ and the manifold $\mathcal{Q}$ includes all $\mathbf{q}_\Delta$ points. By analytic continuation, as $\zeta$ increases, manifold $\mathcal{Q}$ will form a closed hypersurface around $\mathbf{q}_\Delta$ points. This is the same reasoning as before for the apparition of half moons, except that the $\mathbf{q}_\Delta$ points are not attached to pinch points, and the resulting patterns in equal-time structure factor will be model dependent. 

Remarkably, as $\zeta$ increases from $-\infty$ (parent cluster model) to positive values, the ground state of the model starts as a fragile topological spin liquid (using the terminology of Ref.~\cite{Yan_2024_long}), goes through a nonsingular band touching at $\zeta_c(\Delta)$, then orders magnetically at wavevector $\mathbf{q}_\Delta$ up to $\zeta_\Delta$ where it becomes a spiral spin liquid. And if the lowest dispersive band of the parent system $\lambda_{ld}$ has one (or more) saddle point(s), a Lifshitz transition(s) shall occur when $\lambda_c(\zeta)$ exceeds a saddle energy $E_s$, separating distinct spiral spin liquids.\\

As an illustration, let us consider the hexagonal kagome model from Table~\ref{tab: cluster systems}, known to be gapped~\cite{Davier_2023,Yan_2024_long}. Choosing the simplest constrainer
\begin{equation}
    \bm{\mathcal{C}}_{\hexagon}=\sum_{i\in\hexagon}\mathbf{S}_i,
\end{equation}
places the model in the class satisfying conditions (i)–(iii), with $\Xi=\xi=6$, and, since the lattice is corner–sharing, $\Omega=\omega=1$. For this system $\lambda_c = 3\left(1+\zeta/6\right)$ and the three relevant thresholds for $\zeta$ are thus
\begin{equation}
        \zeta_c = -6, \qquad
        \zeta_\Delta = 0, \qquad 
        \zeta_s = 2,
\end{equation}
as $\Delta = 3$ and $E_s = 4$.
The three successive regimes are clearly visible in the structure factors of Fig.~\ref{fig: gapped topo transition}. 
For $\zeta<\zeta_c$ the structure factor closely matches that of the non-interacting parent model ($\eta=0$), showing only smooth features as the ground state is the flat–band manifold that contains no singularities.
For $\zeta_c<\zeta<\zeta_\Delta$, bright spots appear and broaden as $\zeta \to \zeta_\Delta$, reflecting that the ground state now consists only of the points $\mathbf{q}_\Delta$ located at the $K$ high–symmetry points in this model.
At $\zeta = \zeta_\Delta$, the ground–state set becomes a family of triangular contours surrounding the $K$ points, producing triangular bright lines in the structure factor. 
These contours grow with $\zeta$ until they touch at the $M$ points when $\zeta=\zeta_s$, after which they reconnect into circular contours encircling the maxima of $\lambda$ at the $\Gamma$ points for $\zeta>\zeta_s$. 
These circular contours are expected to shrink into bright regions, as when $\zeta$ becomes large enough so that $\lambda_c(\zeta)$ equates the maxima of the dispersive band $\lambda$, the ground state manifold $\mathcal{Q}$ becomes the set of points $\mathbf{q}_m$ such that $\lambda(\mathbf{q}_m) = \mathrm{Max}\left[\lambda(\mathbf{q})\right]$.

\section{Discussion}

Somewhat counter-intuitively, the tools recently developed for the classification of classical spin liquids \cite{Benton_Moessner_2021, Yan_2024_short, Yan_2024_long, Davier_2023,Fang_2024} prove useful to explore the physics of systems beyond spin liquids. In order to destabilize the spin-liquid ground state, we used methods of flat-band engineering coming from graph theory, in particular connectivity matrices that are well suited to account for interaction between spin clusters \cite{katsura10a,Essafi_2017,Mizoguchi_2018,Mizoguchi_Masafumi_2019_flat_bands}. The two frameworks, connectivity matrix and constrainers of spin liquids, are actually two faces of the same coin (see Eq.~(\ref{eq:hL}) and Ref.~\cite{Fang_2024}).

Here we have built a generic theory for multifold half moons, encompassing the evolution of pinch lines into extended surfaces as signatures of spiral spin liquids whose magnetic texture hosts higher-rank gauge charges in the ground state. The manifold of minimum energy, forming a closed contour in reciprocal space, behaves as an effective Fermi surface of the parent noninteracting system and undergoes topological changes at Lifshitz transitions, when the dispersive band of the parent cluster Hamiltonian possesses saddle points. In most, if not all, three-dimensional models, thermal order by disorder will take place at very low temperature and long-range order the system. As was the case on pyrochlore \cite{Mizoguchi_2018}, this magnetic ordering can be seen as the crystallization of the gauge charges that populate the dispersive band. In the case of higher rank spin liquids, one might thus be able to observe a crystallization of fractons. When the parent cluster Hamiltonian is gapped, half moons disappear but spiral spin liquids and Lifshitz transitions remain.

The present work is primarily aimed at establishing a generic and unifying theoretical framework for interacting cluster systems, rather than proposing specific microscopic Hamiltonians tailored for immediate experimental realization. In particular, the cluster--cluster interactions required for the emergence of half-moon patterns may effectively generate couplings between sites at relatively long distances, which render the models discussed here challenging to realize in conventional magnetic materials. Nevertheless, cluster-based descriptions are a well-established and physically motivated route for stabilizing classical spin liquids, and the mechanisms identified in this work are expected to apply broadly to any system whose low-energy physics admits an effective cluster formulation.

The present framework offers an infinite variety of theoretical models supporting a dense phase of gapless gauge charges at intermediate temperature, stabilized by cluster interaction.
Their non-local dynamics is an open question, and their connection with local momentum vortices found in spiral spin liquids \cite{Yan_2022} -- equivalent to quadrupoles of fractons -- is an appealing direction to investigate. The investigation of half moons in quantum spins $S=1/2$ has given interesting physics on kagome \cite{buessen16a,Lugan22b,Kiese23a} and can now be applied to higher-rank and pinch-line spin liquids.

\textit{Acknowledgements:} 
The authors thank Paul McClarty and J\'er\^ome Cayssol for useful discussions. They acknowledge financial support from Grant No. ANR-23-CE30-0038-01.


\appendix

\section{Band structure and structure factor relation}
\label{Appendix : Band structure and Gauss laws}

Both the Hamiltonian matrix $H^{(1)} = h^{v \leftarrow c}\left[h^{v \leftarrow c}\right]^\dagger$ and the matrix $M = \left[h^{v \leftarrow c}\right]^\dagger h^{v \leftarrow c}$ ruling the projector into the flat bands manifold are built from the same connectivity matrix $h^{v \leftarrow c}$. This implies a deep connection between the band structure and the structure factor that we detail in this Appendix.

\subsection{Equivalence between contact point and pinch point}
\label{Appendix C.1}
Here we show that for cluster Hamiltonians there is an equivalence between the presence of a band touching between dispersive and flat bands, and the presence of pinch point in the static structure factor. The rectangular matrix $h^{v \leftarrow c}$ admits a singular value decomposition 
\begin{equation}
    h^{v \leftarrow c} = U \Sigma V^\dagger.
\end{equation}
where $U$ is an $n_s \times n_s$ unitary matrix, $V$ is an $n_c \times n_c$ unitary matrix and $\Sigma$ is an $n_s \times n_c$ diagonal matrix of singular values $\sigma_\nu$, 
\begin{equation}
\Sigma = 
    \begin{bmatrix}
        \sigma_1 & 0 & \dots & 0 \\
        0 & \sigma_2 & \dots & 0 \\
        \vdots & \vdots & \ddots & \vdots \\
        0 & 0 & \dots & \sigma_{n_c} \\
        0 & 0 & \dots & 0 \\
        \vdots & \vdots & \ddots & \vdots \\
        0 & 0 & \dots & 0
    \end{bmatrix}.
\end{equation}
This expression allows to get that 
\begin{equation}
    H^{(1)} = U \Sigma \Sigma^\dagger U^\dagger, \qquad M = V \Sigma^\dagger \Sigma V
\end{equation}
where $\Sigma \Sigma^\dagger$ is the diagonalized Hamiltonian, composed of $n_c$ trivial eigenvalues and $n_s-n_c$ dispersive eigenvalues $\lambda_i \equiv (\sigma_i)^2$, and $\Sigma^\dagger \Sigma$ is the diagonal matrix with diagonal elements $\lambda_i$. This imposes that the determinant of the matrix $M$, that is equal to the product of its eigenvalues, is equal to the product 
\begin{equation}
    \det M(\mathbf q) = \prod_{i=1}^{n_c}\lambda_i(\mathbf q)
\end{equation}
of the dispersive bands dispersions $\lambda_i(\mathbf{q})$. Hence, at a contact point $\mathbf q^\star$ where the lowest dispersive band vanishes, $\det M(\mathbf q^\star)=0$ and the flat-sector projector
\begin{equation}
    \Pi = I - h^{v \leftarrow c} M^{-1} \left[h^{v \leftarrow c}\right]^\dagger
\end{equation}
becomes singular. This nonanalyticity appears in the equal-time structure factor as a
pinch-type singularity. Thus in cluster systems any contact point in the structure factor induces a pinch point in the static structure factor.

\subsection{Contact points in cluster models are necessarily singular}
\label{Appendix C.2}
The matrix $M(\mathbf q)=h^\dagger h$ is the Gram matrix of the constraint
vectors $\{\mathbf L_X(\mathbf q)\}_{X=1}^{n_c}$ (columns of $h$):
\begin{equation}
    M_{XY}(\mathbf q)=\mathbf L_X(\mathbf q)^*\!\cdot \mathbf L_Y(\mathbf q).
\end{equation}
A basic property of Gram matrices is that $\det M(\mathbf q)=0$ if and only if
the family $\{\mathbf L_X(\mathbf q)\}$ loses linear independence, i.e.\ the
span of dispersive bands eigenvectors drops rank. Consequently, any gap closing (contact
point) requires a rank drop, which can occur only if either
\begin{itemize}
\item[(i)] a constraint vector vanishes: $\mathbf L_X(\mathbf q^\star)=0$, or
\item[(ii)] a nontrivial linear combination vanishes:
there exists a linear combination $\mathbf{L}_c(\mathbf{q})$ of the constraint vectors $\mathbf{L}_X(\mathbf{q})$ such that $\mathbf{L}_c(\mathbf{q}^\star) = 0$.
\end{itemize}
Both mechanisms are \emph{singular} in the sense that some dispersive bands eigenvectors (or
linear combination thereof) vanishes at $\mathbf q^\star$. Therefore, in the
cluster class, every contact point is singular.

The same mechanism can produce singular gap closings along
higher-dimensional manifolds in momentum space (pinch lines/planes), corresponding
to rank loss on sets of codimension $1$ or higher. Note that in these cases this does not correspond to the emergence of a pinch point in the structure factor if the dimension of the contact manifold is the system dimension minus one\cite{Davier_2023}, but that it does always imply the existence of non localized flat band states\cite{Yan_2024_long, Davier_2023}.

\subsection{Order of the Gauss law and contact point dispersion}
\label{Appendix C.3}

Assume first that a single constrainer vanishes at $\mathbf q^\star$:
$\mathbf L_c(\mathbf q^\star)=0$. Let $V(\mathbf q)$ be the span of the
remaining constraint vectors at $\mathbf q$, and $P_V$ the orthogonal projector onto
$V(\mathbf q)$. The Gram determinant factorizes as
\begin{equation}
\det M(\mathbf q)=
    \bigl\|\mathbf L_c(\mathbf q)-P_V\mathbf L_c(\mathbf q)\bigr\|^2\;
\det M_{V}(\mathbf q),
\end{equation}
where $M_V$ is the Gram matrix on $V(\mathbf q)$. If the first nonvanishing
term in the multivariate Taylor expansion of
$\mathbf L_c(\mathbf q)$ about $\mathbf q^\star$ has total degree $n$,
\begin{equation*}
    \mathbf L_c(\mathbf q)=\mathbf\Phi\,\bigl(\delta\mathbf q\bigr)^{(n)}+\mathcal O\!\bigl(\|\delta\mathbf q\|^{\,n+1}\bigr),
    \qquad \delta\mathbf q=\mathbf q-\mathbf q^\star,
\end{equation*}
then
$\bigl\|\mathbf L_c(\mathbf q)-P_V\mathbf L_c(\mathbf q)\bigr\|^2=\mathcal O(\|\delta\mathbf q\|^{2n})$
while $\det M_V(\mathbf q)=\mathcal O(1)$. Thus
\begin{equation}
\det M(\mathbf q)=\mathcal O\!\bigl(\|\delta\mathbf q\|^{2n}\bigr).
\end{equation}

More directly, because $M(\mathbf q)$ is positive semidefinite,
the Courant–Fischer theorem gives
\begin{equation}
\lambda_1(\mathbf q)=\min_{\|\mathbf x\|=1}\mathbf x^\dagger M(\mathbf q)\mathbf x.
\end{equation}
At $\mathbf q^\star$ one has $\bm\alpha^\dagger M(\mathbf q^\star)\bm\alpha=0$
for $\bm\alpha=e_c$ (the unit vector selecting $\mathbf L_c$), so near
$\mathbf q^\star$ the minimizer remains $\bm\alpha$ to leading order, and
\begin{equation}
\lambda_1(\mathbf q)=\bm\alpha^\dagger M(\mathbf q)\bm\alpha
=\bigl\|\mathbf L_c(\mathbf q)\bigr\|^2
=\mathcal O\!\bigl(\|\delta\mathbf q\|^{2n}\bigr).
\end{equation}
Hence the lowest dispersive band has local dispersion of order $2n$, which
matches a Gauss-law of differential order $n$ (Maxwell for $n=1$, higher-rank
for $n>1$).

Now consider the linear-dependence mechanism (ii). There exists a unit vector
$\bm\alpha\in\mathbb C^{n_c}$ such that
\begin{equation}
\mathbf L_c(\mathbf q)\equiv h^{v \leftarrow c}(\mathbf q)\bm\alpha
=\mathbf\Phi\,\bigl(\delta\mathbf q\bigr)^{(n)}+\mathcal O\!\bigl(\|\delta\mathbf q\|^{\,n+1}\bigr).
\end{equation}
Since $M=h^\dagger h$,
\begin{equation}
\lambda_1(\mathbf q)
\le \bm\alpha^\dagger M(\mathbf q)\bm\alpha
=\bigl\|\mathbf L_c(\mathbf q)\bigr\|^2
=\mathcal O\!\bigl(\|\delta\mathbf q\|^{2n}\bigr),
\end{equation}
and positivity implies this is the actual leading scaling. Thus, again, the
lowest dispersive band softens with order $2n$, and the emergent Gauss law is
of order $n$.

\emph{Degenerate contact point.} If the rank drops by $r>1$ at $\mathbf q^\star$,
then $r$ independent combinations $\mathbf{L}_c^{(a)}$ vanish simultaneously, and
the $r$ lowest eigenvalues soften. The product of these $r$ eigenvalues scales
as $\mathcal O\!\bigl(\|\delta\mathbf q\|^{2(n_1+\cdots+n_r)}\bigr)$, with each
$n_a$ the lowest total degree for the corresponding critical combination. The
single band touching case above is recovered for $r=1$.

\emph{Visibility in $S(\mathbf q)$.} While form-factor weights can redistribute
intensity and obscure a pinch point at special symmetry points, the singular
character (rank loss) of the projector is unaffected. Hence the contact
point~$\Longleftrightarrow$~pinch-type singularity equivalence remains valid.

\section{Self Consistent Gaussian Approximation}
\label{Appendix : SCGA}

Consider any isotropic Heisenberg Hamiltonian that can be expressed as
\begin{equation}
    \begin{split}
        \mathcal{H} &= \frac{1}{2}\sum_{i,j} H_{i,j} \mathbf{S}_i \cdot \mathbf{S}_j \\ 
        &=\frac{1}{2N}\sum_\mathbf{q} \sum_{\mu, \nu} H_{\mu \nu}(\mathbf{q}) \mathbf{S}_\mu(\mathbf{q}) \cdot \mathbf{S}_\nu(-\mathbf{q})
    \end{split}
\end{equation}
with $N$ the number of unit cells. Further consider there are $n_s$ sublattices, and denote the $n_s$ eigenvalues and eigenvectors from the Hamiltonian $H(\mathbf{q})$ as
\begin{equation}
    \varepsilon_\kappa(\mathbf{q}), \qquad \bm{\psi}_\kappa(\mathbf{q}).
\end{equation}
The self consistent Gaussian approximation\cite{Garanin_SCGA} corresponds to take into account the spin length constraint $|\mathbf{S}_i| = 1$ only in average, enforcing only the constraint $\langle |\mathbf{S}_i| \rangle = 1$. Using that the three spin components are equivalent, the constraint that is really enforced in practice is
\begin{equation}
    \langle (S_i ^\alpha)^2 \rangle = \frac{1}{3}.
\end{equation}
This can be done using a single Lagrange multiplier $\lambda$ to enforce the constraint in the Hamiltonian, while enforcing the hard constraint for every spin would amount to introduce $Nn_s$ Lagrange multipliers. Also shifting the energy origin by defining $\varepsilon_0 =$ Min$(\varepsilon_\mu)$ and considering the positive semi definite matrix $K(\mathbf{q}) = H(\mathbf{q}) - \varepsilon_0 I$, the Hamiltonian can finally be expressed as 
\begin{equation}
    \mathcal{H} = \frac{1}{2N}\sum_\mathbf{q} \left( K_{\mu \nu}(\mathbf{q}) + \frac{\lambda}{\beta} \delta_{\mu \nu} \right) \mathbf{S}_\mu(\mathbf{q}) \cdot \mathbf{S}_\nu(\mathbf{q}).
\end{equation}
The eigenvalues of $K$ are equal to the ones of $H$ but shifted of $\varepsilon_0$ and are therefore positive, while its eigenvectors are simply identical. For a semi-positive definite matrix $A$ the general formula
\begin{equation}
    \frac{\int \prod_i  dx_i \; x_\mu x_\nu e^{-\frac{1}{2} \mathbf{x}^t A \mathbf{x}}}{\int \prod_i  dx_i e^{-\frac{1}{2} \mathbf{x}^t A \mathbf{x}}} = \left(A^{-1}\right)_{\mu \nu}
\end{equation}
can be used to write that, defining $M = \beta K + \lambda I$,
\begin{equation*}
    \begin{split}
        \langle S^\alpha_\mu(-\mathbf{q})  S^\alpha_\nu(\mathbf{q}) \rangle &= \frac{\int \prod_{i,}  dS_i^\alpha \; S^\alpha_\mu S_\nu^\alpha e^{-\frac{ 1}{2 N} \sum_{\mathbf{q}'}\sum_{\eta, \xi} M_{\eta \xi} S_\eta^\alpha  S_\xi^\alpha }} 
    {\int \prod_{i}  dS_i^\alpha e^{-\frac{ 1}{2 N} \sum_{\mathbf{q}'}\sum_{\eta, \xi} M_{\eta \xi} S_\eta^\alpha  S_\xi^\alpha }} \\
    &= N \left(M^{-1}\right)_{\mu \nu} = N \left[ \lambda I + \beta K(\mathbf{q})\right]^{-1}_{\mu\nu}.
    \end{split}
\end{equation*}
This allows to self consistently fix the Lagrange multiplier $\lambda$ by enforcing the constraint 
\begin{equation}
    \begin{split}
        \langle (S_i ^\alpha)^2 \rangle &= \langle S_i^\alpha S_i^\alpha \rangle 
    = \frac{1}{N} \sum_i\langle S_i^\alpha S_i^\alpha \rangle \\
    &= \frac{1}{N^2} \sum_\mathbf{q} \sum_{\mu = 1}^{n_s} \langle S^\alpha_\mu(-\mathbf{q})   S^\alpha_\mu(\mathbf{q}) \rangle \\
    &= \frac{1}{N}\sum_\mathbf{q} \sum_{\mu = 1}^{n_s} \left[ \lambda I + \beta K(\mathbf{q})\right]^{-1}_{\mu\mu} 
    = \frac{1}{3},
    \end{split}
    \label{Eq: self eq for lambda}
\end{equation}
where we used that all spins in the system are equivalent by symmetry to write the first line. This also allows to express the static structure factor as 
\begin{equation*}
    \begin{split}
        \mathcal{S}(\mathbf{q}) &= \frac{1}{3N}\sum_{\mu, \nu} \langle \mathbf{S}_\mu(-\mathbf{q}) \cdot \mathbf{S}_\nu(\mathbf{q}) \rangle 
        = \frac{1}{N}\sum_{\mu, \nu} \langle S^\alpha_\mu(-\mathbf{q}) \cdot S^\alpha_\nu(\mathbf{q}) \rangle \\
        &= \sum_{\mu, \nu} \left[ \lambda I + \beta K(\mathbf{q})\right]^{-1}_{\mu\nu} 
        = \sum_{\kappa = 1}^{n_s}\sum_{\mu, \nu} \frac{\left[\psi_\kappa ^*(\mathbf{q})\right]_\mu \left[\psi_\kappa(\mathbf{q})\right]_\nu }{\lambda + \beta \varepsilon_\kappa(\mathbf{q})}
    \end{split}
\end{equation*}
because the three spin components are equivalent, and then the spin components can be treated as three independent variables. 

In the limit of zero temperature $\beta \to \infty$, only flat bands contributions are selected. In this case the static structure factor becomes simply proportional to the sum of the components of the projector into flat band manifold 
\begin{equation}
    \begin{split}
        \Pi_{\mu \nu} &= \sum_{\kappa = 1}^{n_\text{f.b}} \left[\psi_\kappa ^*(\mathbf{q})\right]_\mu \left[\psi_\kappa(\mathbf{q})\right]_\nu \\ 
        &= I_{\mu \nu} - \sum_{\kappa = n_\text{f.b} + 1}^{n_s} \left[\psi_\kappa ^*(\mathbf{q})\right]_\mu \left[\psi_\kappa(\mathbf{q})\right]_\nu.
    \end{split}
\end{equation}
As the manifold spanned by the dispersive bands eigenvectors $\bm{\psi}_\kappa, \; \kappa > n_\text{f.b}$ is identical to the one spanned by the constraint vectors, this projector can be re-expressed as in Eq.~(\ref{Eq : Projector Pi at zero temperature}).

\section{Relevance of the GLT method for clusters systems}

\label{Appendix:LTA}

For cluster systems, being able the express the Hamiltonian as a sum of squares 
\begin{equation}
    \begin{split}
        \mathcal{H} &= \sum_{i,j} H_{ij} \, \mathbf{S}_i \cdot \mathbf{S}_j \\
        &= \sum_{n,X} |\bm{\mathcal{C}}_{n,X}|^2 - N_{u.c}\sum_X \sum_{i \in X} \left(\gamma_i^X \right)^2
    \end{split}
    \label{Eq: appendix cluster Hamiltonian}
\end{equation}
corresponds in fact to identify that the Langrange multipliers allowing to find a real space physical ground state are $\gamma_i^X$, as this allows to express the Hamiltonian as a positive definite term which is minimized by configurations satisfying the ground state constraint $\bm{\mathcal{C}}_{n,X} = 0 \;\forall \; n,X$. Fourier transforming this Hamiltonian leads to 
\begin{equation}
    \begin{split}
        \mathcal{H} =  \sum_{\mathbf{q}} \sum_X \sum_{\mu, \nu} \left[h_\mu^X(\mathbf{q}) \mathbf{S}_\mu(\mathbf{q}) \right] \cdot \left[h_\nu^X(-\mathbf{q}) \mathbf{S}_\nu(-\mathbf{q}) \right] \\
        - \lambda N_{u.c} \sum_\mu \alpha_\mu^2
    \end{split}
    \label{Ex: appendix H in function of h(q)}
\end{equation}
where we defined the parameters $\lambda, \alpha_\mu$ such that 
\begin{equation}
    \lambda \sum_\mu \alpha_\mu^2 = \sum_X \sum_{i \in X} \left( \gamma_i^X\right)^2
    \label{Eq: choice of alpha_mu}
\end{equation}
where $h^X_\mu(\mathbf{q})$ is the $\mu$ component of Fourier transform of the constrainer $\bm{\mathcal{C}}_{0,X}$ as explained in the main text. Under this form the Hamiltonian corresponds to the Generalized Luttinger Tisza (GLT) method \cite{kaplan_2007_LTA}. This procedure encodes a soft spin length constraint 
\begin{equation}
    \sum_{n,\mu} \alpha_\mu^2 |\mathbf{S}_{n,\mu}|^2 = \sum_\mathbf{q} \sum_\mu \alpha_\mu^2 |\mathbf{S}_\mu(\mathbf{q})|^2 = N_{u.c} \sum_\mu \alpha_\mu^2,
\end{equation}
thanks to a unique Lagrange multiplier $\lambda$, associated with a set of parameters $\alpha_\mu$, specific to the different sublattices, that can be cleverly chosen by a analyzing the symmetries of the problem. For a generic system, encoding this soft constraint would lead to a Hamiltonian
\begin{equation}
    \begin{split}
        \mathcal{H} =  \sum_{\mathbf{q}} \sum_{\mu, \nu} \left( H_{\mu \nu}(\mathbf{q}) - \alpha_\mu \alpha_\nu \delta_{\mu \nu} \right) \mathbf{S}_\mu(\mathbf{q}) \cdot \mathbf{S}_\nu(-\mathbf{q}) \\
        + \lambda N_{u.c} \sum_\mu \alpha_\mu^2.
        \label{Eq: LTA Hamiltonian}
    \end{split}
\end{equation}
Therefore, solving the ground state would amount to diagonalize the matrix
\begin{equation}
    F_{\mu \nu} = \frac{1}{\alpha_\mu \alpha_\nu } H_{\mu \nu}(\mathbf{q})
\end{equation}
as the configuration associated with its lowest eigenvalue will naturally minimize the Hamiltonian (\ref{Eq: LTA Hamiltonian}). Finally, one would need to check the resulting configuration fulfills the strong condition $|\mathbf{S}_i| = 1$ to ensure the built state is indeed a physical state. 

For cluster systems, however, the story is different. Indeed, because the real-space structure of the Hamiltonian imposes a macroscopic number of ground-state configurations, one should not expect a unique configuration associated with the lowest eigenvalue of $F$, but rather a macroscopic number, since the lowest eigenvalues of $F$ form a flat band in reciprocal space. This macroscopic degeneracy originates from local constraints encoded in the cluster structure. Consequently, for these systems, the appropriate choice of parameters $\alpha_\mu$ is given by Eq.~(\ref{Eq: choice of alpha_mu}) and corresponds to selecting parameters such that the matrix $F$ admits flat bands at the bottom of its spectrum. In this setting, checking that a single obtained configuration satisfies the strong constraint $|\mathbf{S}_i|=1$ should be replaced by demonstrating that there exists a macroscopic family of superpositions of flat-band states that fulfill this constraint. Since flat bands are generically associated with the existence of compact localized states~\cite{Rhim_2019, Yan_2024_long}, these can typically be superposed to form physical ground-state configurations, even in systems with inequivalent sublattices. This scenario has been confirmed, for example, by Monte Carlo simulations for the centered pyrochlore lattice~\cite{nutakki23b}.

The GLT method aims to determine the ground state of spin systems, but it does not address the excited energy bands. When discussing the upper part of the spectrum of the Hamiltonian matrix, one should therefore refer to the large-$\mathcal{N}$ approximation, in which the strong constraint $|\mathbf{S}_i|=1$ is relaxed and the spin degrees of freedom are treated as scalar fields. While, for the reasons discussed above in the context of the GLT, this approximation provides excellent results for studies focusing on the properties of the ground-state manifold, there is no \emph{a priori} reason to consider it reliable for higher, dispersive energy bands. However, in the particular case of spiral spin liquids discussed in the present manuscript, the ground-state manifold obtained within the large-$\mathcal{N}$ approximation also displays a large degeneracy, suggesting again the existence of superpositions of these states that realize physical ground states. This explains the excellent agreement between Monte Carlo simulations and large-$\mathcal{N}$ predictions in previous observations of half-moon patterns for interacting cluster versions of the kagome and pyrochlore lattices~\cite{Mizoguchi_2018}. Since the ground-state manifold realizes a cut of the non-interacting band-system band structure, such an agreement indicates that the band structure obtained within the large-$\mathcal{N}$ approximation is reliable, at least for systems with equivalent sublattices. A complementary way to understand the success of the large-$\mathcal{N}$ approximation for cluster spin liquid systems is that the extensive entropy associated with the massively degenerate ground-state manifold allows for strong thermal fluctuations even at low temperature, thereby washing out small discrepancies between the true ground state and the states predicted within the approximation when considering observables such as the structure factor.

\section{General case of non equivalent interacting clusters }

\label{Appendix:Generalized_cluster_systems}

The main text derivation of Sec. \ref{Sec:unifcluster} can in fact be generalized to the case of cluster systems composed of inequivalent types of clusters and mixing different type of geometrical intersections between clusters, as for example the two versions of the square kagome lattice depicted in Table. \ref{tab: cluster systems}. This requires to generalize the Hamiltonian (\ref{Eq: general H(2) Cm + Cn}) to allow for more complex cluster-cluster connectivity, setting it as
\begin{equation}
    \mathcal{H} = \alpha \sum_n|\bm{\mathcal{C}}_n|^2 + 2\eta \sum_{\langle m,n \rangle} \Gamma_{mn} \bm{\mathcal{C}}_m\cdot \bm{\mathcal{C}}_n.
\end{equation}
This Hamiltonian can also be well expressed in the framework of connectivity matrices, defining the connectivity matrix $h^{c \leftarrow c}$ as a $N_c \times N_c$ connectivity matrix linking cluster centers to centers of neighboring clusters. Its coefficients $h^{c \leftarrow c}_{mn}$ are equal to $\Gamma_{mn}$ if the clusters $m$ and $n$ are neighboring clusters, and zero otherwise. Now one can choose the cluster-cluster links weights $\Gamma_{mn}$ such that 
\begin{equation}
        h^{c \leftarrow v} h^{v \leftarrow c} = h^{c \leftarrow c} + \Xi I_{N_c} .
\end{equation}
This corresponds in fact to represent the link between a pair of clusters $(m,n)$ as a sum of the different two move paths going from the cluster center-$m$ to a vertex shared by the clusters $m$ and $n$ and then to the center of the cluster $n$. With this construction the Hamiltonian matrix expresses as 
\begin{equation}
    H =  \left(\alpha - \eta  \Xi \right) H^{(1)} + \eta \left(H^{(1)}\right)^2
\end{equation}
and the band structure of the interaction cluster model can therefore be expressed as 
\begin{equation}
    \Lambda_\mu(\mathbf{q}) = \left(\alpha - \eta \Xi \right) \lambda_\mu(\mathbf{q}) + \eta \lambda_\mu(\mathbf{q})^2. 
\end{equation}
For systems satisfying the set of conditions (i) to (iii) the present choice cluster-cluster interaction would correspond to encapsulate the coefficient $\Omega$ directly within $\eta$.

\bibliographystyle{apsrev4-2}
\bibliography{Biblio}

\end{document}